# On the internal architecture of lightweight negative Poisson's ratio (auxetic) metastructures: a review

Ali Rahimi-Lenji[1], Mohammad Heidari-Rarani[1,*], Mohsen Mirkhalaf[2,*], Mohammad Mirkhalaf[3]

[1]Department of Mechanical Engineering, Faculty of Engineering, University of Isfahan, Isfahan, Iran

[2]Department of Physics, University of Gothenburg, Gothenburg, Sweden

[3]School of Mechanical, Medical & Process Engineering, Queensland University of Technology, Brisbane, Australia

[*]Corresponding authors:

m.heidarirarani@eng.ui.ac.ir (M. Heidari-Rarani)

mohsen.mirkhalaf@physics.gu.se (M. Mirkhalaf)

**Abstract**

Development of lightweight materials with enhanced mechanical properties has been a long-standing challenge in science and engineering. Lightweight auxetic metastructures (AMSs) provide attractive solutions to this problem. AMSs' ngative poisson's ratio is a unique characteristic which results in very interesting practical properties such as high energy absorption and improved toughness. Different properties of metastrcutures, including anisotropy, are dependent, in addition to their original material, on the unit cell shape and geometrical features which have very high variations. Over the past few years, researchers have developed AMSs with various unit cells, either introducing new internal architecture or enhancing and optimizing the existing ones. Despite the progress made in this field, there is no comprehensive review of the AMS unit cells. This review describes the cellular AMSs associated with a classified set of metastructures, comprising more than 100 distinct auxetic repetitive unit cells. Besides, achievable range of negative Poisson's ratios are provided for different categories of AMSs. Finally, the future perspective of the research field and potential developments and applications in this field are discussed.

## 1. Introduction

Decreasing weight while maintaining mechanical properties has been of great interest in the science and engineering of structural materials [1]. Over the past 2-3 decades, tailoring the internal architecture of materials has led to interesting pathways to achieve this goal in a paradigm commonly identified as mechanical metamaterials or mechanical metastructures [2–6]. We prefer using the term mechanical metastructure in this work, to emphasize that the unit cell size in the constructs discussed here is not necessarily orders of magnitude smaller than the size of the component [7,8]. Auxetic structures are a subclass of metastructures with a negative Poisson's ratio (NPR) [9], leading to other interesting properties and functionalities. Materials in the elastic region contract in a transverse direction when loaded in uniaxial tension and expand transversely when compressed – a phenomenon known as Poisson's effect [10]. The same elastic material can be architectured in certain ways to show an opposite behavior [11]: expansions transversely under tension and contraction transversely under compression [12,13], a behavior leading to an "NPR" [14].

The NPR was initially noted in natural materials like tendons, ligaments, and certain spongy structures. Subsequent studies recognized the potential of these materials or materials/structures with similar internal architectures in areas such as impact protection [15,16], energy absorption, and vibration damping [17]. When pressure is applied to an auxetic metastructure, it moves inward instead of outward from the periphery, requiring more force and energy for crushing to progress [8,18,19]. Likewise, in tension and the crack tip, the metastructure expands [20,21], requiring additional force and energy to deform the meta-structure further or advance the crack tip [22,23]. Both experimental and theoretical approaches have shown the greater toughness and energy absorption of AMS compared to conventional meta-structures with the same density [24–33] Because of the increase in the absorption of specific energy in these structures, there is a significant focus on these structures today [34–40]. The development of advanced modeling techniques based on artificial intelligence and 3D/4D printing technologies has significantly increased the speed of research and innovation in the field of AMSs.[41–43]. With the increasing development of AMS using 3D/4D printing techniques, researchers have studied various aspects of the physics of these metastructures such as energy absorption, vibrations, thermal, and optical behaviors. However, a significant challenge in 3D/4D printing is yet low consistency and high prices compared to mass production techniques such as molding [44].

The recent reviews on the metastructures cover diverse internal architectures, ranging from regularly repeating unit structures to non-repeating configurations with unusual behaviors that can manifest peculiar and advantageous responses when subjected to various stimuli such as force, heat, magnetism, electromagnetism, and vibration[45,46]. For example, thermal metastructures that combine negative thermal expansion with the NPR can contract when heated while simultaneously expanding laterally. This unique coupling of thermal and mechanical responses paves the way for adaptive materials with enhanced performance in applications like thermal management and energy absorption [47,48]. Examples include reviews on tubular AMS [49], and the applications of these metastructures [50]. However, there is currently no review article rationally classifying auxetic metastructure based on their internal architectures or providing a comparative analysis of Poisson's ratio and energy absorption capacity. This article addresses this gap in the literature and provides researchers in the field with a comprehensive overview of the AMS with a single unit cell, Poisson's ratio, and auxetic behavior. The status of AMS among metastructures is stated first. More than 100 auxetic cell models are then examined and reviewed by categorizing their not-varying unit cells. Our logical classification of AMS based on geometry can be a basis for creating a common point of view for research, development, and industrial translation in this field.

## 2. Classification of auxetic structures

A conventional and widely employed method for categorization involves extracting various features under examination and organizing them based on commonalities and differences within the subject [51,52]. Auxetic structures can similarly be categorized through different criteria, such as geometry, auxetic behavior mechanism, or fabrication methods. Classifying based on the mechanism of auxetic behavior may lead to confusion due to the greater differences than similarities among structures, requiring numerous defined categories, and may ultimately lack effectiveness. Similarly, classifying based on manufacturing methods may not sufficiently differentiate between various categories and may lack guiding principles [53].

Another approach involves classifying based on how unit cells are assembled, including two-dimensional (2D), three-dimensional (3D), cylindrical, etc, and in the categories, 2D and 3D arrangements have been tried in each section. However, this classification can be more comprehensive when combined with the geometry of single cells. Consequently, an effective classification for auxetic structures is based on their geometry. The classification identifies fundamental geometries and incorporates developed geometries within those categories. Essentially, auxetic structures are either composed of two or more basic auxetic structures or represent developments of those structures. Each category can further include subcategories, such as 2D normal structures, 2D hierarchical structures, 3D normal structures, and 3D hierarchical structures, as suggested by Yin et. al. [54]. Typical 3D structures may be truss-based, shell-based, or plate-based. In general, AMSs can be divided into two groups: metastructures with repetitive unit cells and non-repetitive (disordered/irregular) AMS. The main focus of this study is on repetitive metastructures. The fundamental auxetic structures with repetitive unit cells encompass various categories as shown in Fig 1:

a) Re-entrant honeycomb

b) Star-shaped

c) Arc-shaped

d) Arrow

e) Chiral

f) Peanut

g) Rotating rigid body

h) Origami

In each section of AMS, the background of the development of each structure and the initial idea of that structure have been discussed. In this discussion, the reason for the classification of auxetic metastructure based on the geometry of the unit cell is based on the attitude of creating each auxetic metastructure, then the primary unit cell of each category, which is the fundamental representative of those AMS, has been examined in detail. In the following, the timeline of initial research and continued geometric developments is presented. In the following, auxetic behavior and developments are discussed.

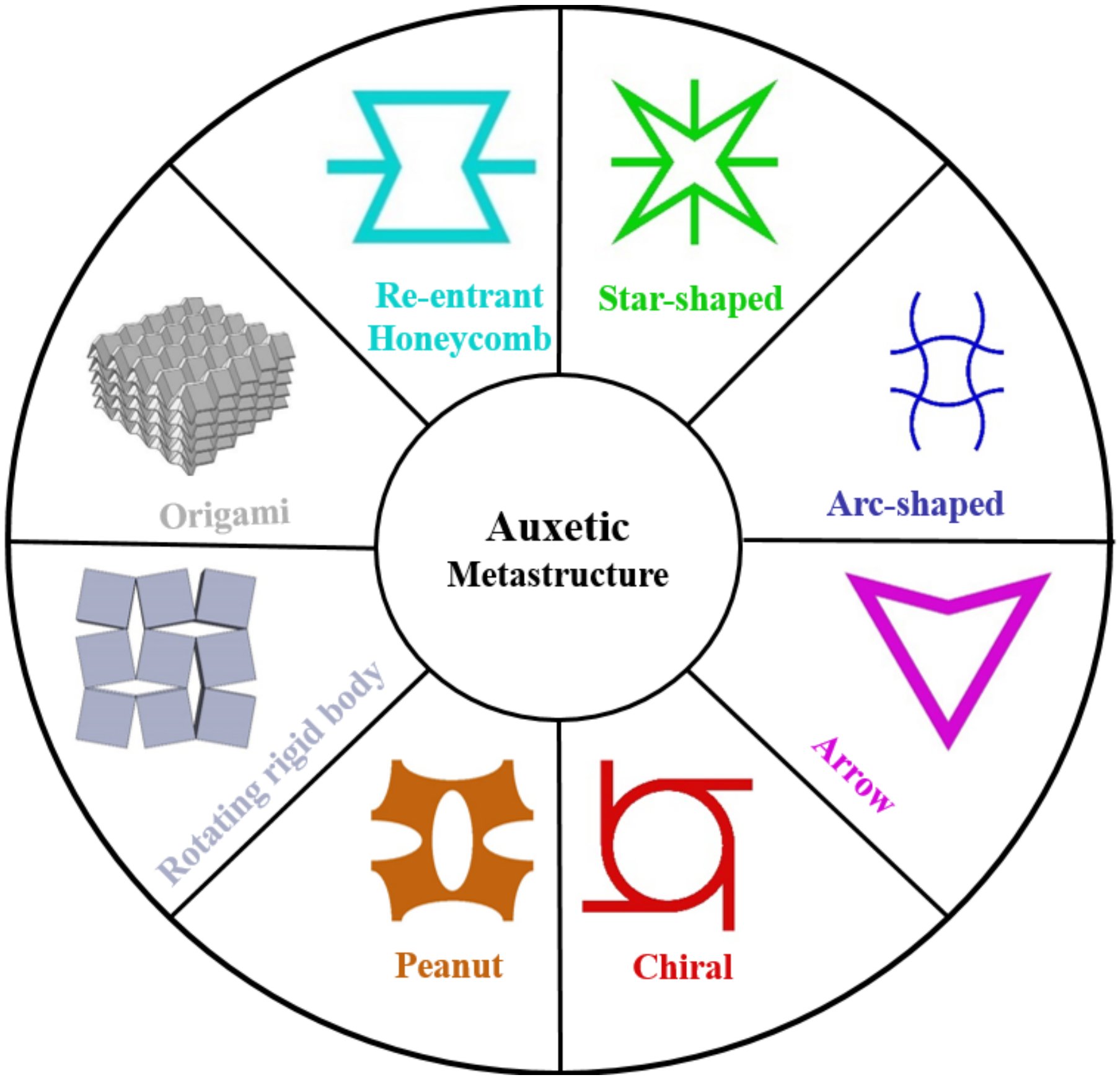

Fig. 1 Eight different AMS categories.

### *2.1. Re-entrant honeycomb*

The re-entrant honeycomb structure is inspired by the traditional honeycomb structure. The traditional honeycomb structure, which consists of regular hexagons, has almost zero Poisson's ratio. The re-entrant honeycomb structure was first introduced by Almgren in 1985 [55]. Almgren developed an isotropic 3D structure with a Poisson's ratio of –1, using a network of rods, hinges, and springs, achieving unique mechanical properties such as uniform compressive and tensile strain responses while preventing shear deformation. After him, Lakes in 1987 [56] experimentally demonstrated an isotropic 3D structure with a Poisson's ratio of –1, showcasing enhanced energy absorption, indentation resistance, and mechanical stability under deformation. Friis et al. [57] developed polymeric and metallic foams with an NPR, based on a structure similar to a re-entrant honeycomb, achieving enhanced energy absorption, impact resistance, and mechanical performance under compression and tension. In the 2000s, research expanded. In 2005, Kimizuka [58] discovered that α-cristobalite exhibits auxetic behavior under high pressure, enhancing its mechanical properties such as an NPR, anisotropic elasticity, and structural stability in extreme conditions. In 2014, Mir et al. [59] reviewed the auxetic structures developed up to that time. Most of their work was focused on re-entrant honeycomb structures. This shows that most investigations in AMS have been on re-entrant honeycombs. This structure is known by various names, including auxetic honeycomb, re-entrant honeycomb, hourglass honeycomb, and bowtie honeycomb. The initial concept behind this structure shown in Fig. 2 involved guiding the two sides of the traditional honeycomb inwards to create the re-entrant honeycomb.

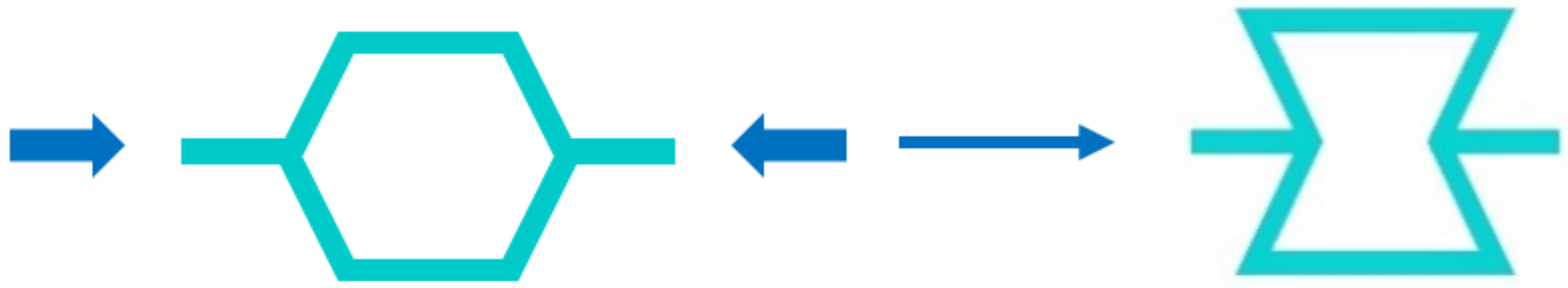

Fig. 2 The main idea of re-entrant honeycomb metastructures

The development ideas of re-entrant honeycomb metastructure have been based on different perspectives [60]. As shown in Fig. 3, one of the methods of re-entrant honeycomb development and providing new metastructures is based on adding different links in the re-entrant honeycomb structure and making graded re-entrant honeycomb structures. Another common method is to change the links between two metastructures and create a new order in the structure. Another way is to add a metastructure next to the re-entrant honeycomb. Another method is the development of hierarchical structures where a re-entrant honeycomb unit cell is made of several grid unit cells. Furthermore, by changing the shape of the links and replacing the curved links, new metastructures have been developed.

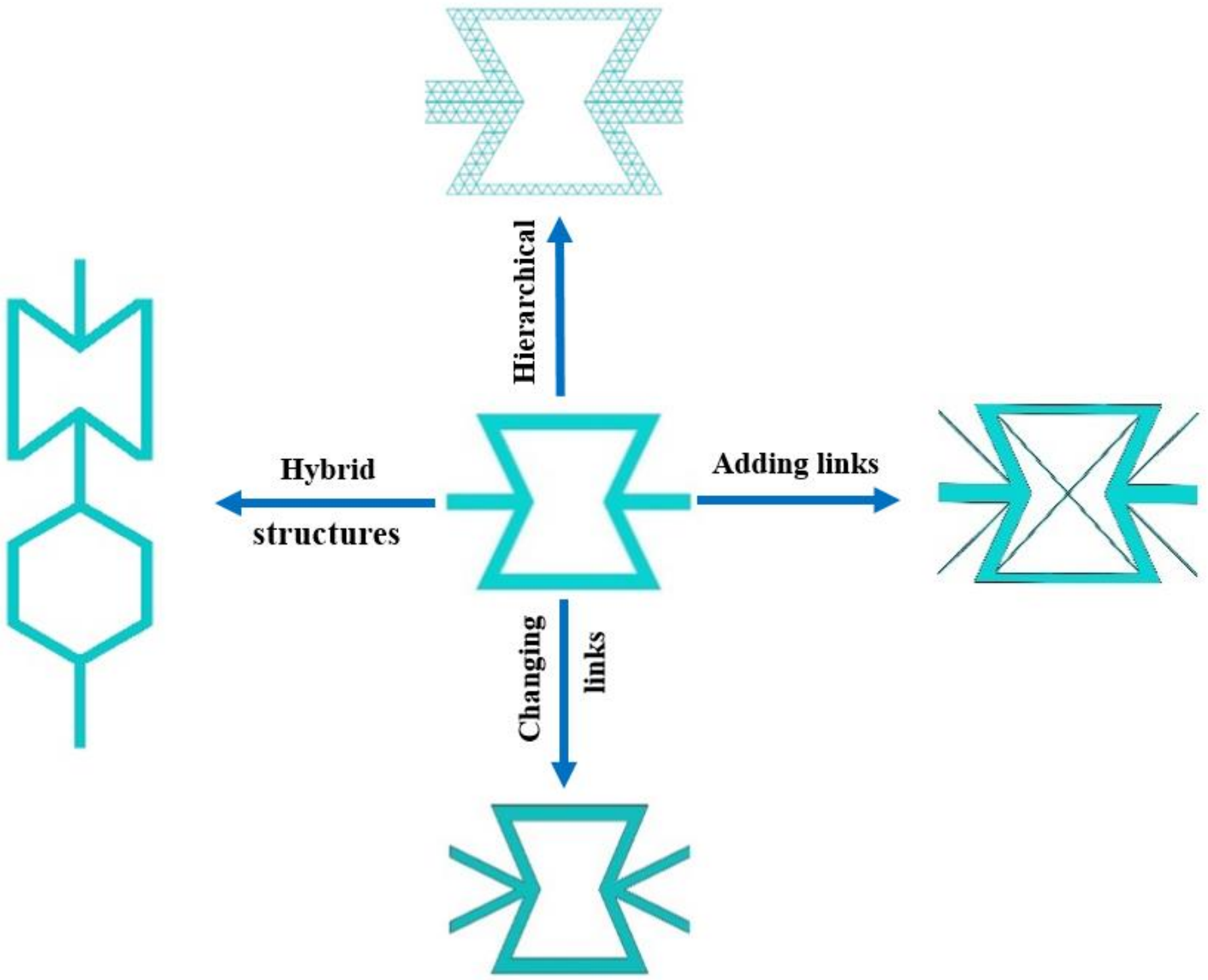

Fig. 3 Developing ideas of re-entrant honeycomb metastructures

For a better understanding of the auxetic behavior of the re-entrant honeycombs metastructure, a schematic of the behavior of this metastructure under axial crushing is presented in Fig. 4. It is worth noting that this behavior is schematic and the collapse behavior of AMS depends on the type, value, and rate of loading as well as the manufacturing conditions of metastructure them.

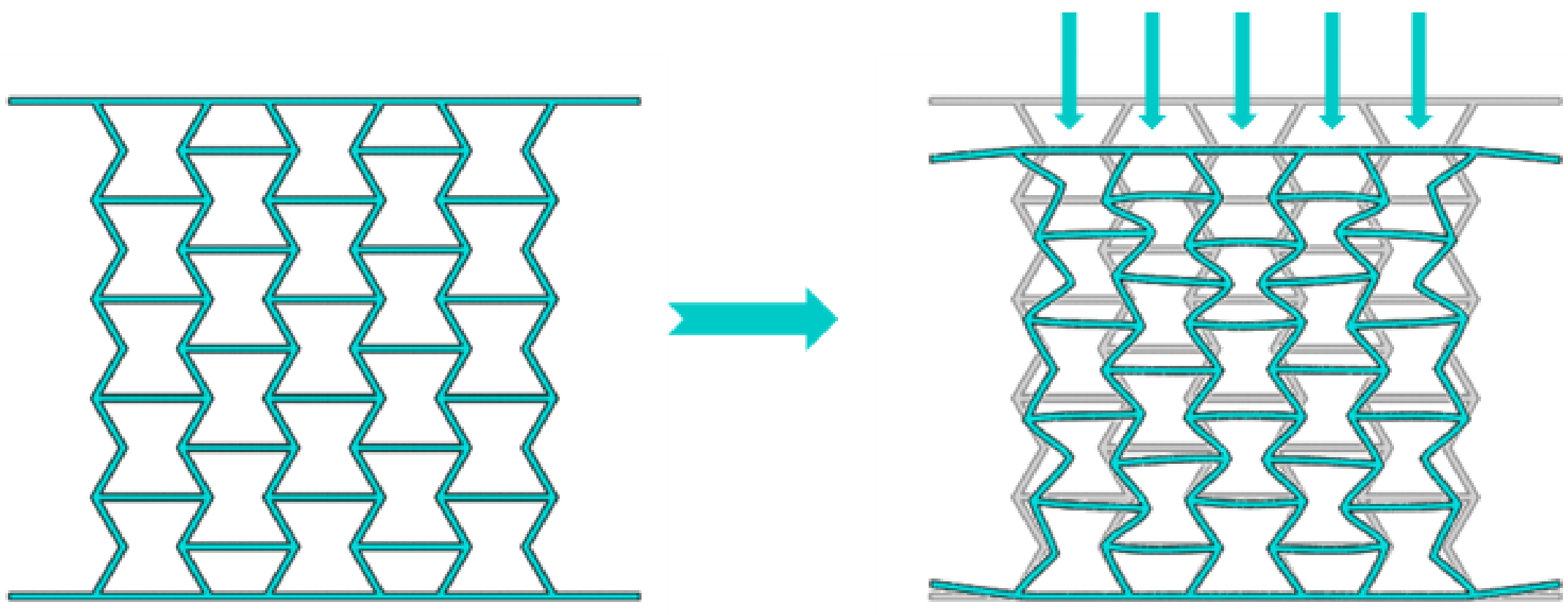

Fig. 4 Schematic of a re-entrant honeycomb structure under axial compression.

Auxetic re-entrant honeycomb structures exhibit outstanding mechanical performance, particularly in energy absorption and efficiency. Specific energy absorption (SEA) for re-entrant honeycombs ranges between 25–30 kJ/kg for aluminum-based materials under quasi-static compression, significantly higher than traditional honeycomb structures with SEA values averaging 15–20 kJ/kg [61]. Energy absorption efficiency under dynamic loading can exceed 90%, far surpassing the 70% efficiency typically observed in conventional honeycombs [62]. Crushing stress values for these structures range between 5–10 MPa, depending on material type and relative density, highlighting their ability to withstand high-impact forces [63]. Additionally, the durability of re-entrant honeycombs is notable, with more than 80% of their initial energy absorption capacity retained after multiple compression cycles [64]. By including a detailed analysis of these parameters, the manuscript can provide a more comprehensive understanding of the energy-absorbing capabilities of AMS, further enhancing its scientific contribution [59].

During the last decade, research in this field expanded greatly. More than thirty unit cells of the re-entrant honeycomb structure, which are given in Table 1, prove much attention to this structure in recent years. Scientists have delved into various aspects of this structure, exploring mechanical properties [65–79], impact resistance [71,80–89], energy absorption [65,90–101], and bending behavior especially in sandwich panels [102–111], fatigue performance [112–116], proposing new improved geometries [117–136], geometry optimization [137–142], failure characteristics [143–145], vibration absorption [146–149], finite element analysis and methods [150–160], different load types [161,162], testing the mechanical properties of structures with different materials [163–179], design based on application [180–186] and 4D printing of re-entrant honeycombs [187]. The main share of research results in recent years has been on Poisson's ratio. Poisson's ratio of 0.3 to 0 is reported in the literature for the hexagonal honeycomb structure. For traditional re-entrant honeycomb structures, the Poisson's ratio was reported between -0.1 and -1. However, the NPR values in recent research for re-entrant honeycomb structures developed are mostly in the range of -0.25 to -0.4, and this is a more accurate range for the NPR performance of this metastructure. In Table 1, the types of introduced and developed unit cells of this structure have been collected and exhibited so far.

Table 1 Classification of re-entrant honeycomb metastructures.

| Structure name | Model dimension | Unit cell | Aggregate of unit cells | Reference |
|---|---|---|---|---|
| Re-entrant honeycomb | 2D | 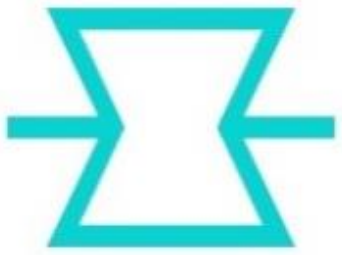 | 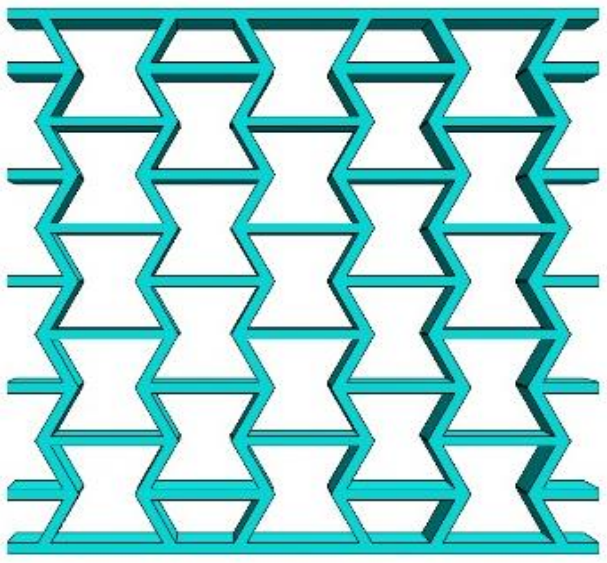 | |
| Re-entrant rhombic honeycomb | 2D | 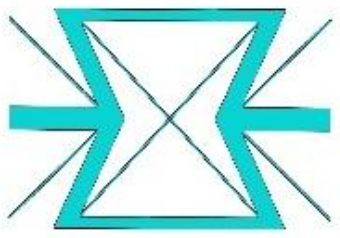 | 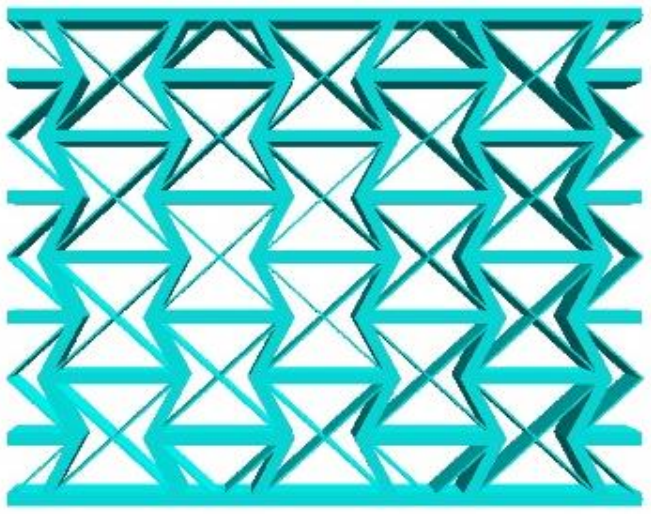 | [188] |
| Re-entrant star-shaped honeycomb | 2D | 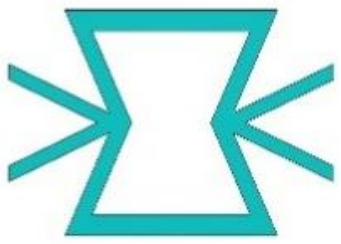 | 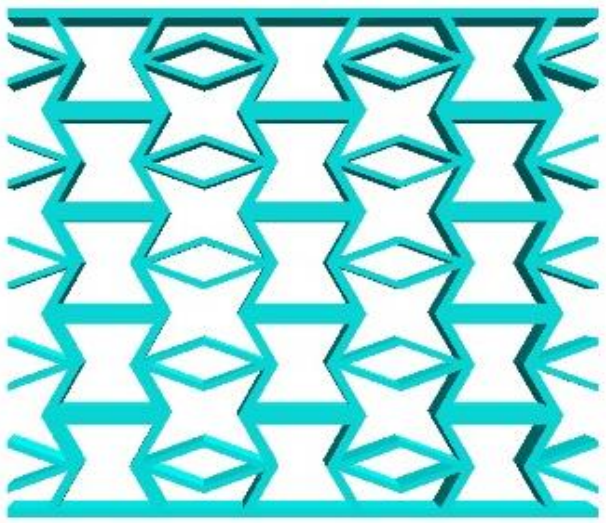 | [189] |
| Optimized re-entrant honeycomb | 2D | 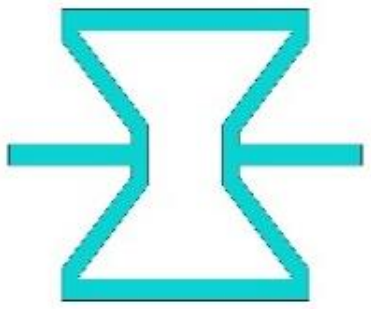 | 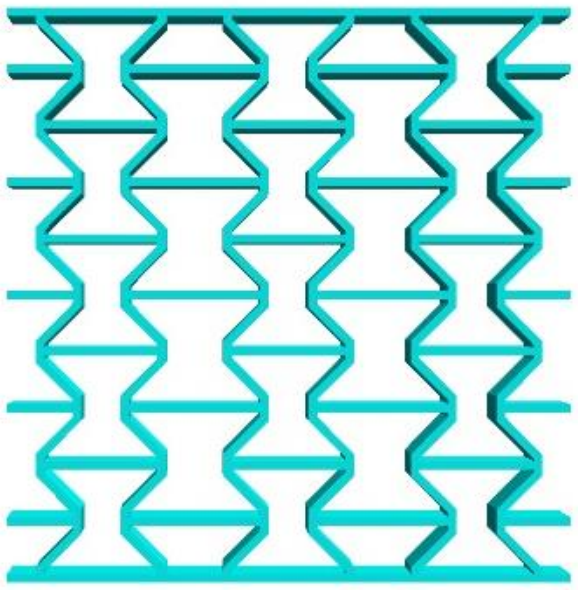 | [190] |
| Hybrid re-entrant honeycomb | 2D | 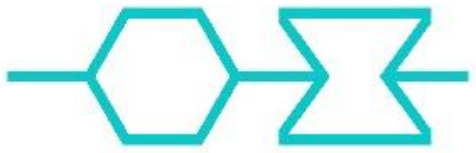 | 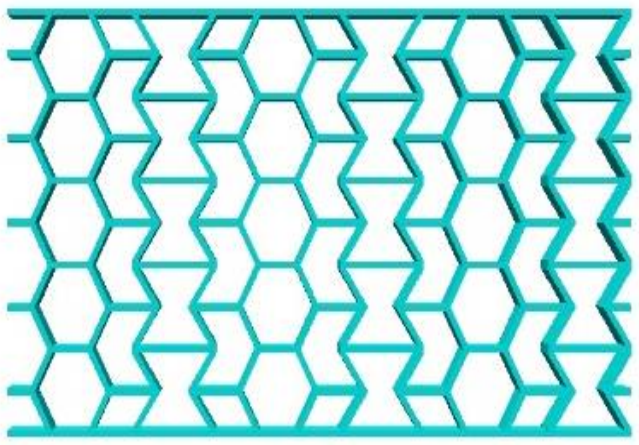 | [191] |
| Developed re-entrant star shaped honeycomb | 2D | 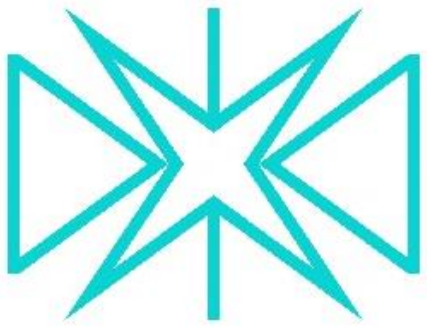 | 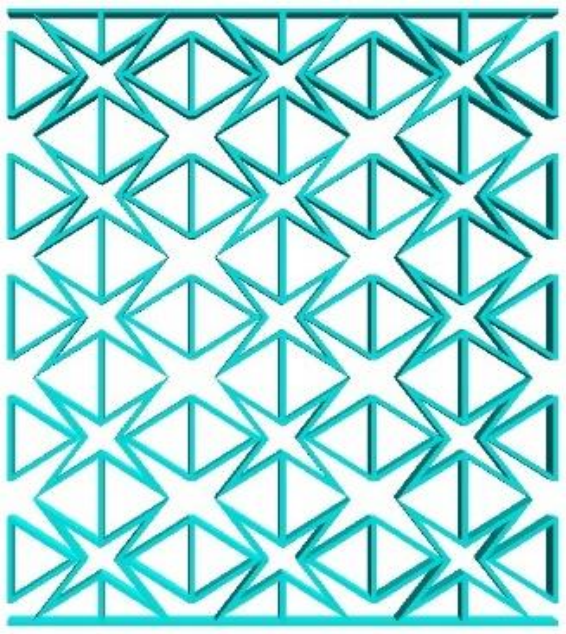 | [192] |
| Optimized re-entrant honeycomb | 2D | 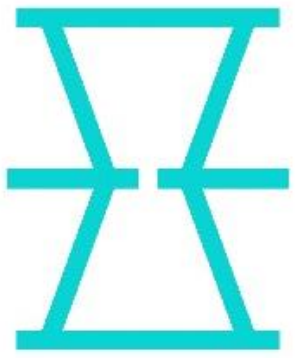 | 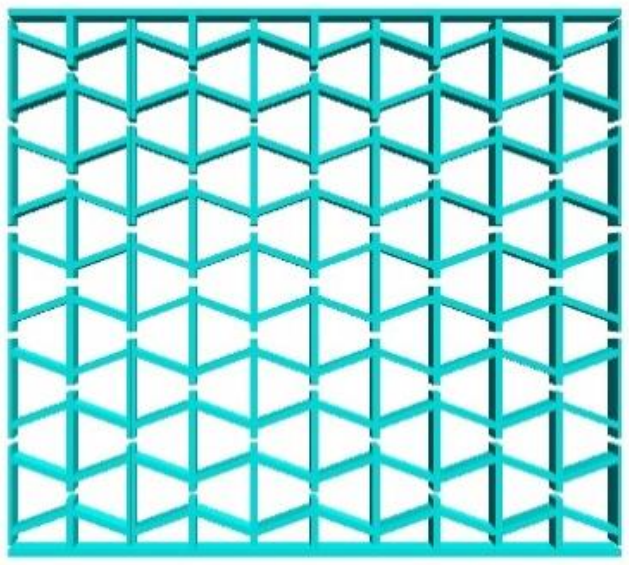 | [193] |
| None re-entrant honeycomb | 2D | 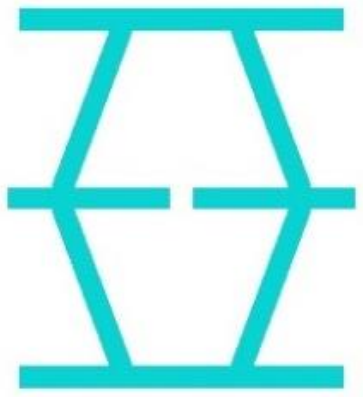 | 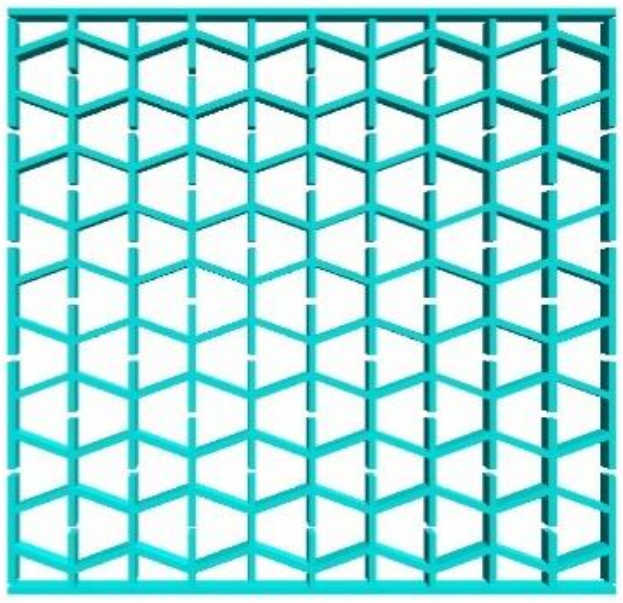 | [193] |
| Improved re-entrant honeycomb | 2D | 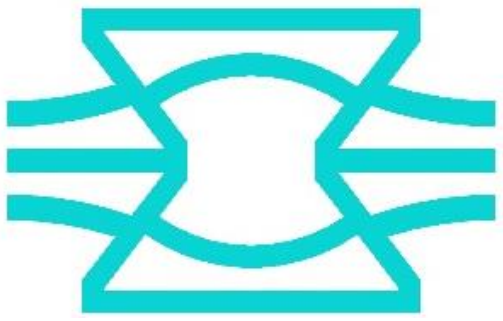 | 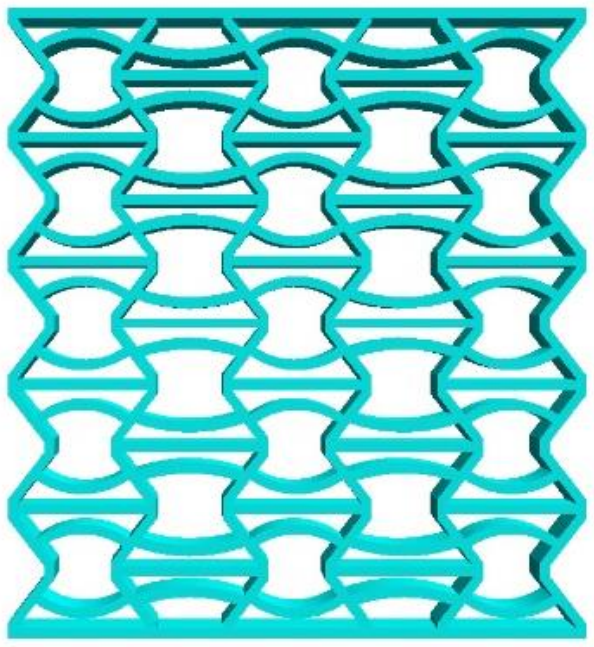 | [194] |
| Re-entrant honeycomb with tunable auxeticity | 2D | 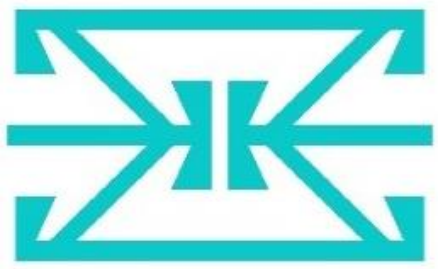 | 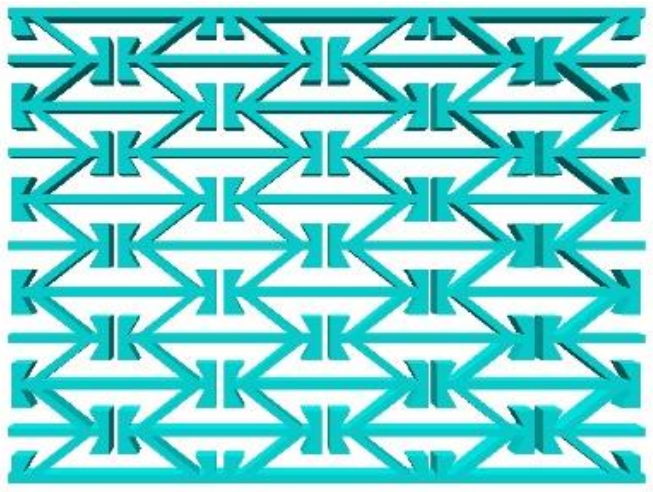 | [195] |
| Re-entrant honeycomb with tunable stiffness | 2D | 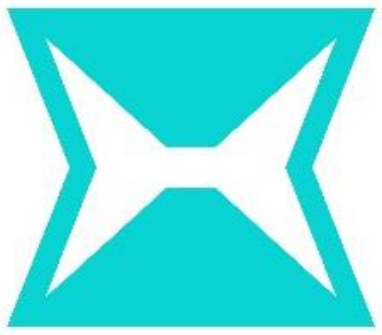 | 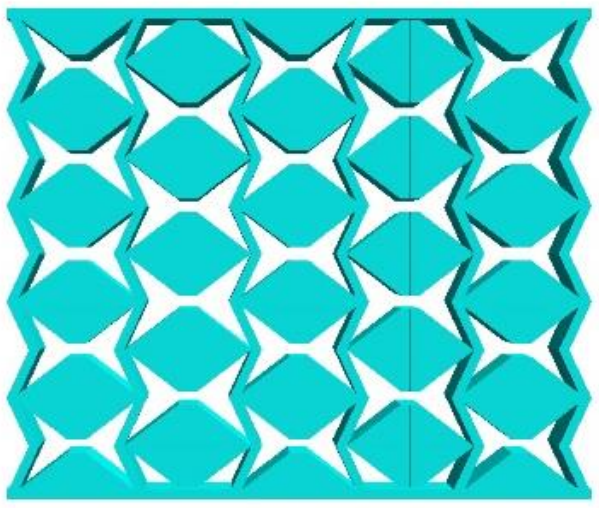 | [196] |
| Re-entrant star-shaped honeycomb | 2D | 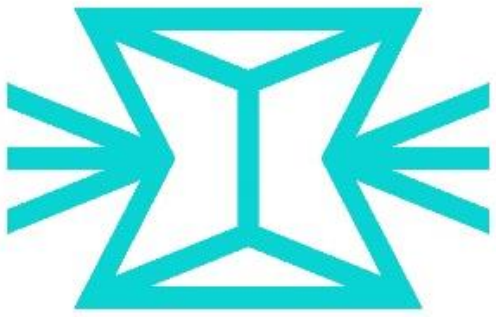 | 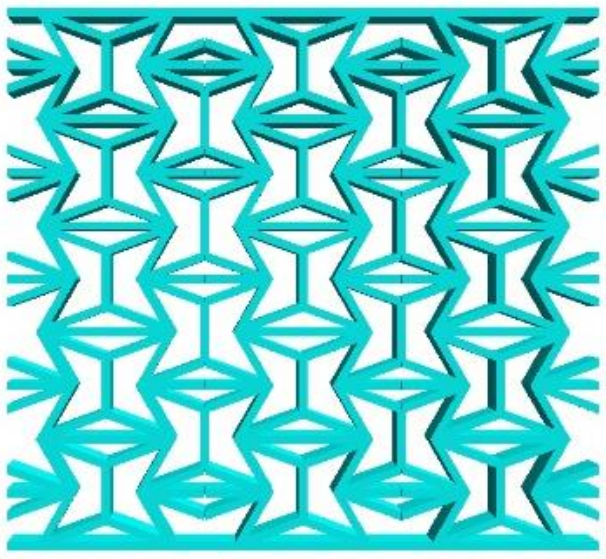 | [197] |
| Re-entrant anti-tetra chiral honeycomb | 2D | 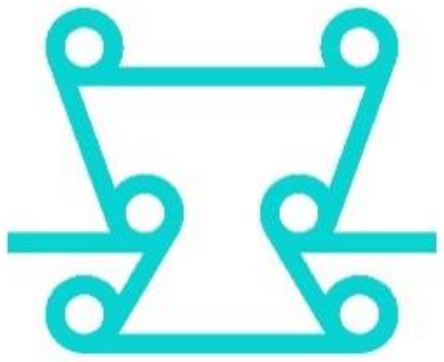 | 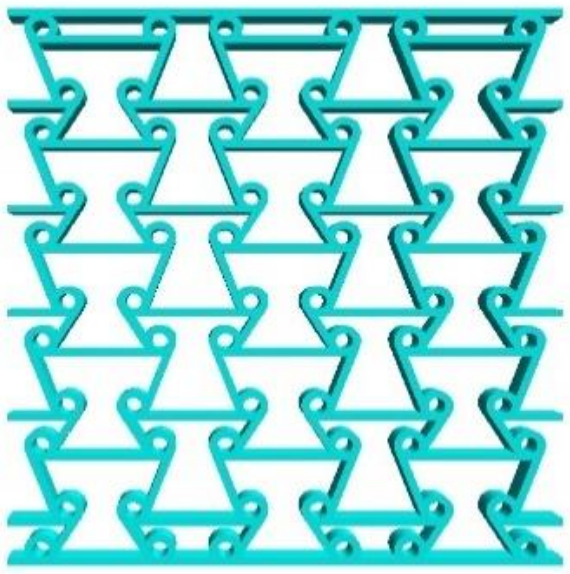 | [127] |
| Re-entrant honeycomb with tree different auxetic | 2D | 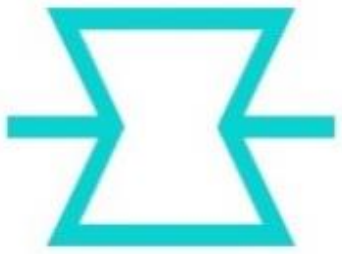 | 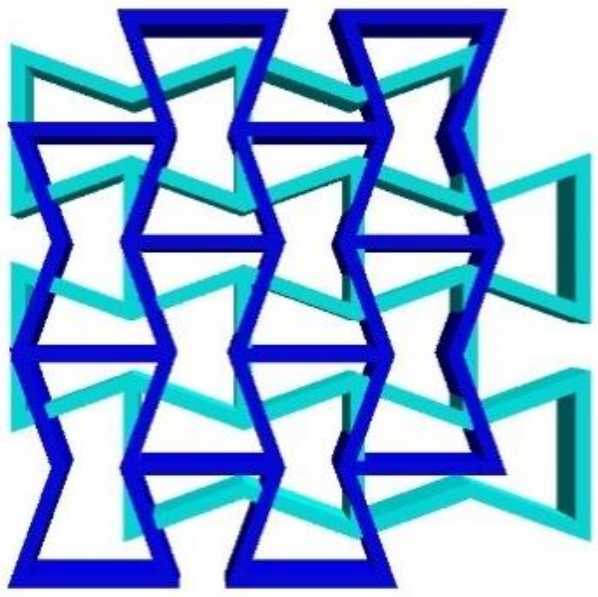 | [198] |
| Re-entrant honeycomb with arc angle gradient | 2D | 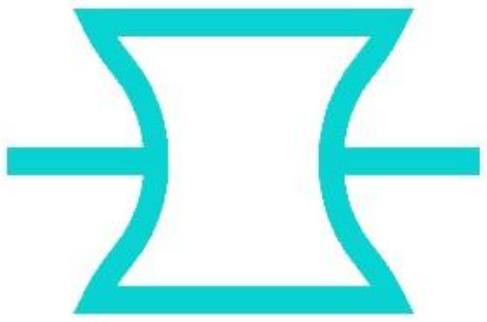 | 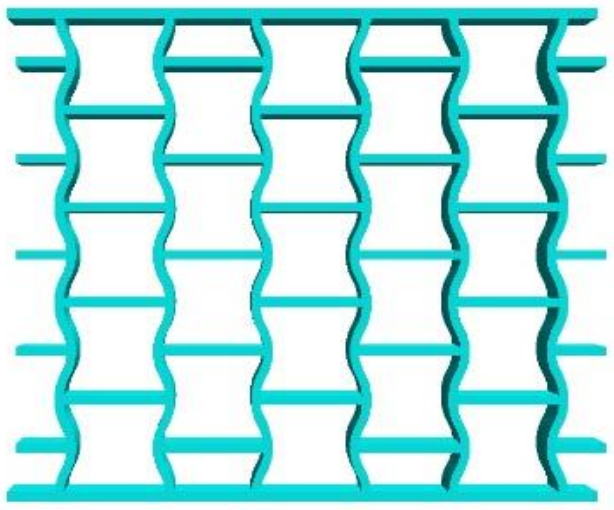 | [199] |
| Self-similar re-entrant honeycomb | 2D | 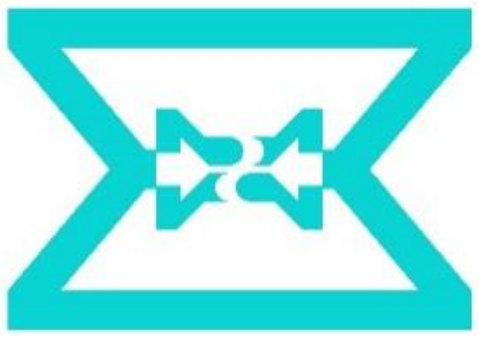 | 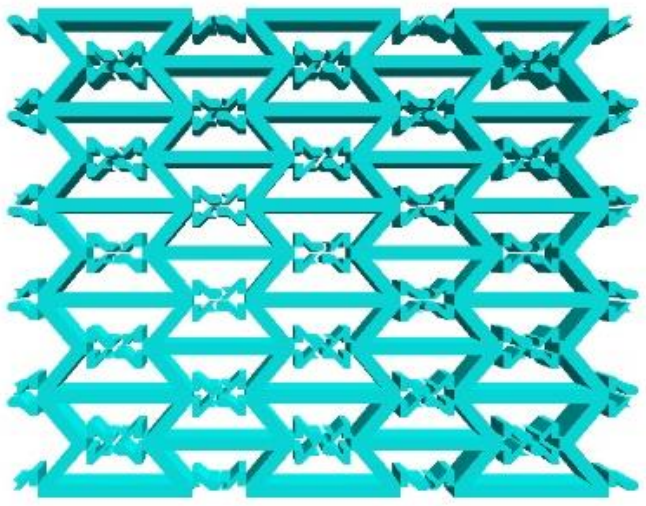 | [200] |
| Re-entrant honeycomb with enhanced auxeticity and tunable stiffness | 2D | 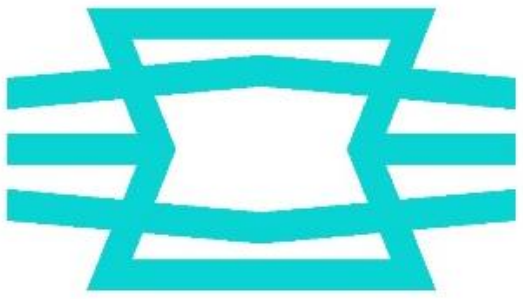 | 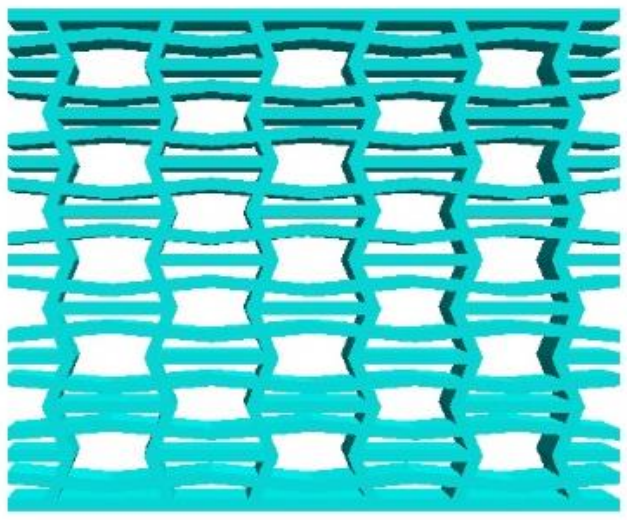 | [201] |
| Hierarchical re-entrant honeycomb | 2D | 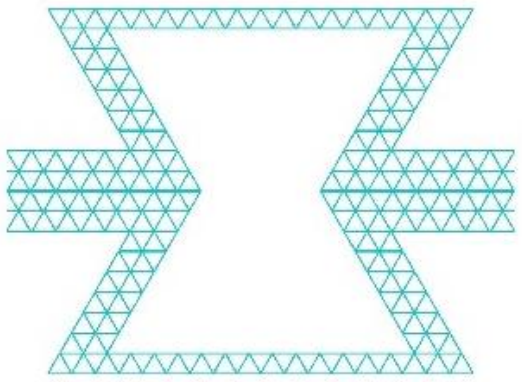 | 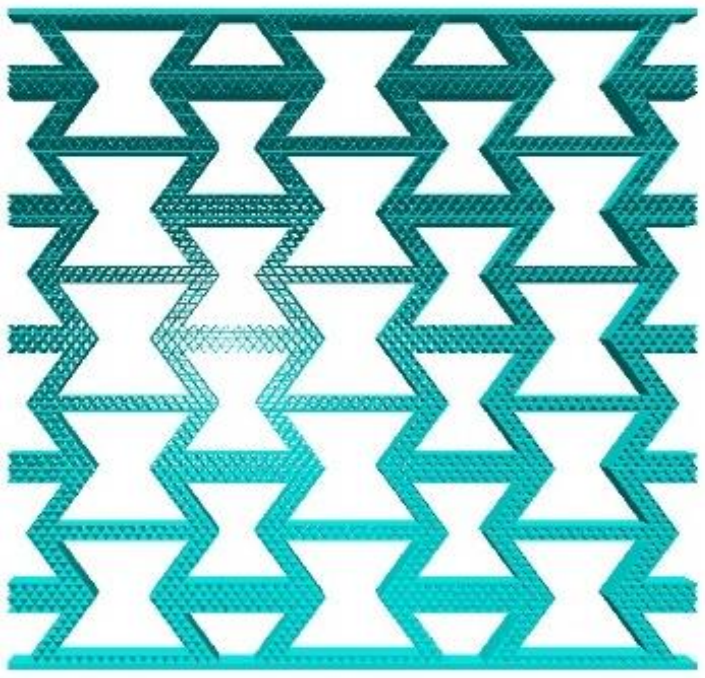 | [202] |
| Developed re-entrant honeycomb | 2D | 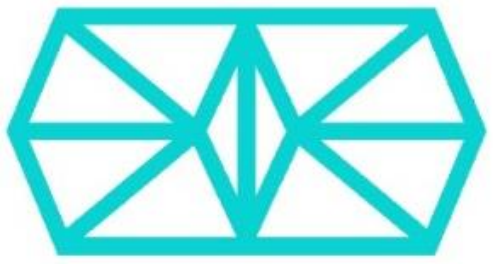 | 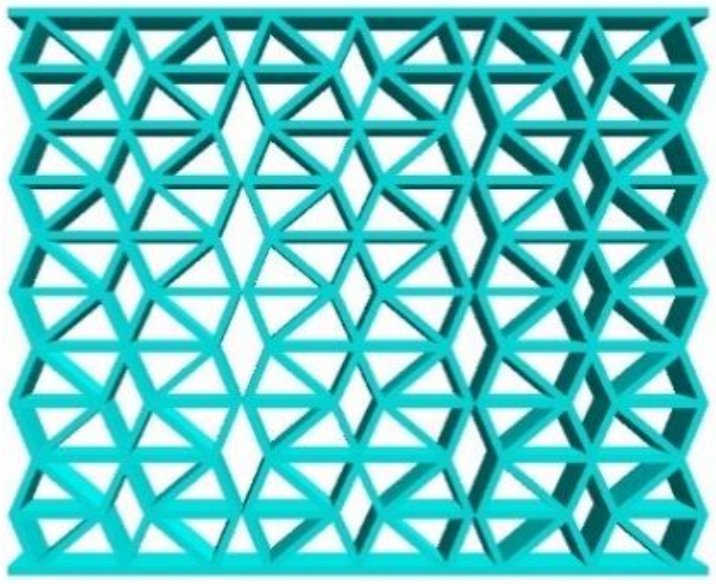 | [203] |
| Butterfly re-entrant honeycomb | 2D | 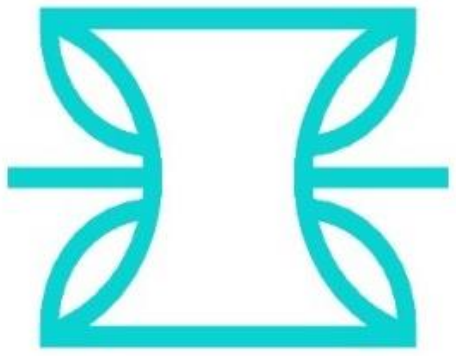 | 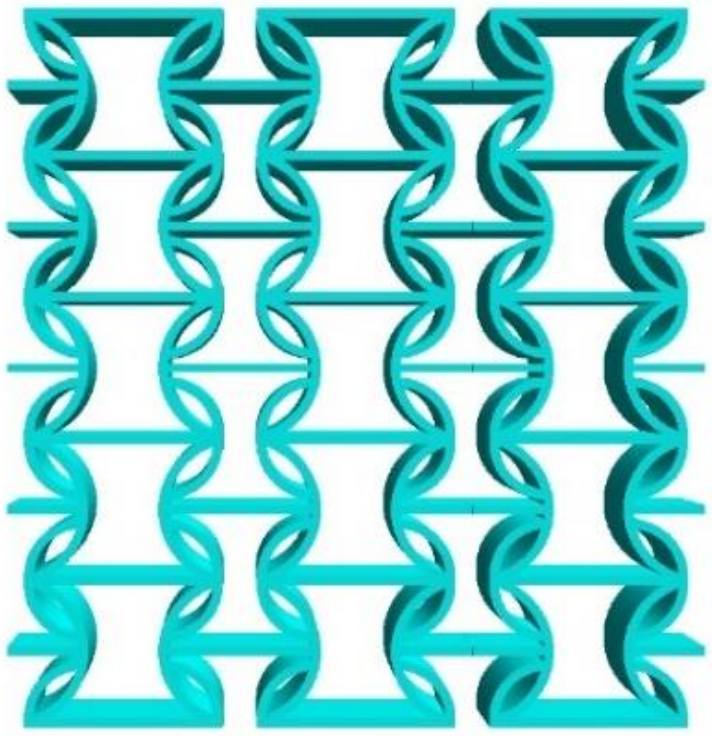 | [204] |
| Developed re-entrant honeycomb | 2D | 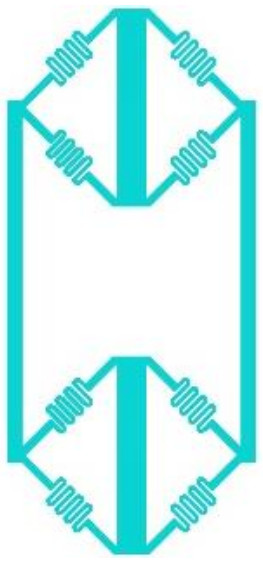 | 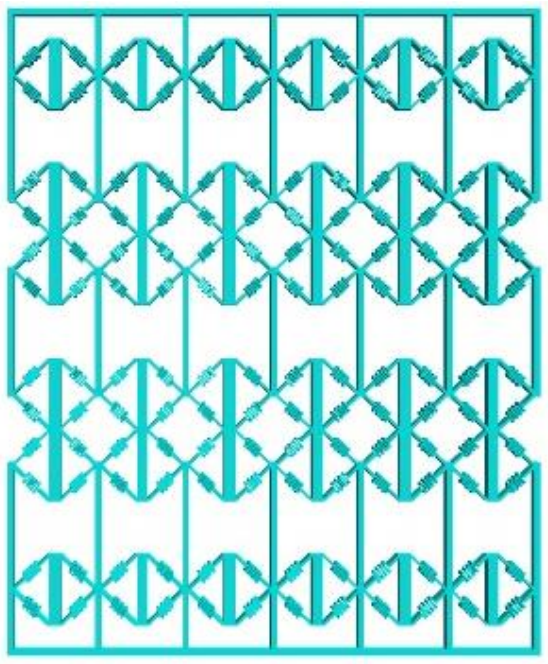 | [205] |
| Novel re-entrant honeycomb | 2D | 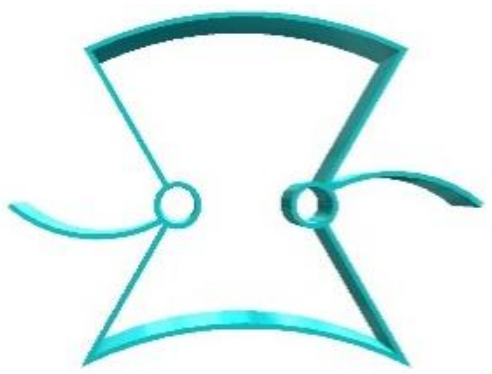 | 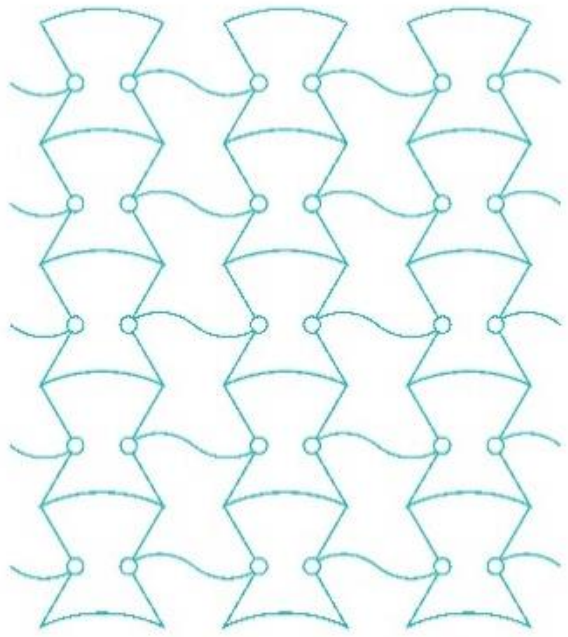 | [206] |
| Developed re-entrant honeycomb | 2D | 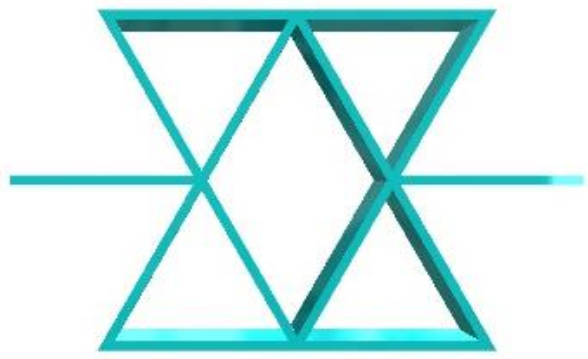 | 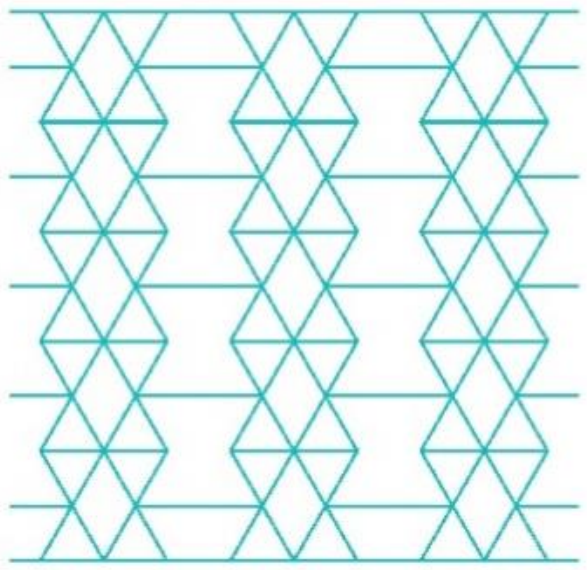 | [207] |
| Novel star re-entrant honeycomb | 2D | 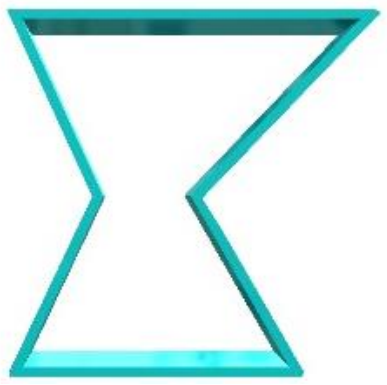 | 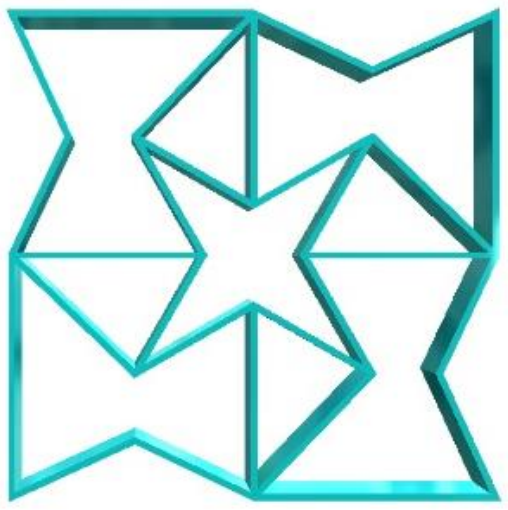 | [208] |
| Developed re-entrant honeycomb | 2D | 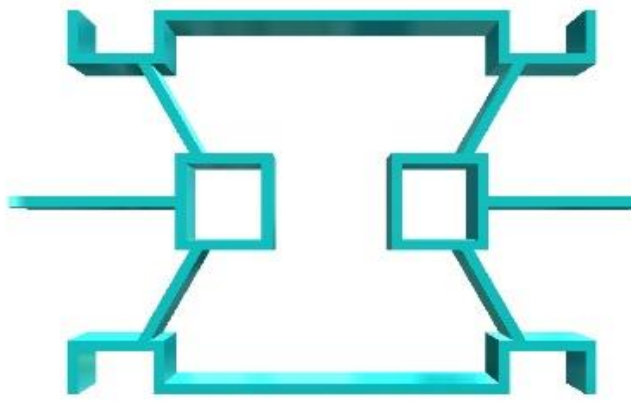 | 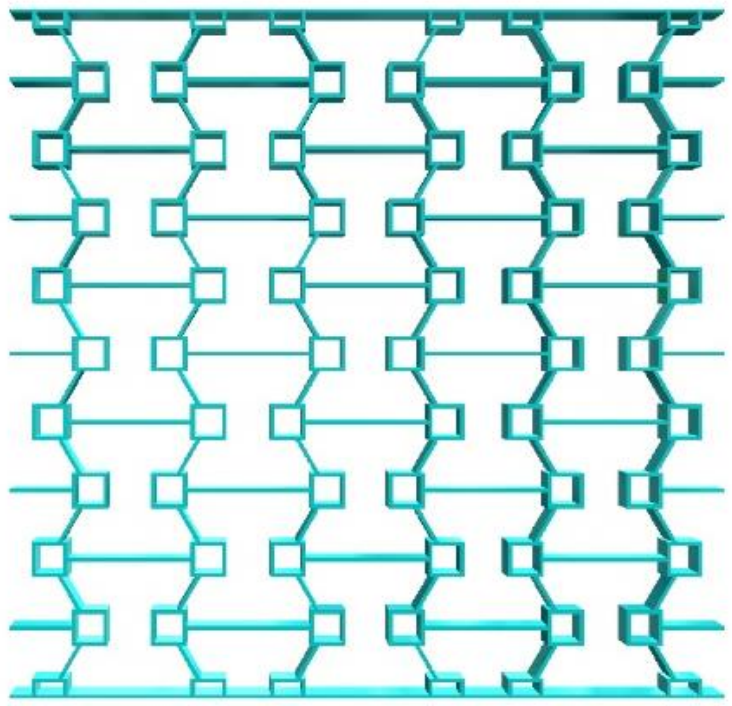 | [209] |
| Developed re-entrant honeycomb | 2D | 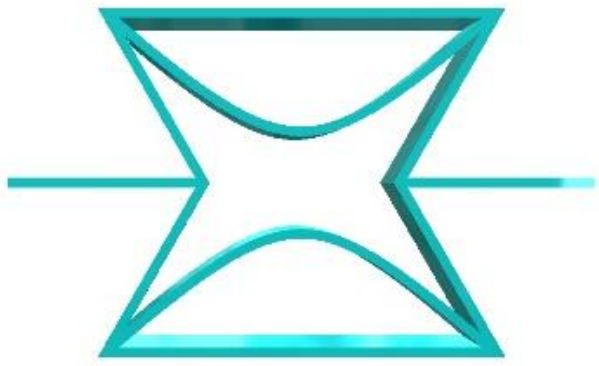 | 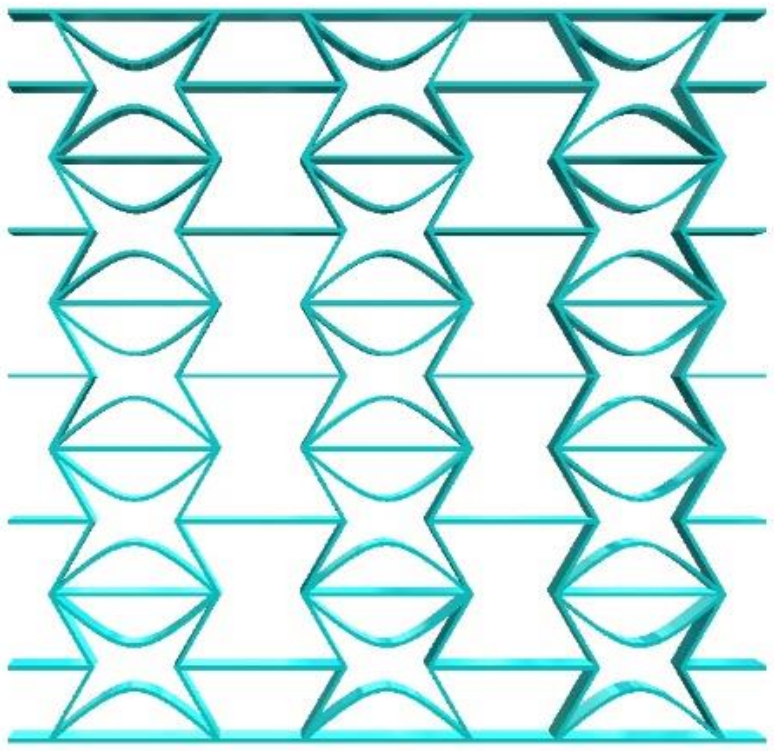 | [210] |
| Novel re-entrant honeycomb | 2D | 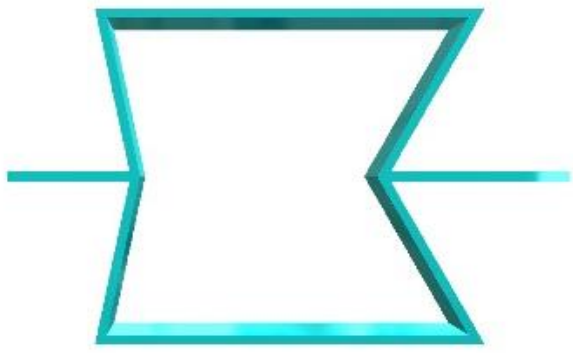 | 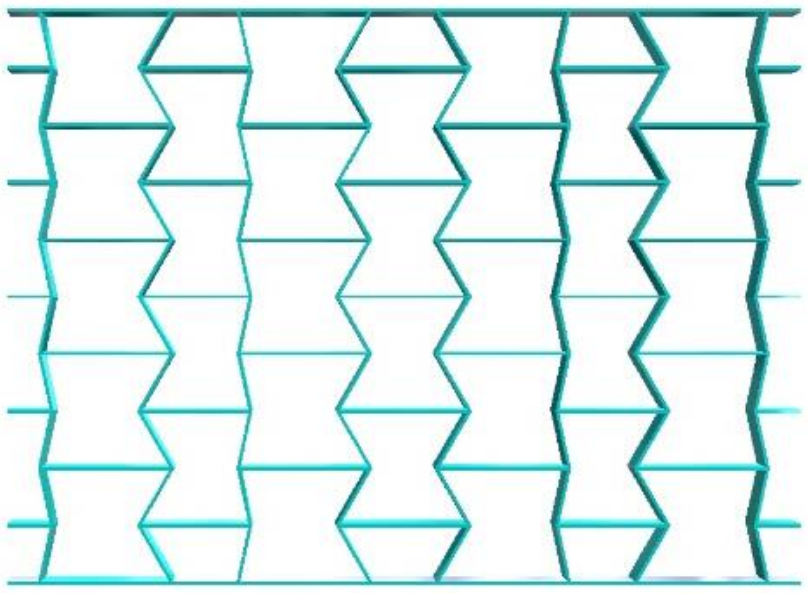 | [211] |
| Novel chiral re-entrant honeycomb | 2D | 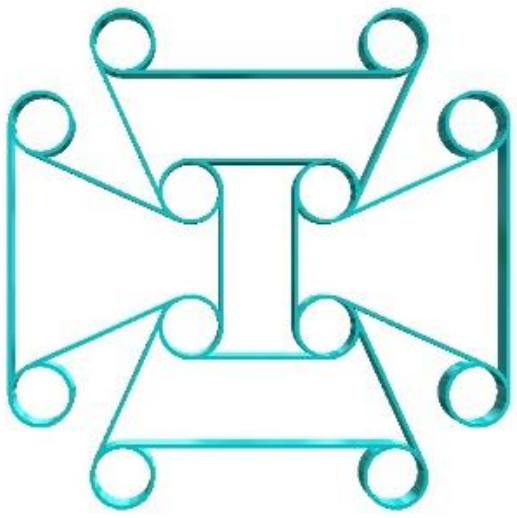 | 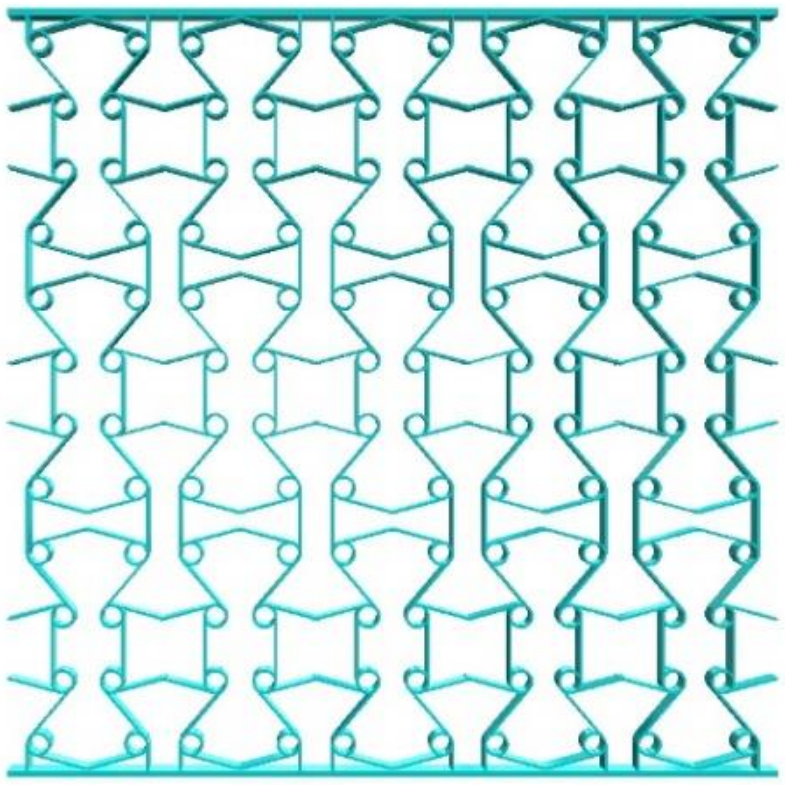 | [212] |
| Developed re-entrant honeycomb | 2D | 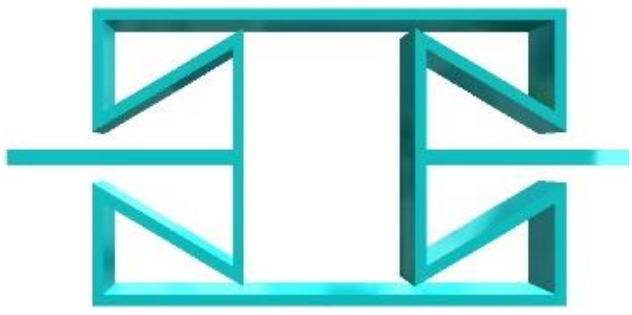 | 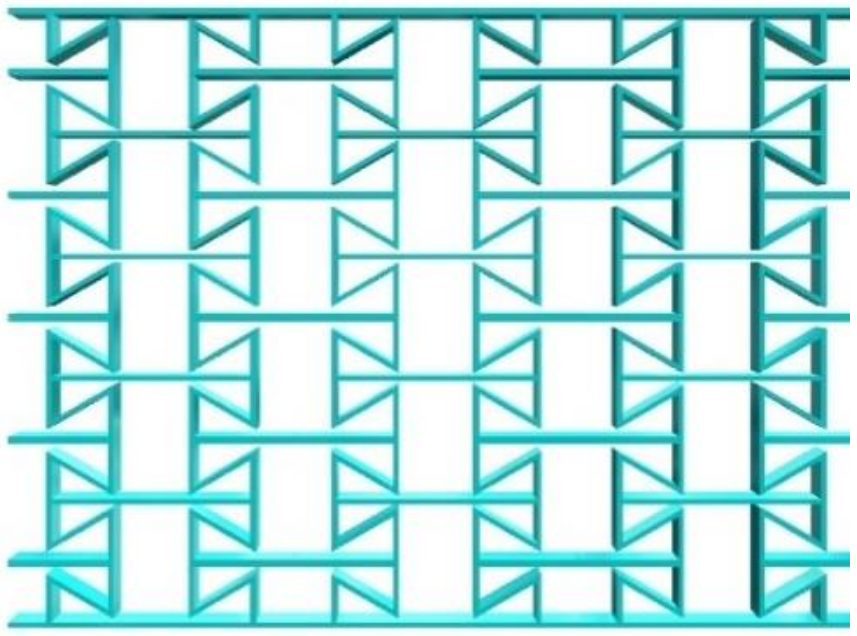 | [213] |
| Novel re-entrant honeycomb | 2D | 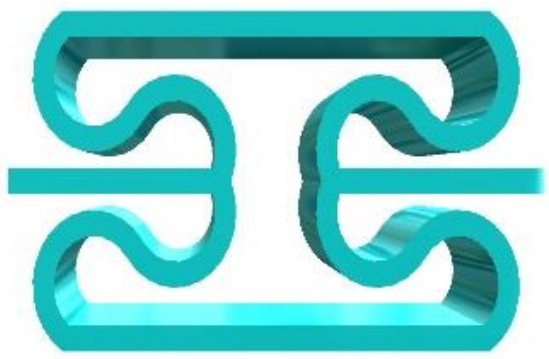 | 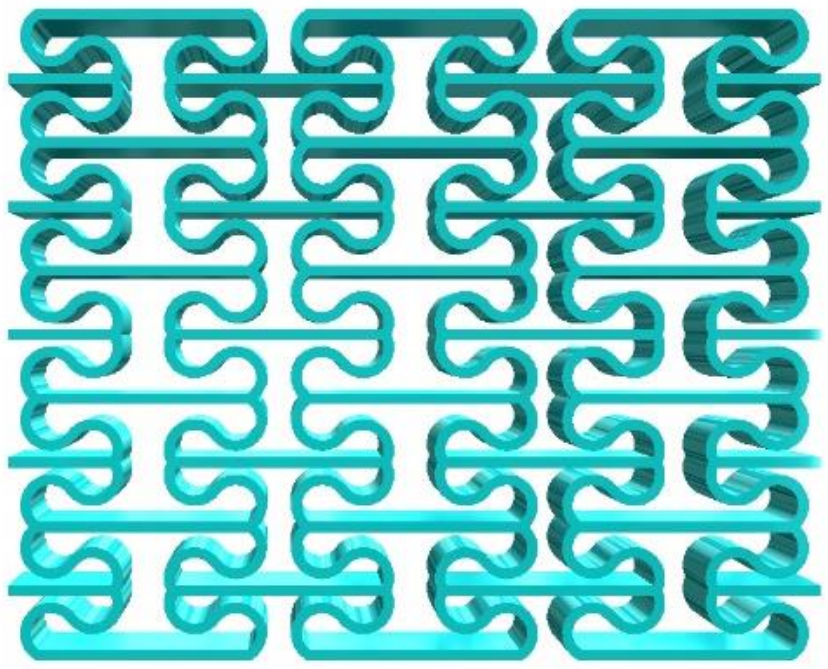 | [214] |
| Novel re-entrant honeycomb | 2D | 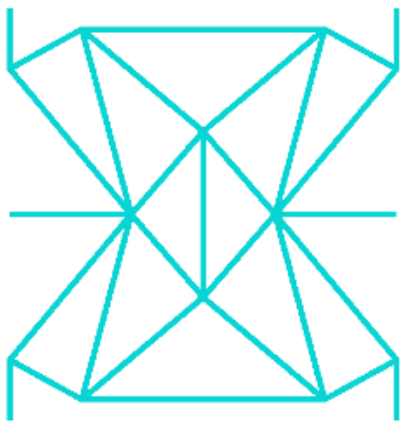 | 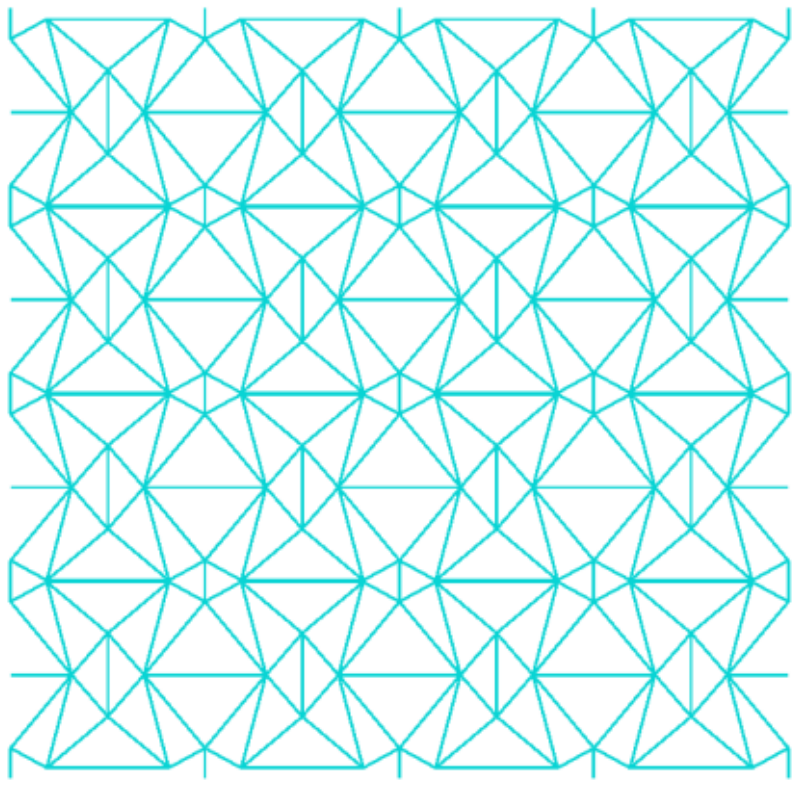 | [215] |
| Novel re-entrant honeycomb | 2D | 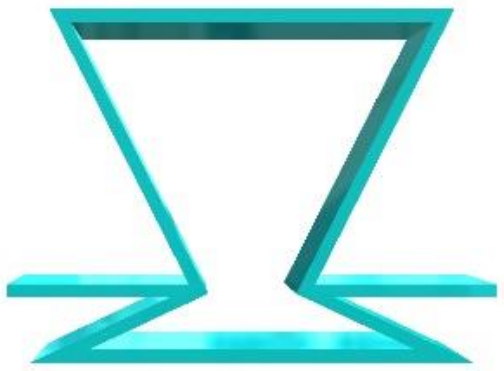 | 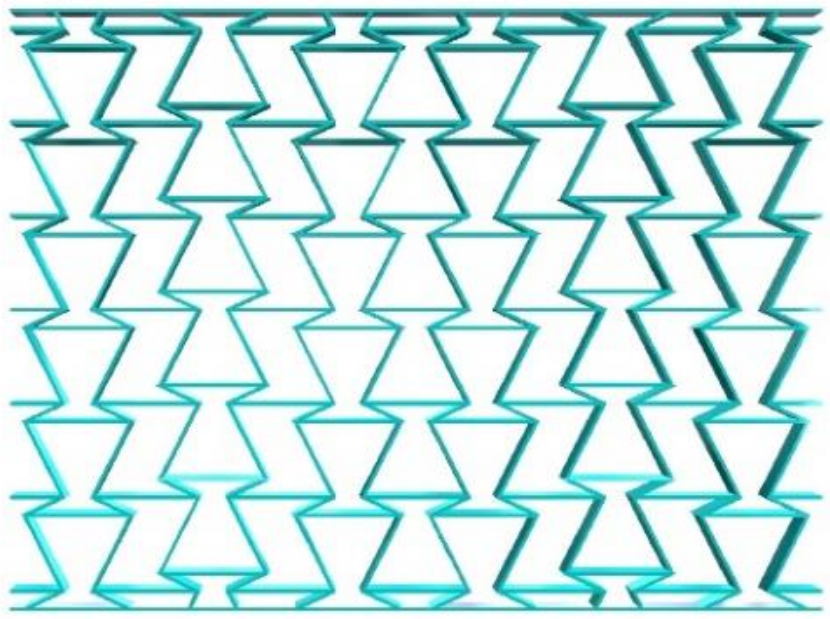 | [216] |
| Novel re-entrant honeycomb | 2D | 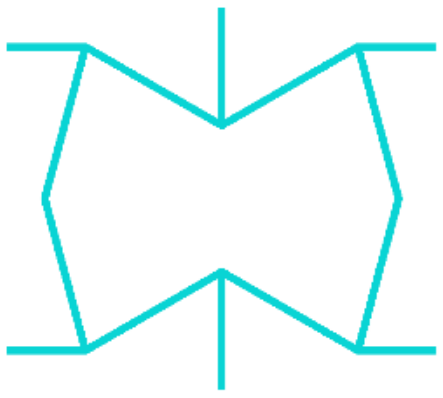 | 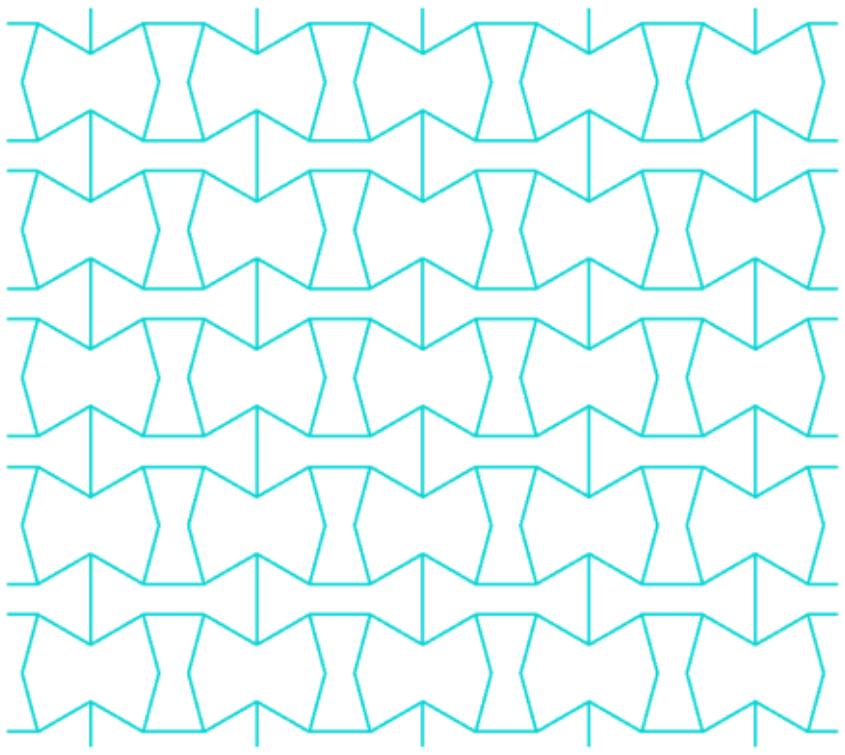 | [217] |
| Novel chiral re-entrant honeycomb | 2D | 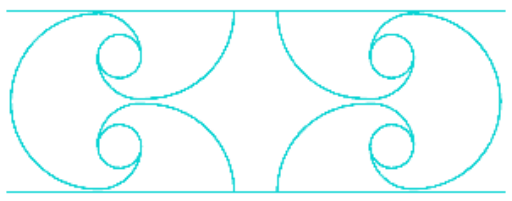 | 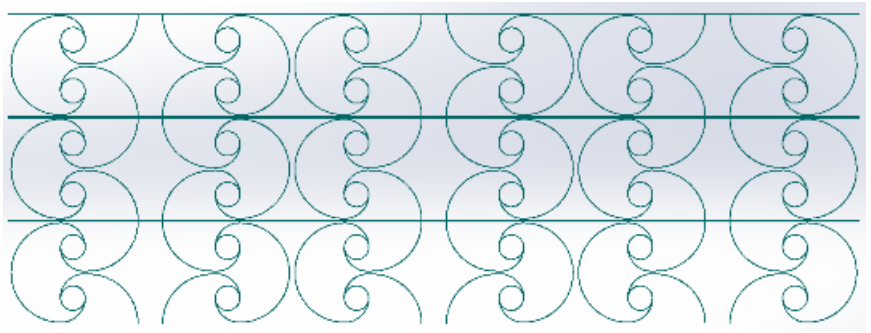 | [218] |
| 3D augmented cellular re-entrant honeycomb | 3D | 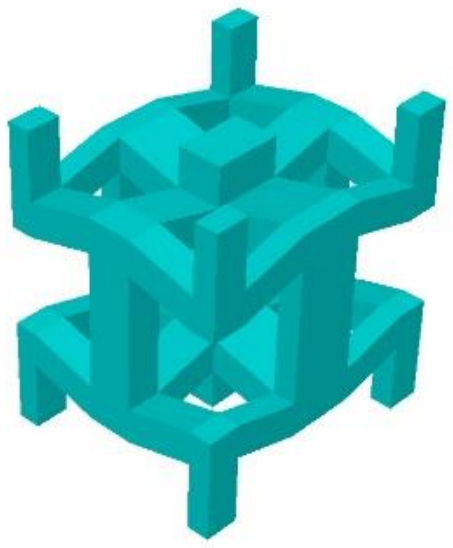 | 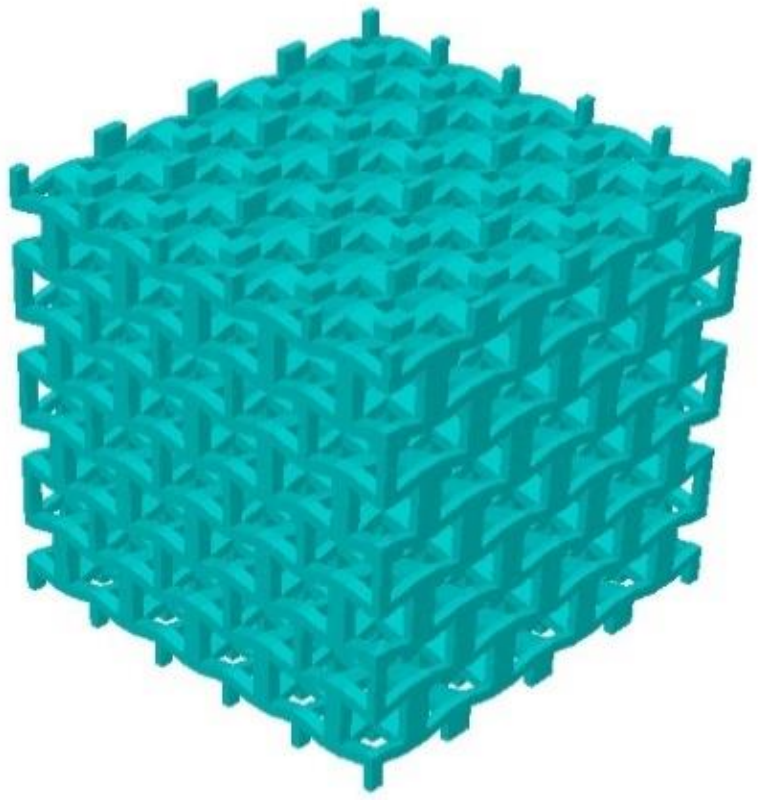 | [219] |
| 3D re-entrant honeycomb shell | 3D | 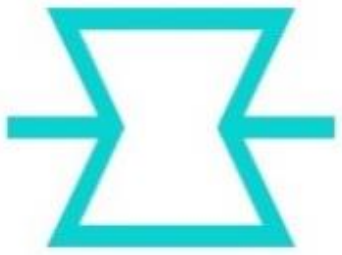 | 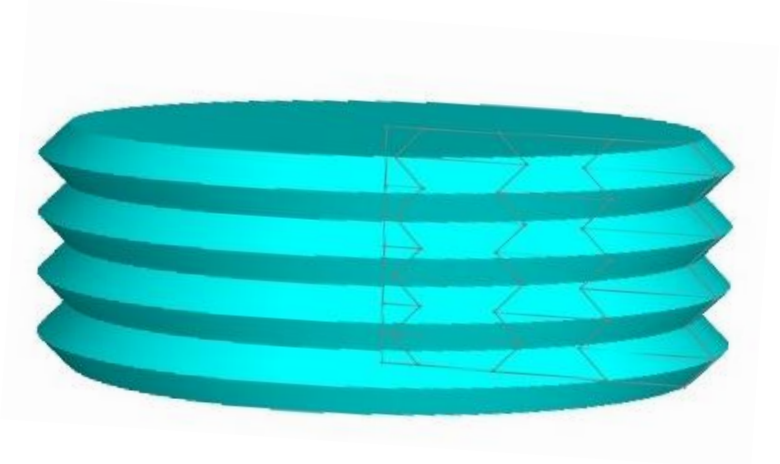 | [220] |
| Cylindrical tube re-entrant honeycomb | 3D | 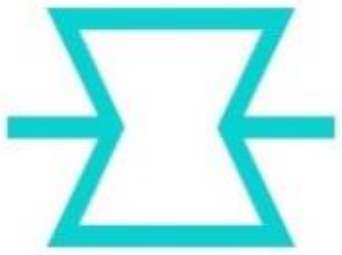 | 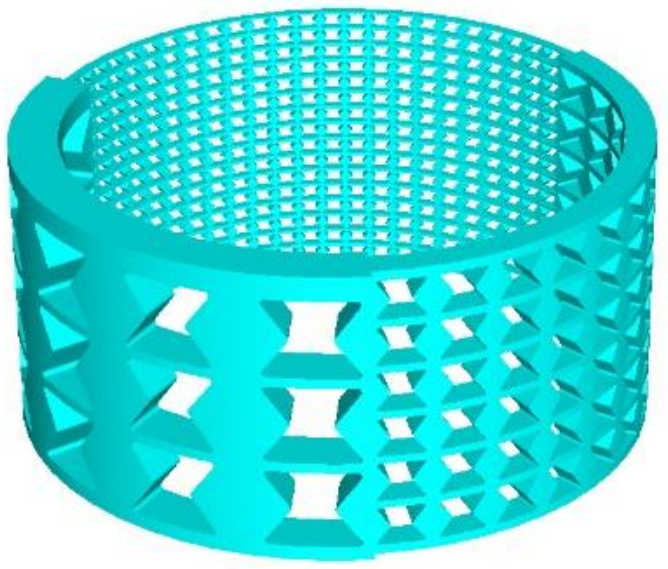 | [221] |
| 3D re-entrant honeycomb | 3D | 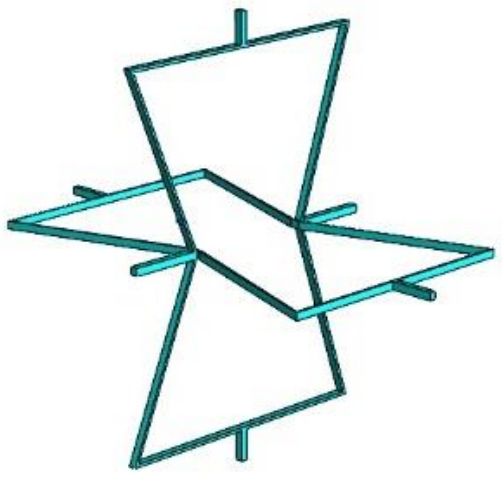 | 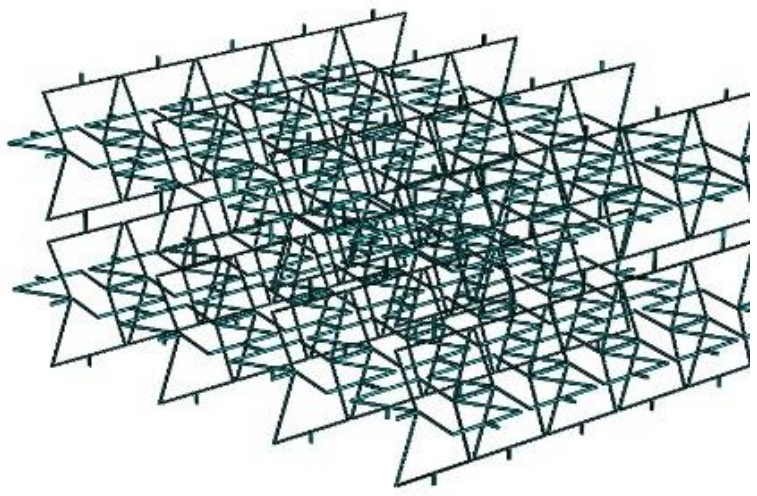 | [222] |
| 3D re-entrant honeycomb | 3D | 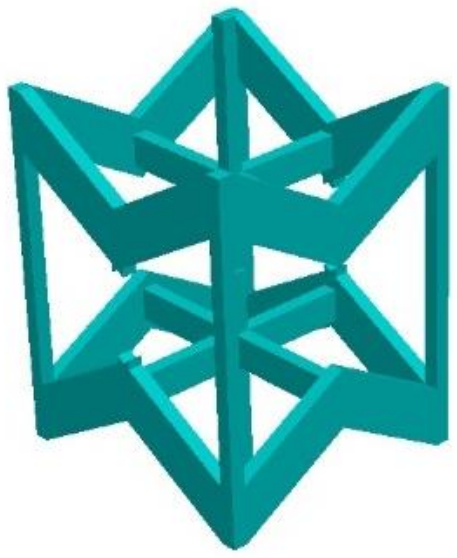 | 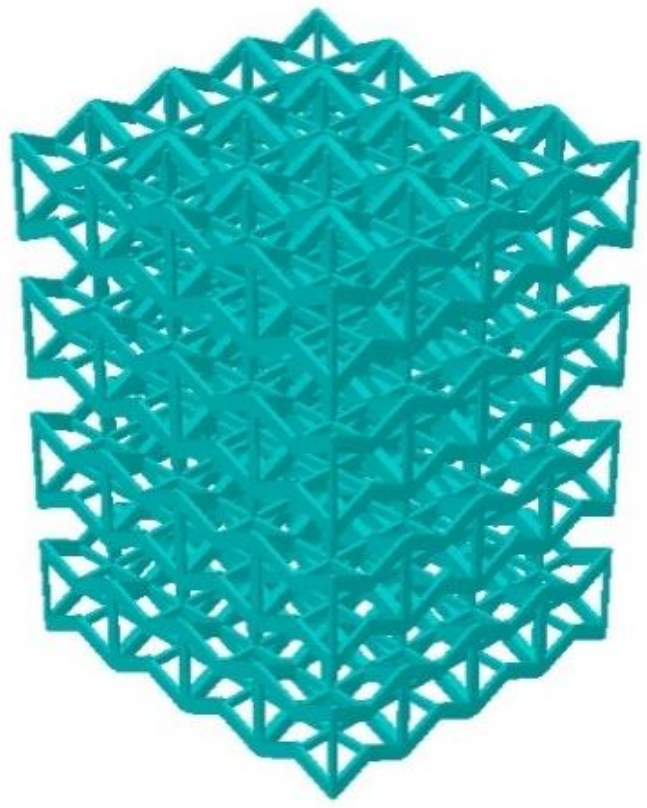 | [222] |
| 3D re-entrant honeycomb | 3D | 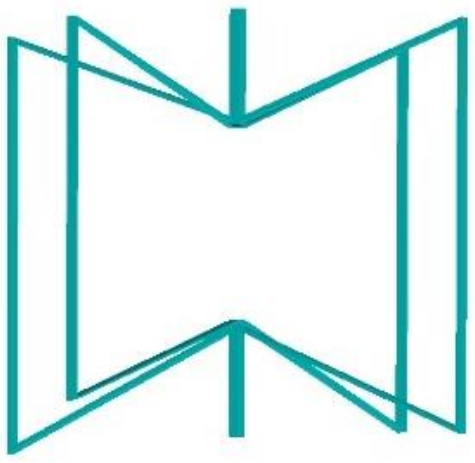 | 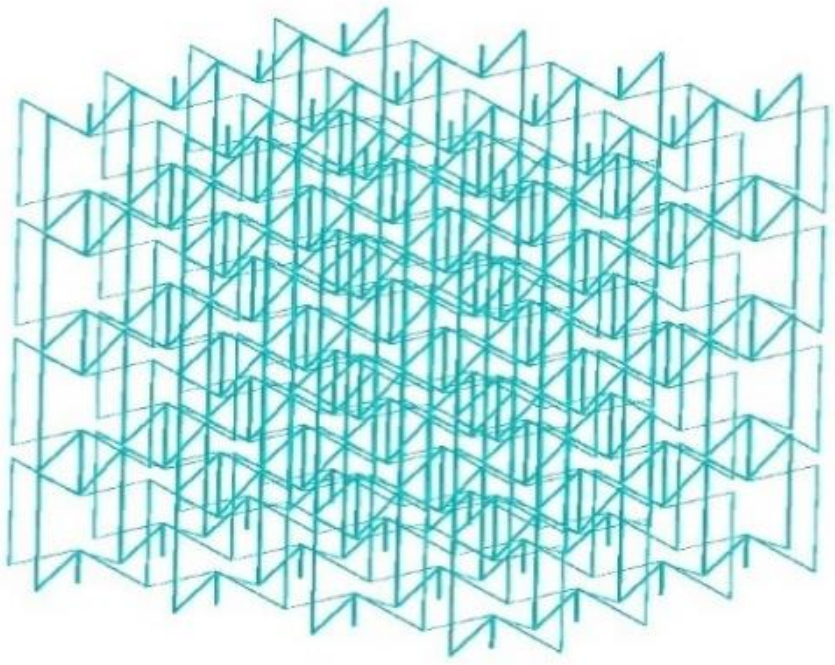 | [223] |
| 3D flat plate re-entrant honeycomb | 3D | 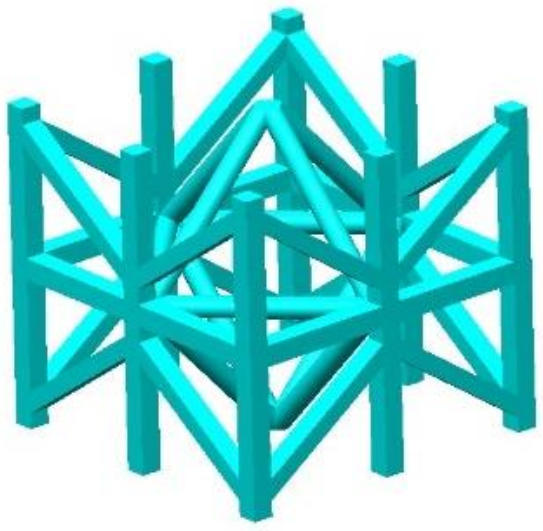 | 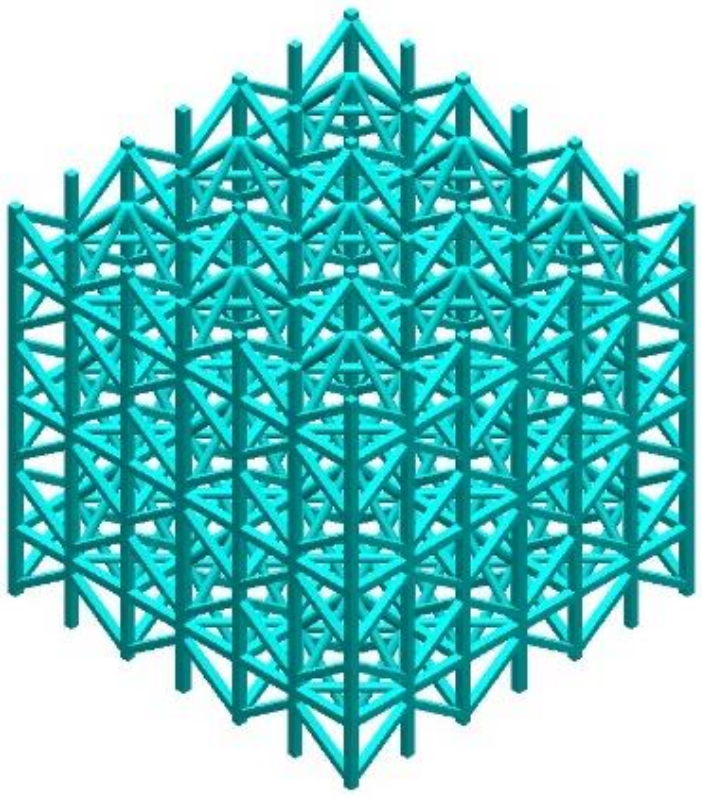 | [224] |
| 3D flat plate re-entrant honeycomb | 3D | 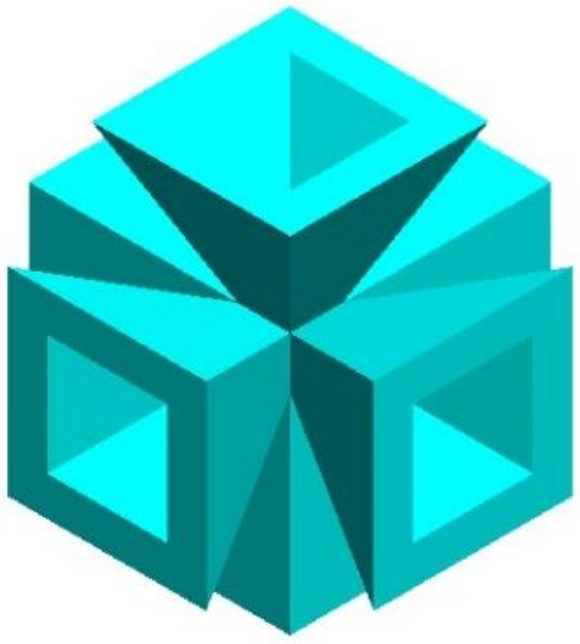 | 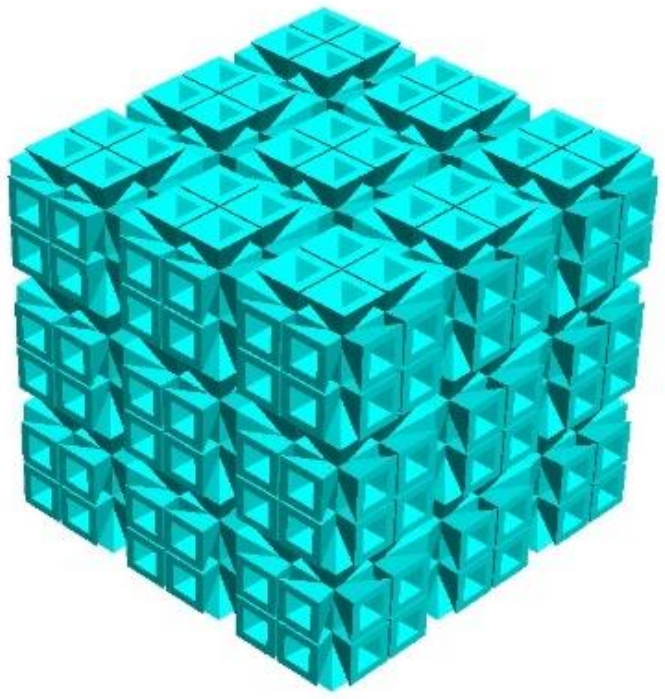 | [224] |
| 3D flat plate re-entrant honeycomb | 3D | 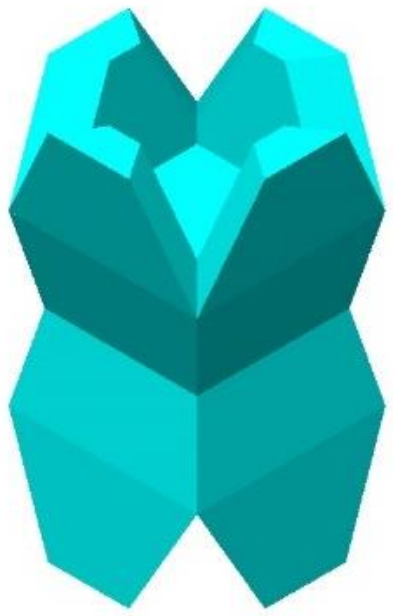 | 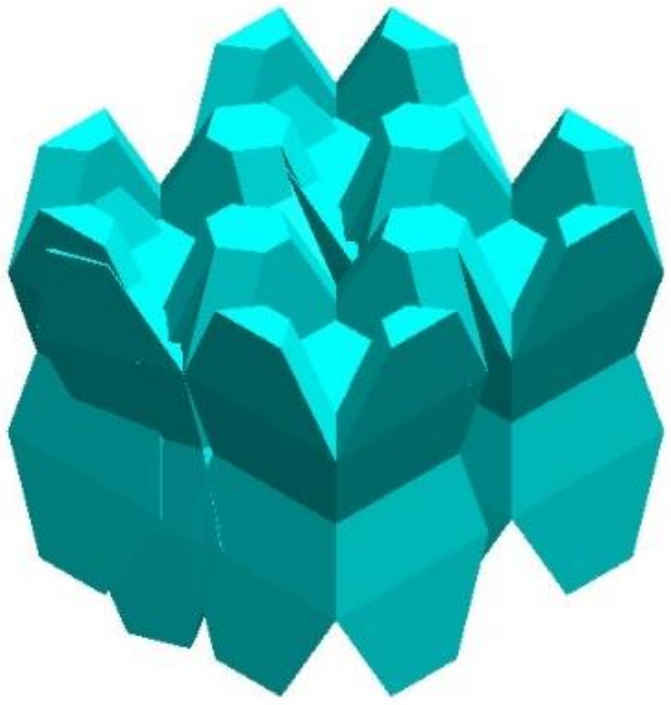 | [224] |
| 3D Re-entrant honeycomb | 3D | 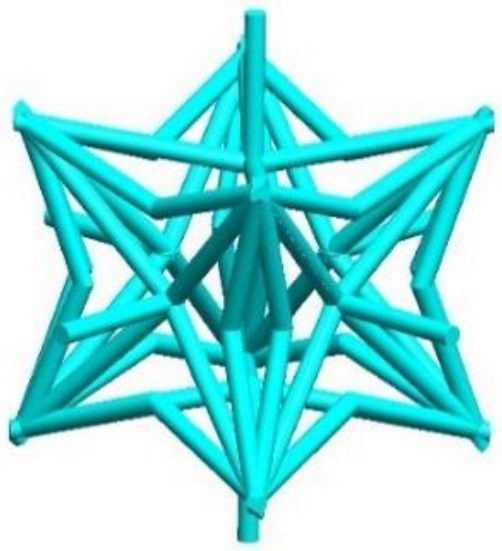 | 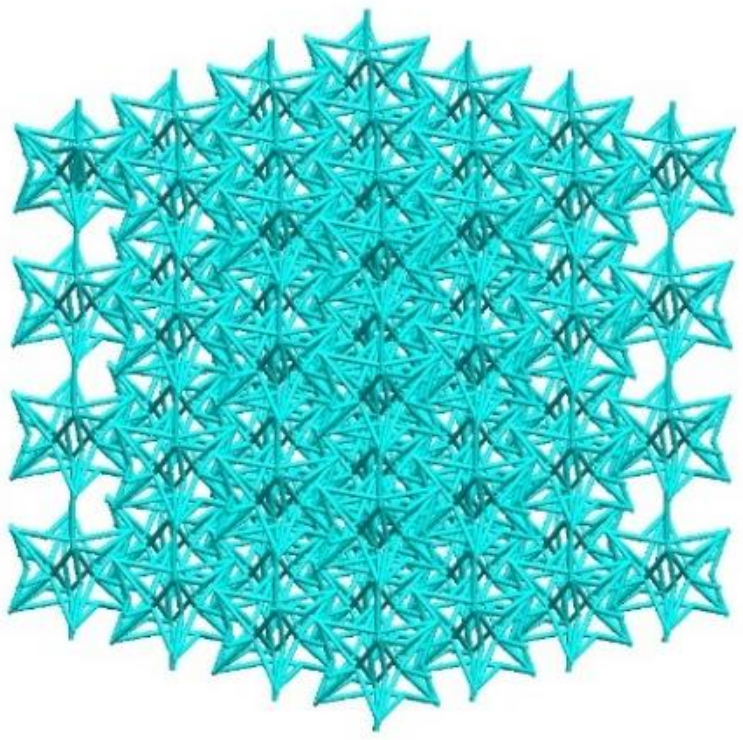 | [59] |
| Cross chiral Re-entrant honeycomb | 3D | 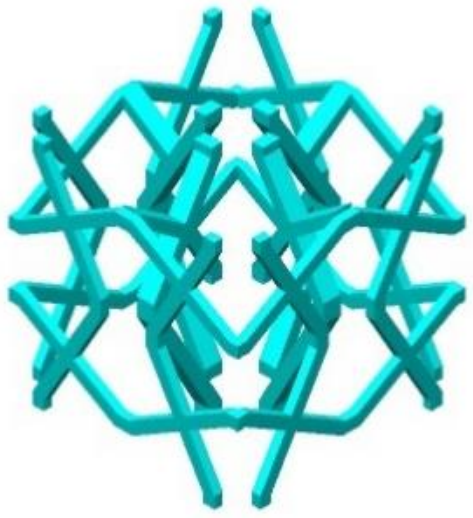 | 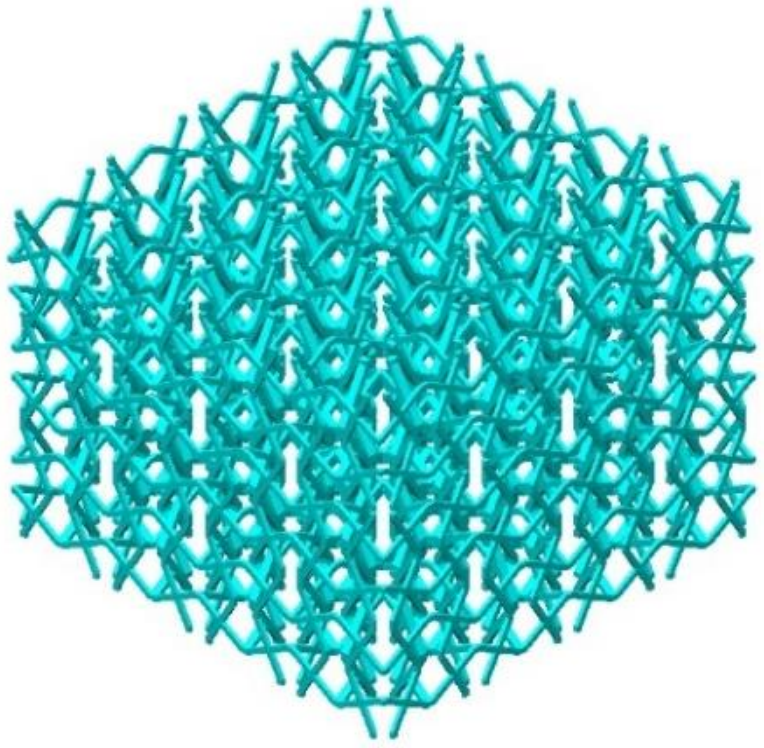 | [225] |
| 3D tubular re-entrant honeycomb | 3D | 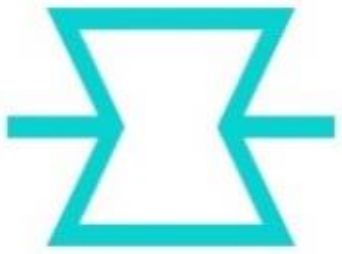 | 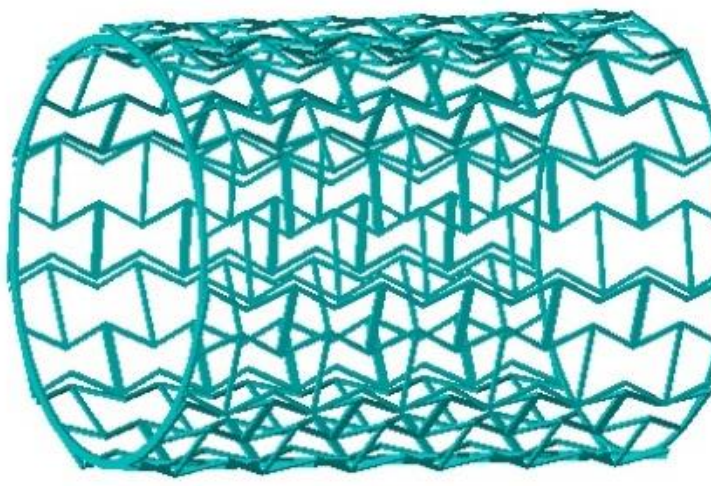 | [226] |

| | | | |
|---|---|---|---|
| 3D re-entrant honeycomb | 3D | 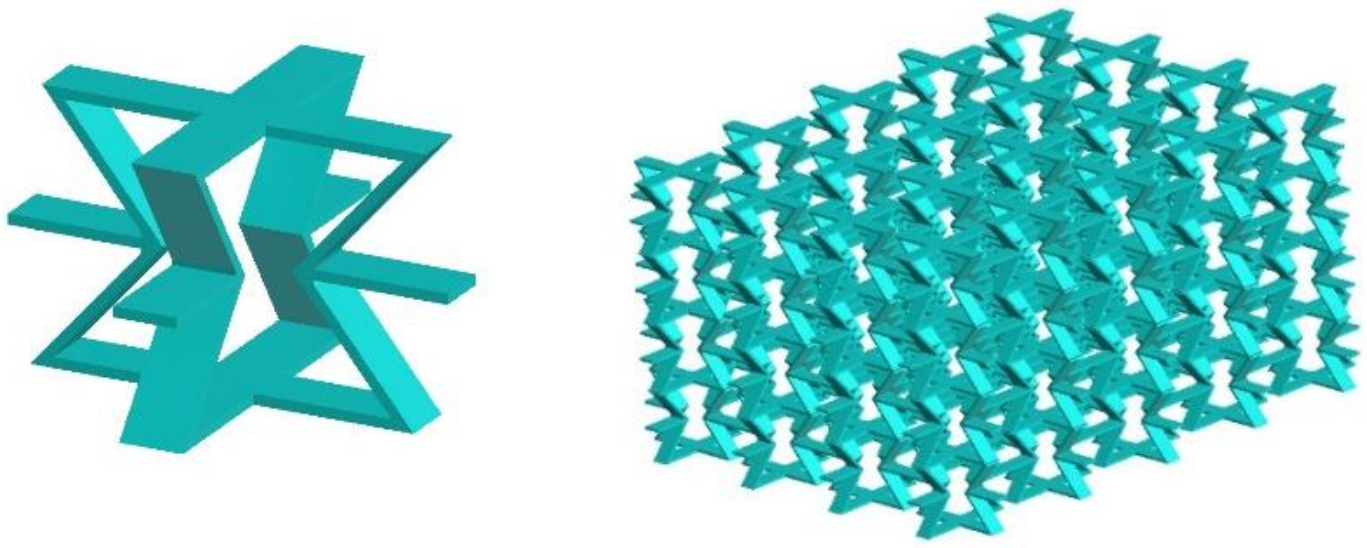 | [227] |
| 3D re-entrant honeycomb | 3D | 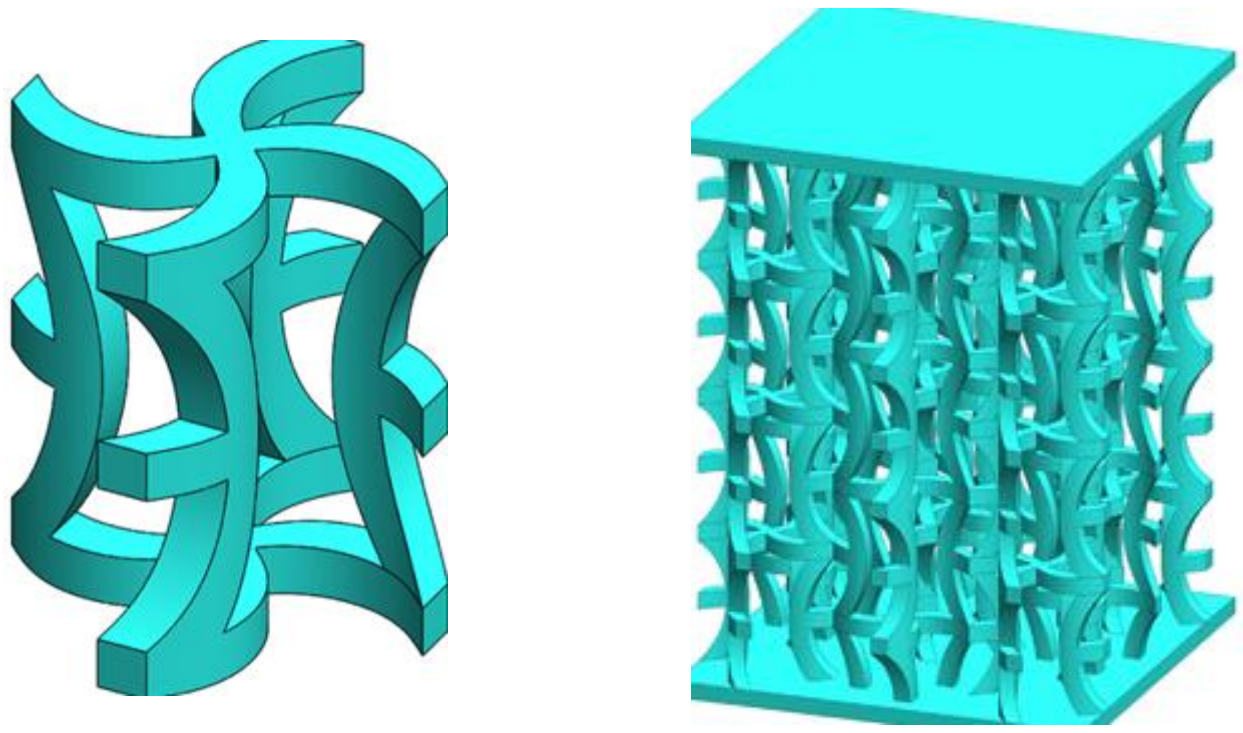 | [228] |
| 3D re-entrant honeycomb | 3D | 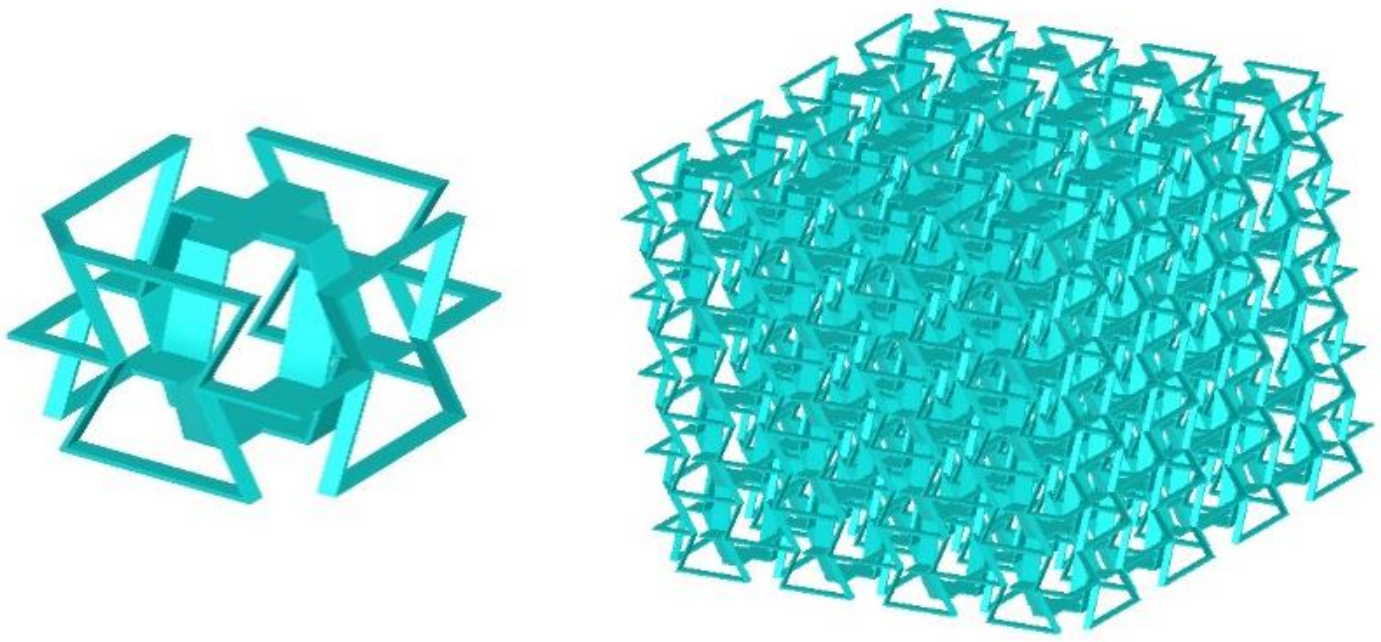 | [229] |

### *2.2. Star-shaped AMS*

The star structure was first developed from the re-entrant honeycomb structure. When Grima and his colleagues looked at the re-entrant honeycomb structure differently. They considered the re-entrant honeycomb structure to be a star structure with a one-star wing and considered the current common star structure with two wings or four corners. Also, their idea for the development of this structure was to increase the number of star feathers. This was the starting point of star-shaped structures. Grima introduced these structures in 2005 [230] as structures with a high potential for auxetic behavior. But over time, research in this field expanded. In 2014, Brighenti [231] investigated the behavior of the star structure and developed numerical relations in this field. In 2017, Hsiang-Wen et al. [232] conducted their research on the aspect ratio of the geometry of this structure.

At first, this structure named star-shaped re-entrant honeycomb appeared in researchers' studies, but with the passage of time and the development of other geometries in this field and considering the expansion of re-entrant honeycombs in recent years and also the potential of this type of structure, stellar structures and its extended family were introduced as a separate category. In recent years, researchers have focused on different aspects of star-shaped structures such as stiffness, Poisson's ratio, compressive strength, vibration behavior, ideal proportions for this structure, and the development of new unit cells with enhanced properties [131]. However, AMSs have gained much attention for their exceptional mechanical properties such as stiffness, compressive strength, and energy absorption. Studies indicate that these structures can achieve specific energy absorption (SEA) values between 30–35 kJ/kg under quasi-static compression, outperforming traditional re-entrant honeycombs with SEA values averaging 25 kJ/kg [233]. Additionally, their energy absorption efficiency has been observed to exceed 92% in dynamic impact tests, making them highly effective for crashworthiness applications [234]. The compressive strength of star-shaped structures varies between 6–12 MPa depending on material properties and aspect ratio, showing superior resistance to crushing compared to conventional honeycomb structures, which typically range between 4–8 MPa [235]. Furthermore, experimental studies highlight that star-shaped auxetic structures retain over 85% of their original energy absorption capacity after repeated loading cycles, emphasizing their durability and long-term performance [236]. These parameters make star-shaped promising candidates for protective structures, biomedical applications, and vibration-damping systems.

For a better understanding of the auxetic behavior of a star-shaped metastructure, the schematic of the axial crushing behavior of this metastructure is presented in Fig. 5. It should be noted that this figure schematically indicates the initiation of collapse mechanism under axial compression while the real collapse depends on other factors such as material type, strain rate, metastructure and processing, etc. However, star-shaped structures show high resistance to axial compressive force due to the existence of links that do not sink into each other and this capability increases the possibility of buckling in these structures. This is why these structures have exhibited a lower NPR in available investigations. Poisson's ratio predicted for these structures is in the range of -0.1 to -0.35. The basic and developed unit cells of this structure have been collected and exhibited in Table 2.

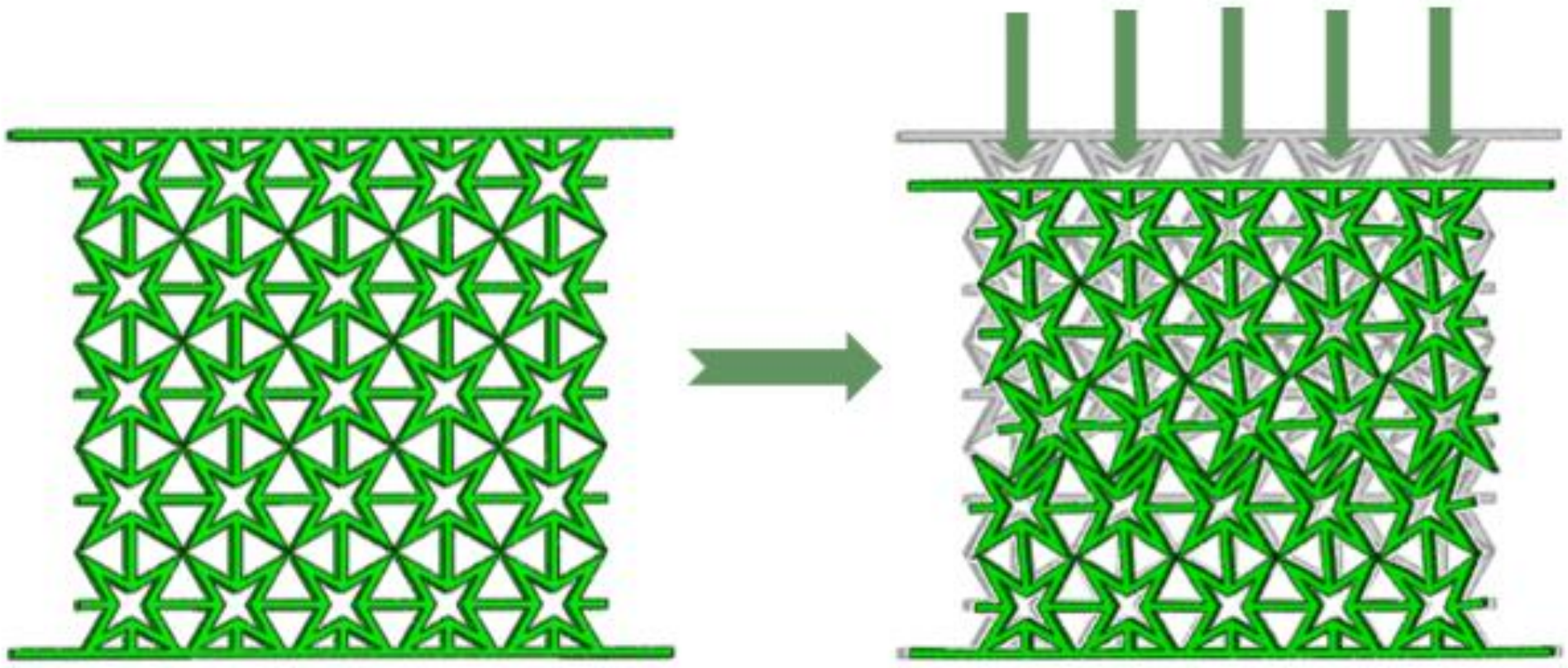

Fig. 5 Schematic of a star-shaped structure under axial compression.

Table 2 Classification of star-shaped AMS

| Structure name | Model dimension | Unit cell | Aggregate of unit cells | Reference |
|---|---|---|---|---|
| Star-shaped auxetic structure | 2D | 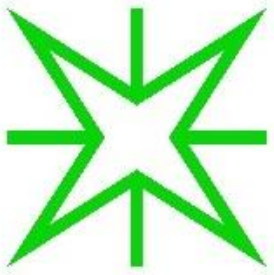 | 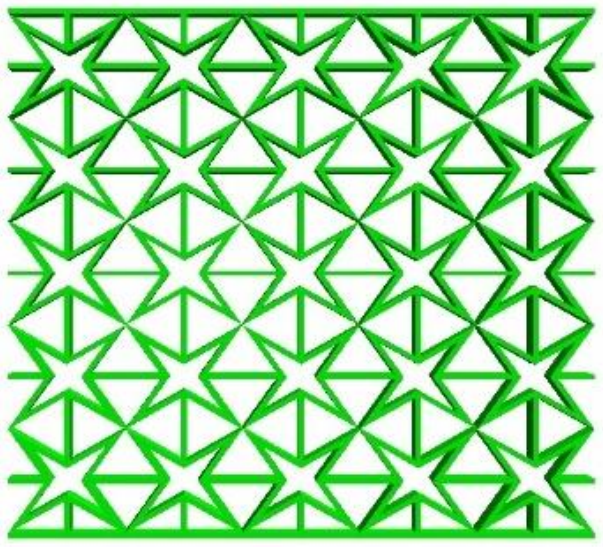 | |
| Star-shaped auxetic structure with three vertices | 2D | 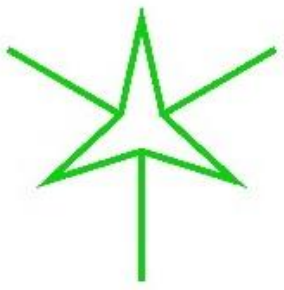 | 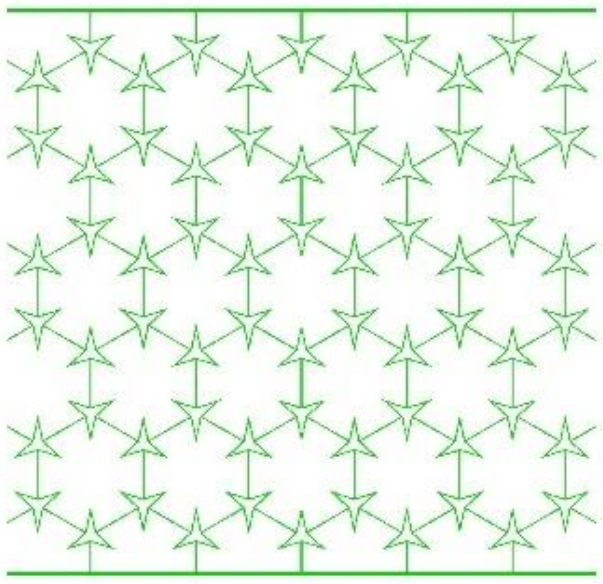 | [230] |
| Star-shaped auxetic structure with six vertices | 2D | 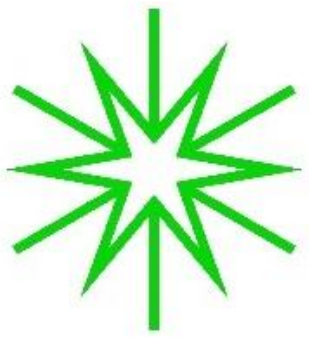 | 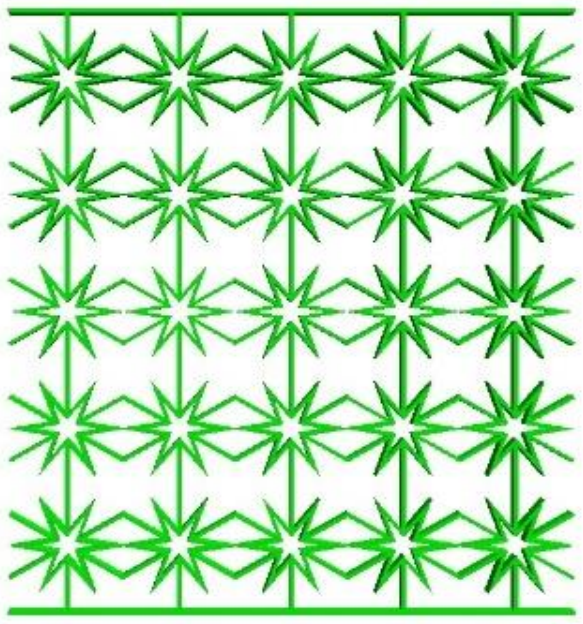 | [230] |
| Petal-shaped auxetic structure with stiffness constraint | 2D | 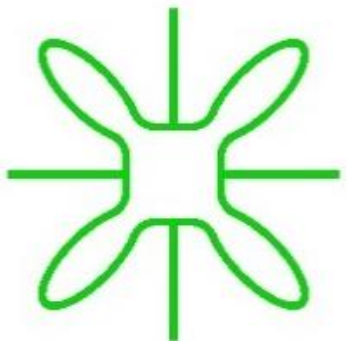 | 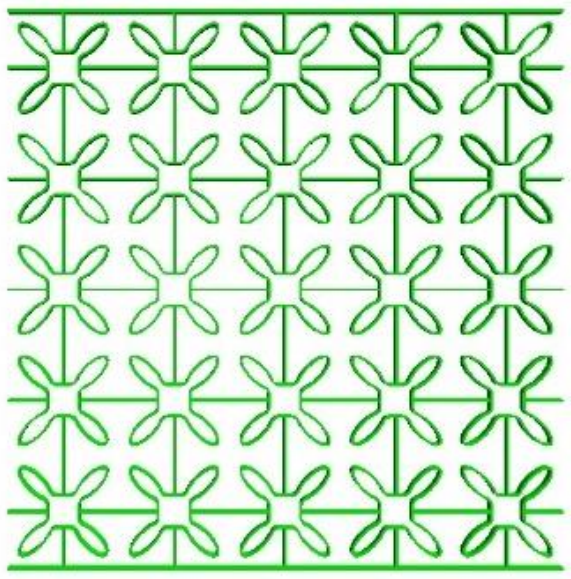 | [237] |
| Petal-shaped auxetic structure with stiffness constraint | 2D | 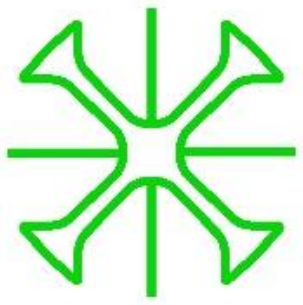 | 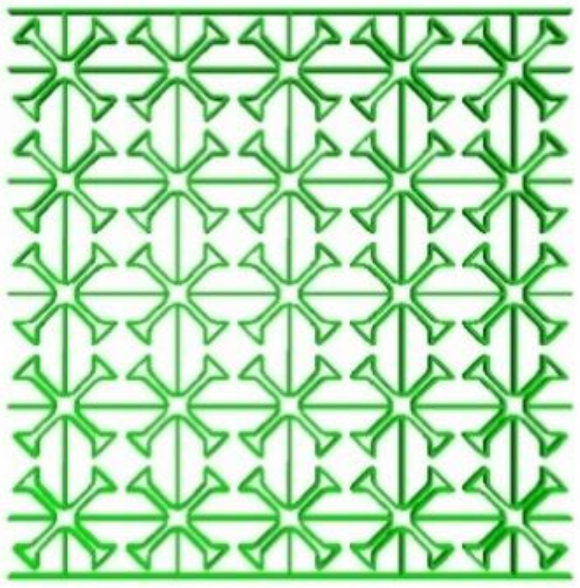 | [237] |
| Petal-shaped auxetic structure with 6 vertices | 2D | 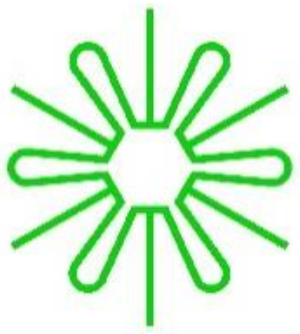 | 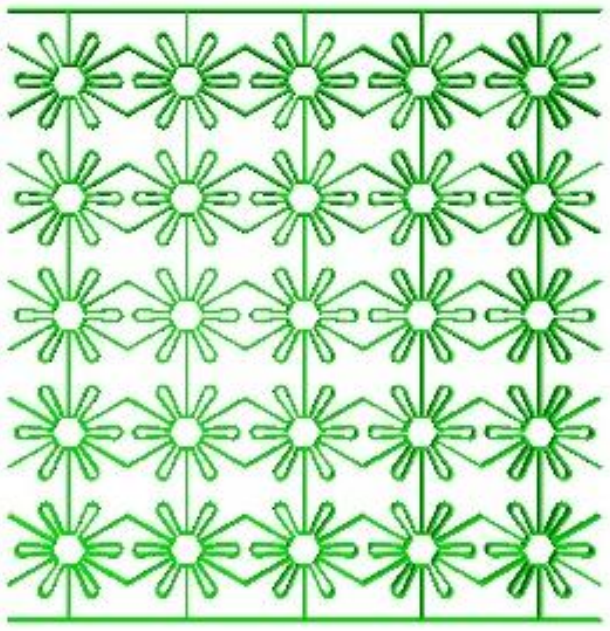 | [222] |
| Star-shaped honeycomb auxetic structure | 2D | 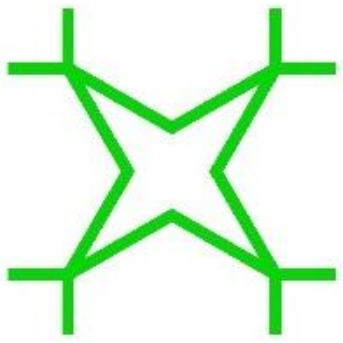 | 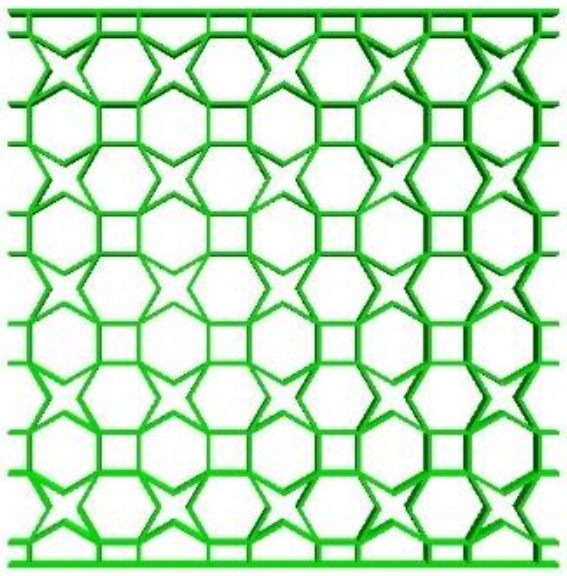 | [238] |
| cookie-shaped auxetic structure | 2D | 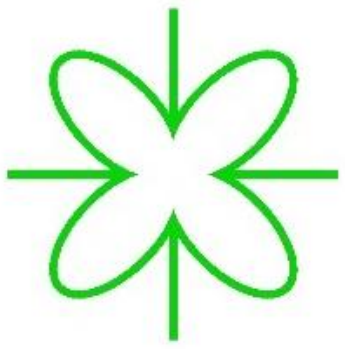 | 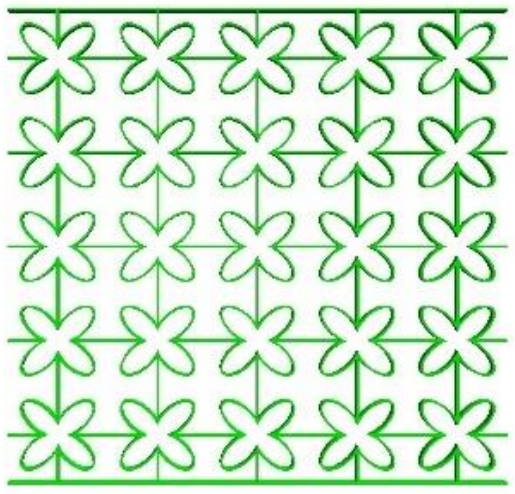 | [239] |
| Star-shaped auxetic structure with enhanced load bearing | 2D | 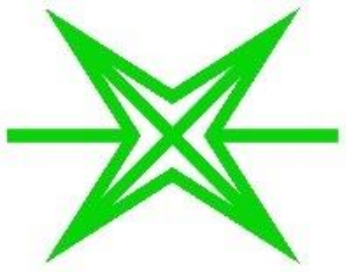 | 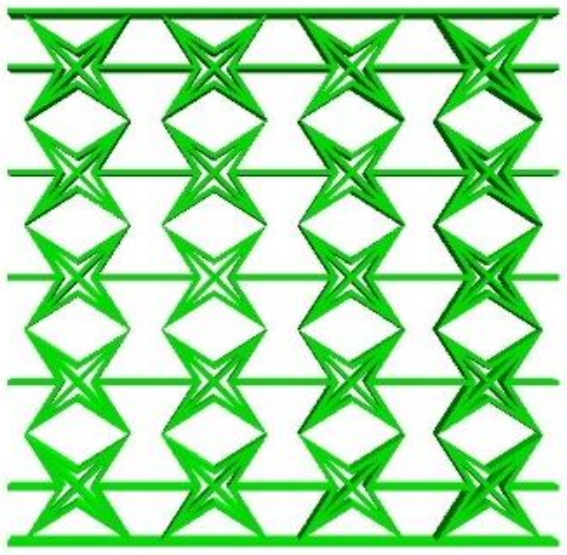 | [236] |
| Star-shaped auxetic structure X-type | 2D | 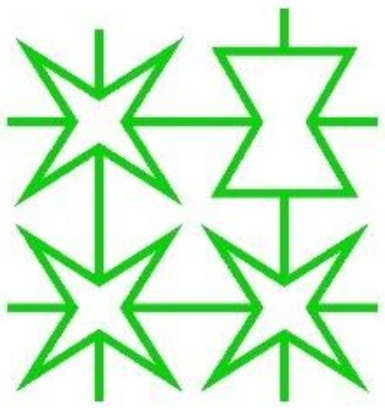 | 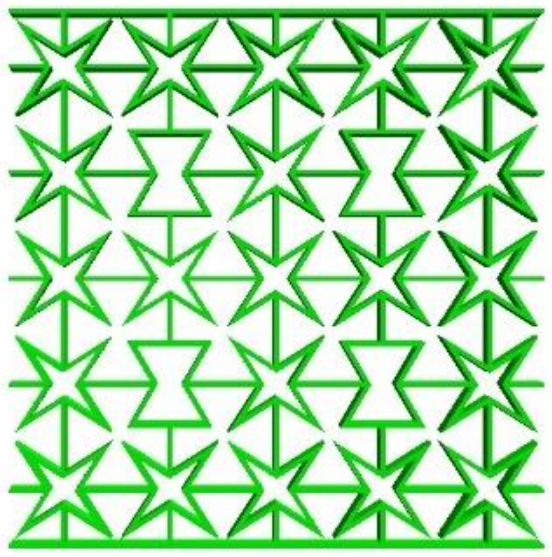 | [119] |
| Star-shaped auxetic structure H-type | 2D | 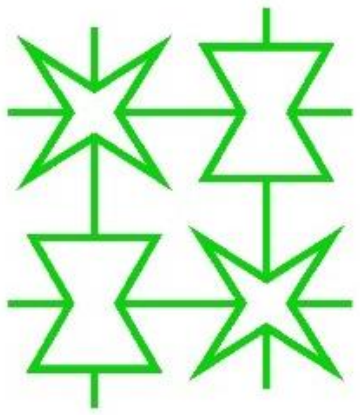 | 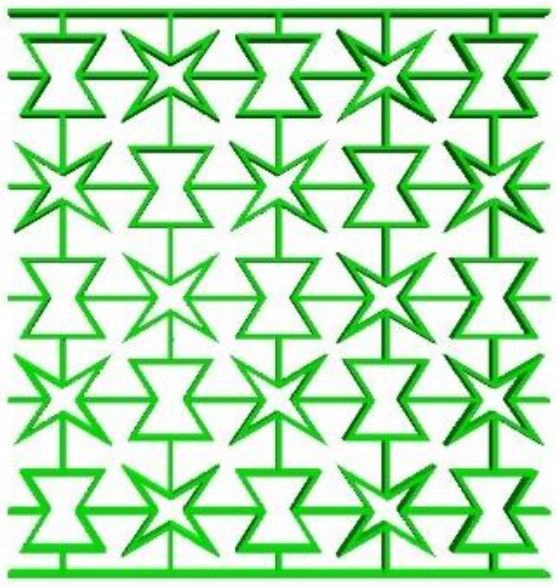 | [119] |
| Star-shaped auxetic structure SD-type | 2D | 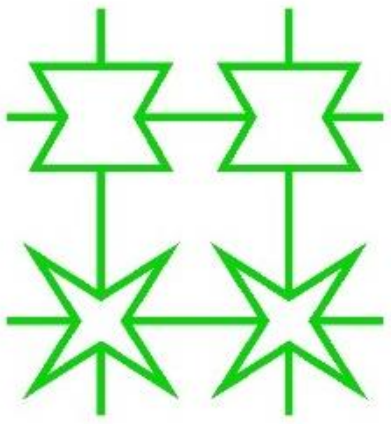 | 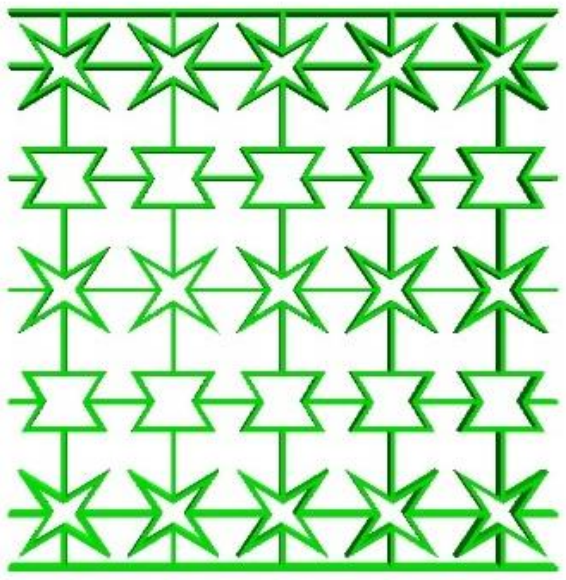 | [119] |
| 3D star-shaped auxetic structure | 3D | 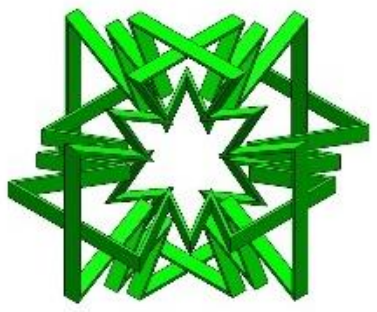 | 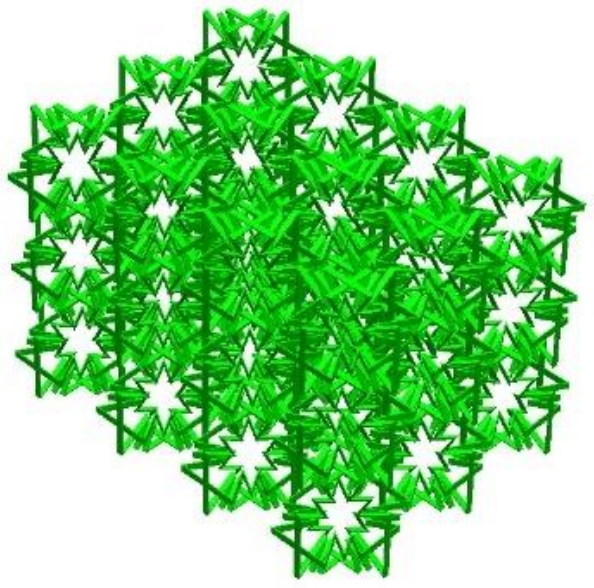 | [240] |

## *2.3. Chiral AMS*

The chiral structure is inspired by the chiral hydrocarbon chemical cell. The arrangement inspired by this chemical cell can be described as a central unit that can be a geometric shape in these structures, and these central units are connected to each other with geometric interfaces with a specific arrangement. The most common types of chiral have a circular geometry as a central geometry and various structures have been created by changing the number of links. The basic nomenclature of chiral structures is also based on this, the position of the links determines whether the chiral structure is conventional or auxetic. Fig. 6 shows the trichiral, tetra chiral, anti-trichiral, and anti-tetra chiral unit cells. For example, "Tetra" indicates the number of links attached to the central circle and has a positive Poisson's ratio. When the arrangement of the links in the structure changes and makes anti-tetra chiral, this structure exhibits auxetic behavior.

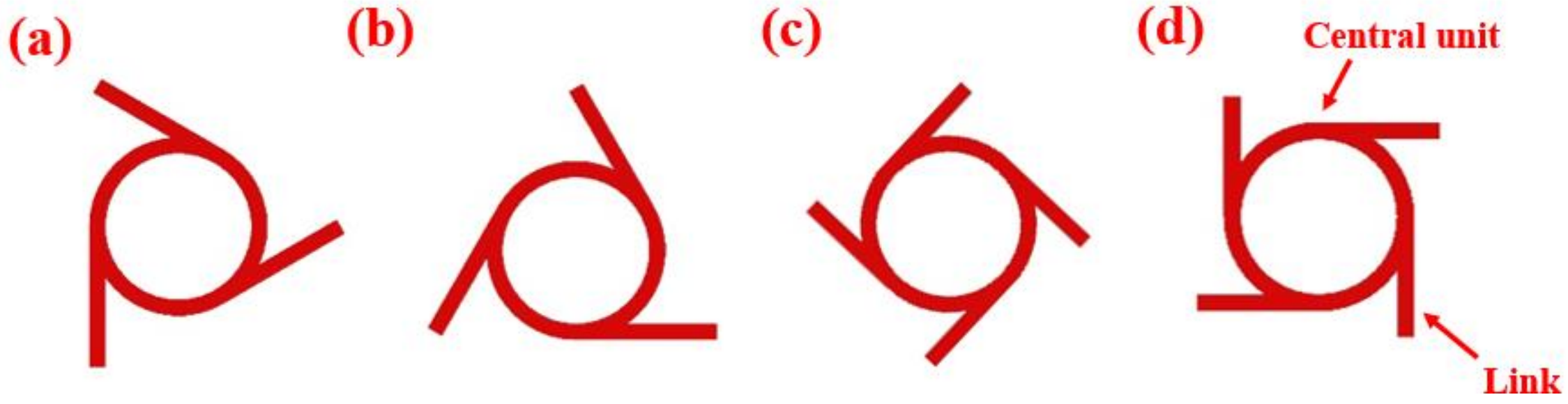

Fig. 6 Basic chiral unit cells, a) trichiral, b) anti-trichiral, c) tetra chiral, and d) anti-tetra chiral.

The beginning of research on chiral structures dates back to 1988. When Friss et al. [36] noticed the auxetic foam and the porous structure of the foam created a new idea for them. Then, Lakes [241] presented schematic research of the chiral structure in 1991, which is now known as three chiral. In 1992, Yeganeh-Haeri [242] considered the chemical structure of the chiral cell, and in 1993, Lakes [243] considered the same chemical schematic of the chiral cell. In 2000, Evans [244] presented a more advanced schematic for the chiral structure. Finally, in 2003, Wojciechowski [245] introduced the structure known as chiral for the first time. In 2014, Mir et al. [59] introduced the structure introduced in 2003 as a chiral cell. In the last decade, research in the field of chiral became more intense. In 2016, Mousanezhad et al. [246] numerically and more accurately investigated the normal chiral and auxetic chiral structures. Behaviors such as compressive behavior, impact behavior, bending behavior, numerical modeling, development of theoretical relationships, and the presentation of new improved structures have been on the agenda of researchers in recent years.

Chiral structures, particularly anti-tetra chiral metastructure among various configurations, exhibit remarkable mechanical properties such as high energy absorption, impact resistance, and tunable stiffness. Experimental and numerical studies indicate that chiral auxetic structures can achieve SEA values in the range of 20–32 kJ/kg [247], significantly higher than conventional honeycomb structures, which typically exhibit SEA values around 15–20 kJ/kg [248]. Under impact loading, energy absorption efficiency for these structures has been measured at 85–92% [249], making them highly suitable for crashworthiness applications [250]. Compressive strength for chiral auxetic structures varies based on material properties and geometric configurations, with reported values ranging from 5 to 12 MPa, while conventional non-chiral structures generally exhibit lower values, typically between 3 and 8 MPa [246]. Additionally, they demonstrate high impact resistance with peak force values recorded between 2–6 kN under quasi-static loading conditions, indicating their superior ability to dissipate energy efficiently. Furthermore, cyclic compression tests reveal that chiral auxetic structures retain over 80% of their original energy absorption capacity after multiple deformation cycles, showcasing their durability and resilience under repeated loading. Given these superior properties, further numerical modeling and experimental validation of chiral structures could significantly enhance their applicability in protective systems, biomedical scaffolds, and aerospace components [247].

For a better understanding of the auxetic behavior of an anti-tetra chiral structure, a schematic of the crushing behavior of this metastructure under axial compression is presented in Fig. 7. It should be noted that this behavior is schematic and the conditions and behavior of AMS collapse depends on the type, amount, speed of loading and construction conditions of these metastructures.

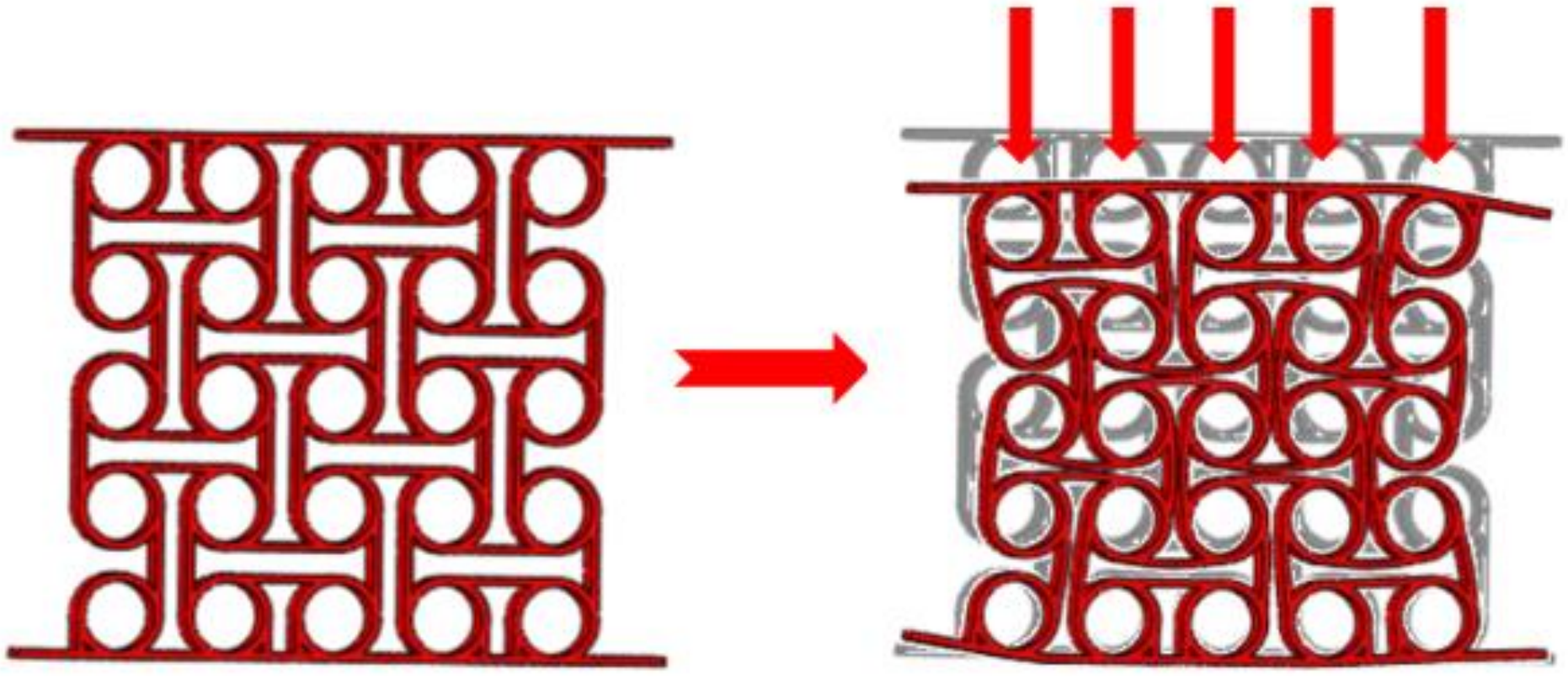

Fig. 7 Schematic of auxetic behavior of anti-tetra chiral structure under axial compression

The central geometry in the conventional chiral unit cell is a circle. In recent years, of course, the central geometries have been changed and other geometries have been proposed as shown in Table 3 [251]. The chiral model has an acceptable behavior under pressure. Researchers have obtained a Poisson's ratio of 0.3 to 0 for conventional chiral and an NPR of -0.1 to -0.3 for chiral auxetic structures. Different types of introduced and developed unit cells of this structure have been collected in Table 3.

Table 3 Classification of chiral AMSs.

| Structure name | Model dimension | Unit cell | Aggregate of unit cells | Reference |
|---|---|---|---|---|
| Anti-tetra chiral auxetic structure | 2D | 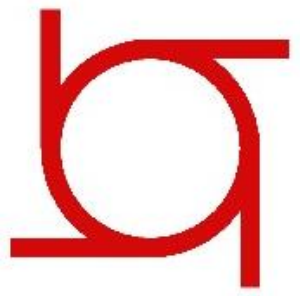 | 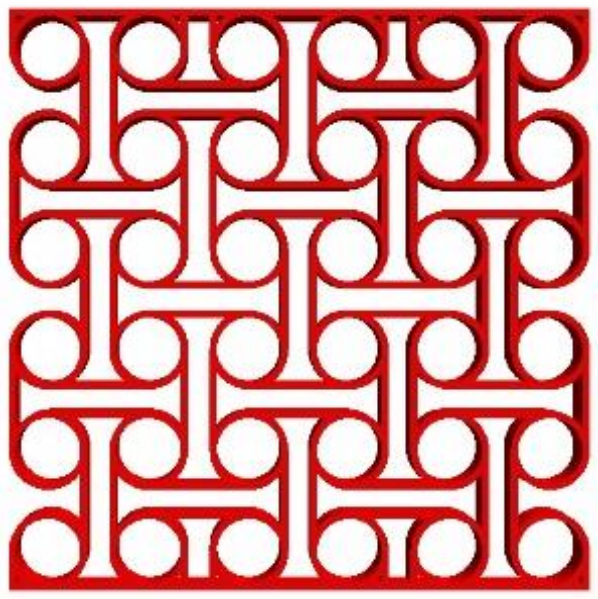 | |
| Tetra chiral auxetic structure | 2D | 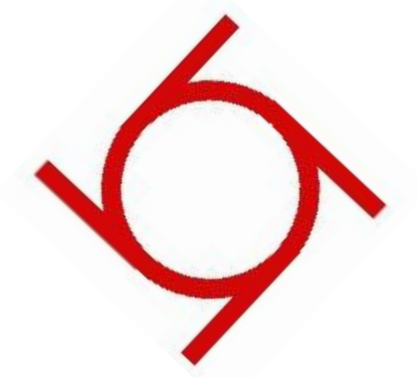 | 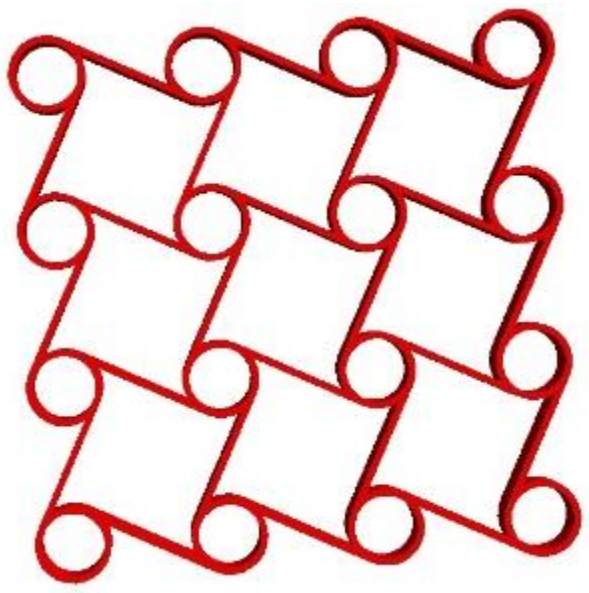 | [247] |
| Anti-trichiral auxetic structure | 2D | 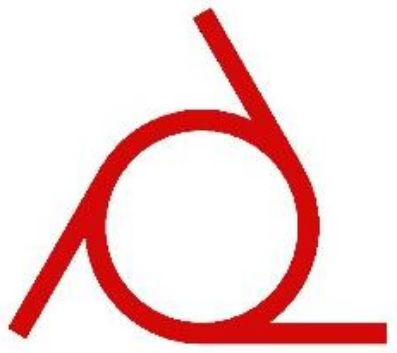 | 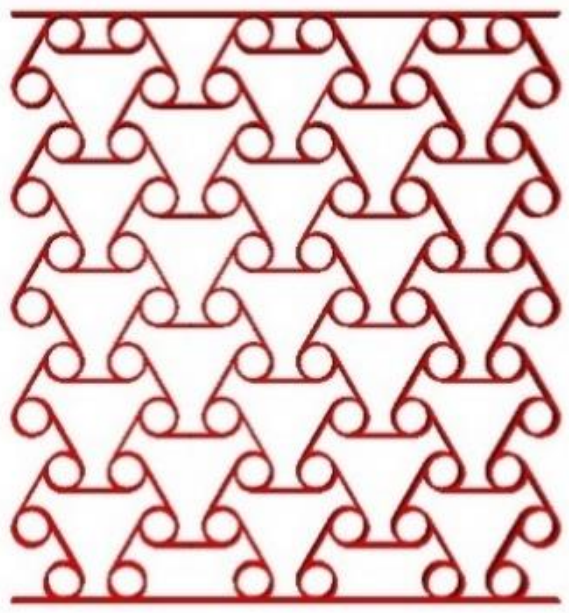 | [247] |
| Hexagonal hierarchical chiral auxetic structure | 2D | 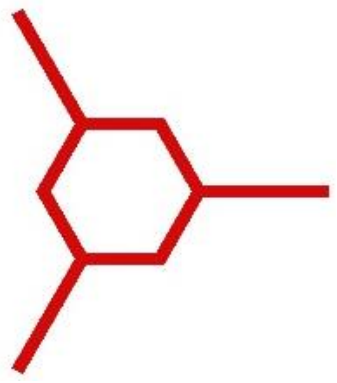 | 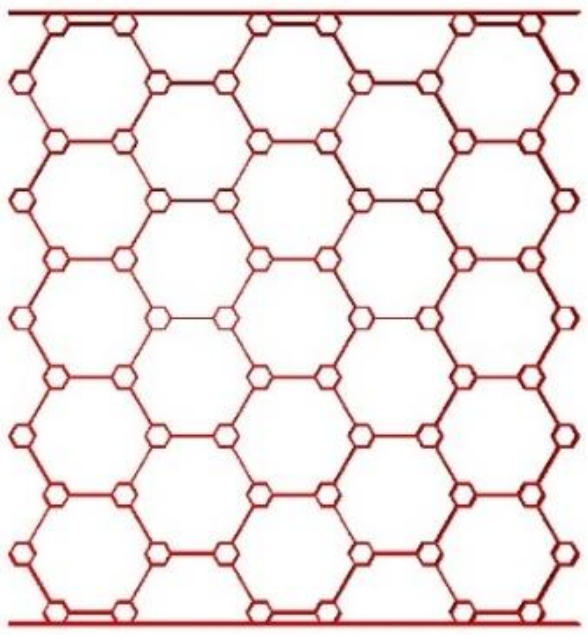 | [247] |
| Hierarchical anti-tetra chiral auxetic structure | 2D | 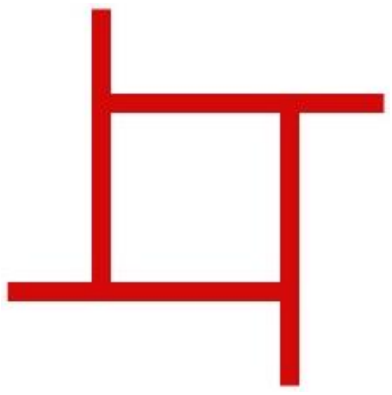 | 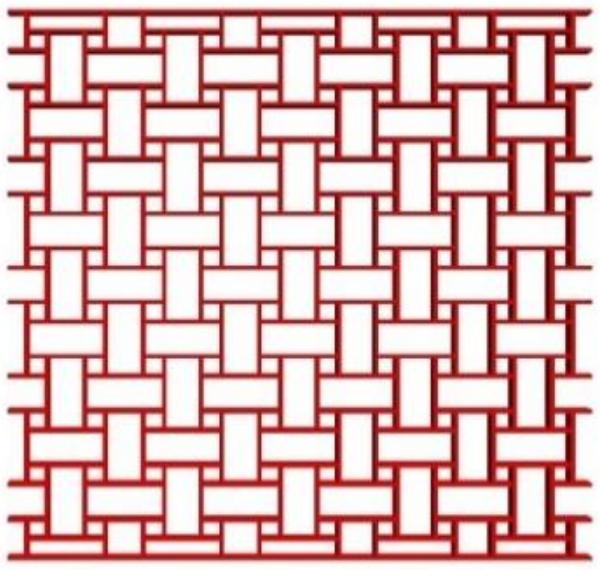 | [252] |
| Wavy anti-tetra chiral auxetic structure | 2D | 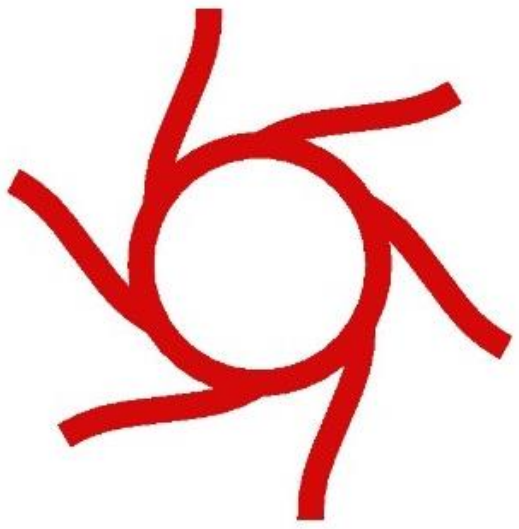 | 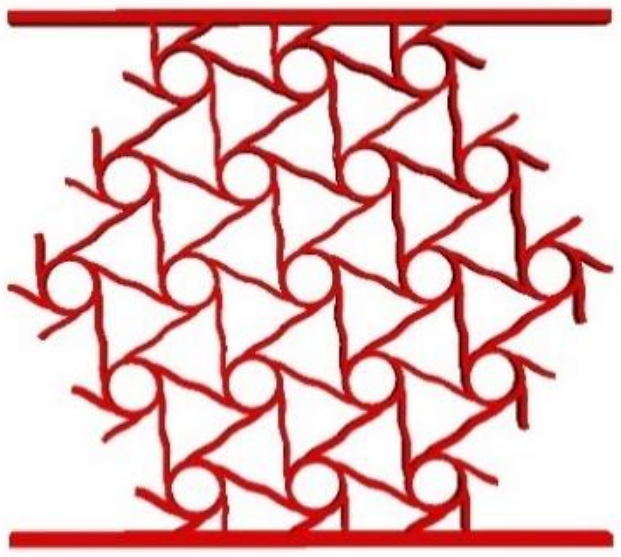 | [248] |
| Developed anti-tetra chiral auxetic structure | 2D | 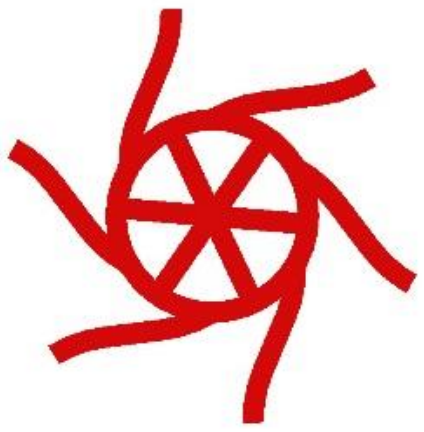 | 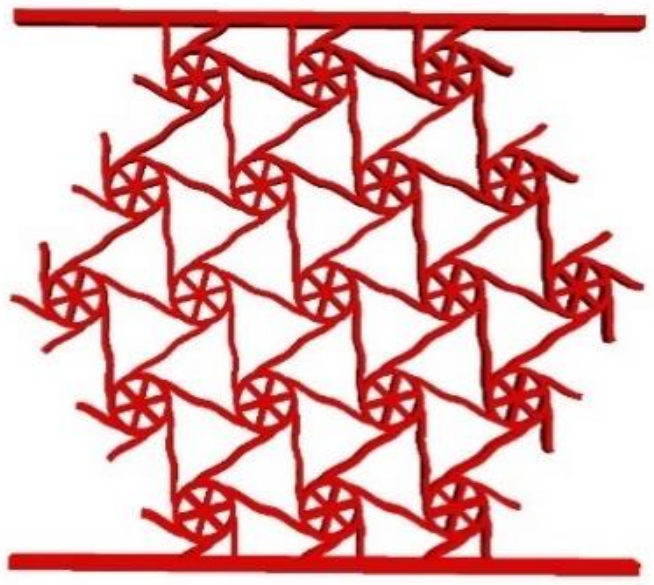 | [253] |
| Missing rib chiral auxetic structure with different core | 2D | 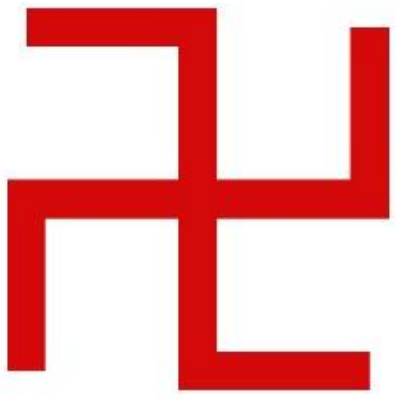 | 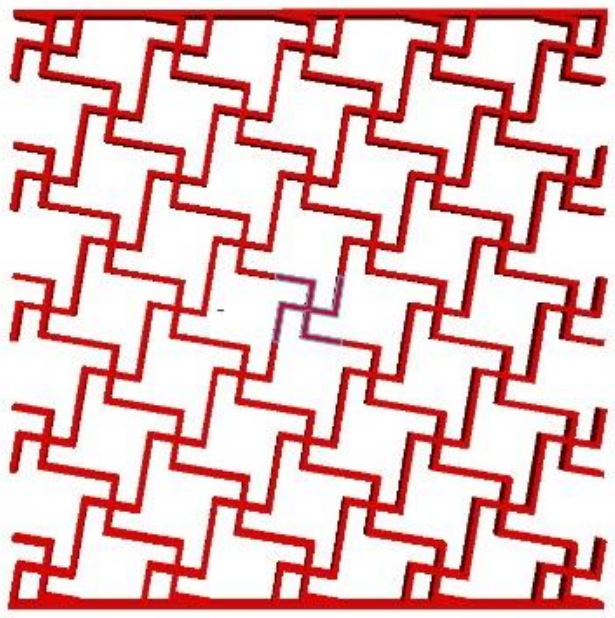 | [254] |
| Enhanced hexa missing rib structure | 2D | 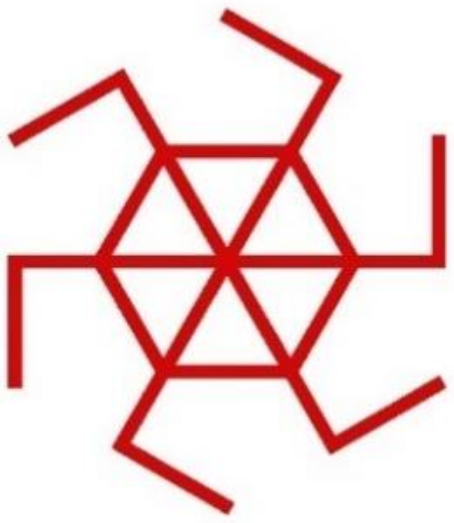 | 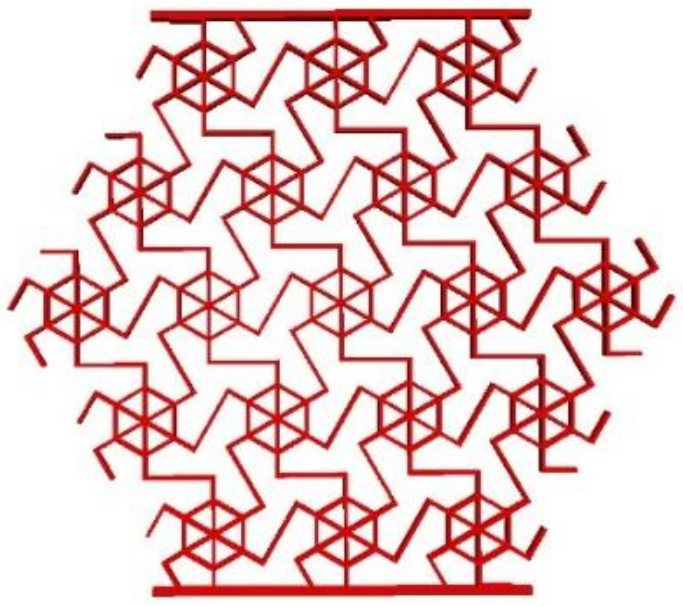 | [255] |
| Enhanced anti-tetra missing rib structure | 2D | 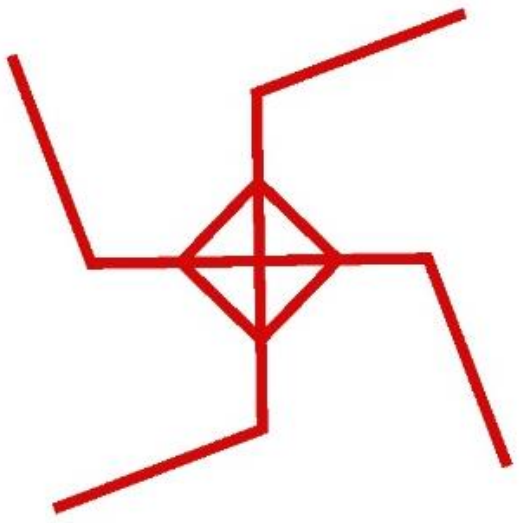 | 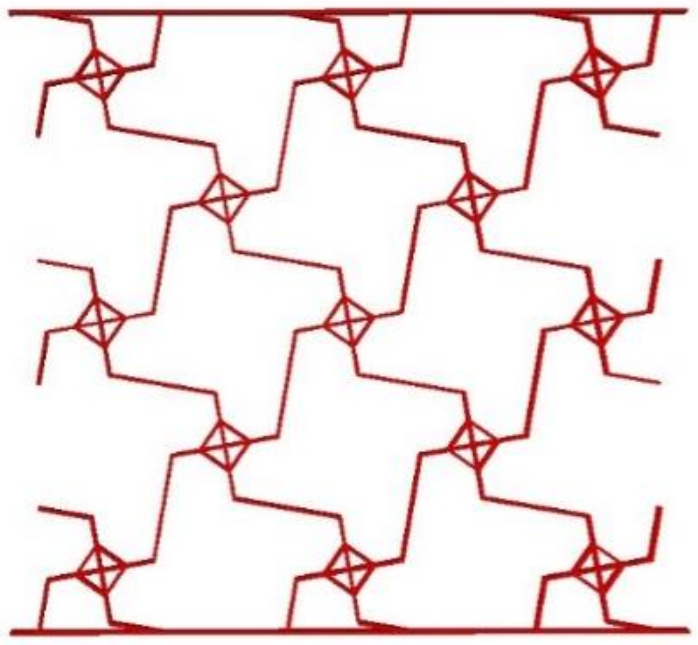 | [255] |
| Developed anti-tetra chiral auxetic structure | 2D | 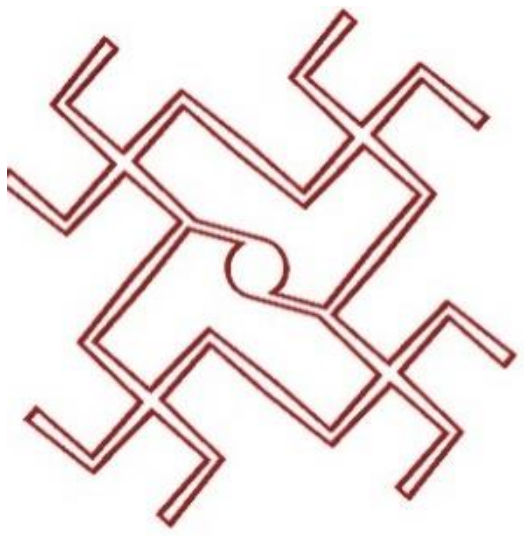 | 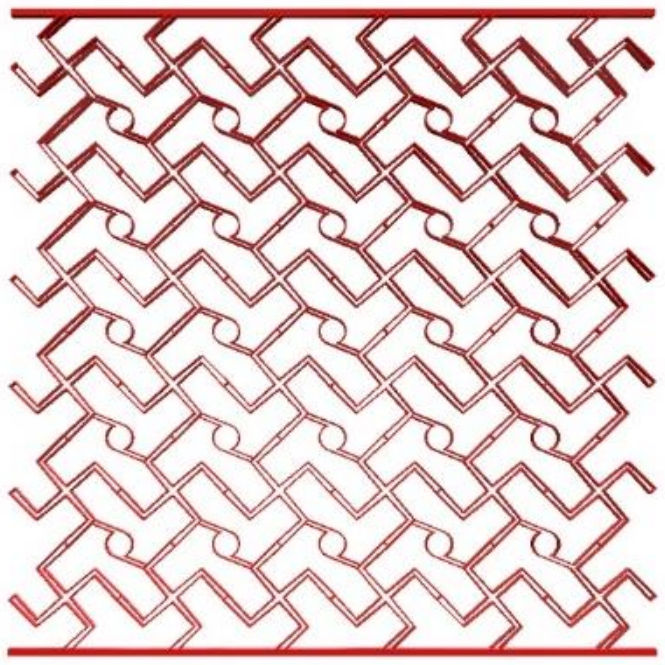 | [256] |
| 3D anti-tetra chiral auxetic structure with tension torsion coupling effect | 3D | 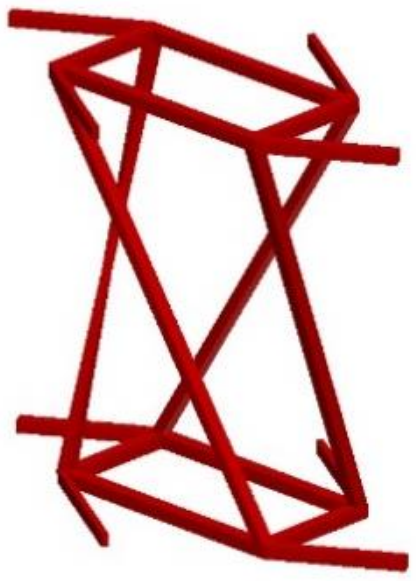 | 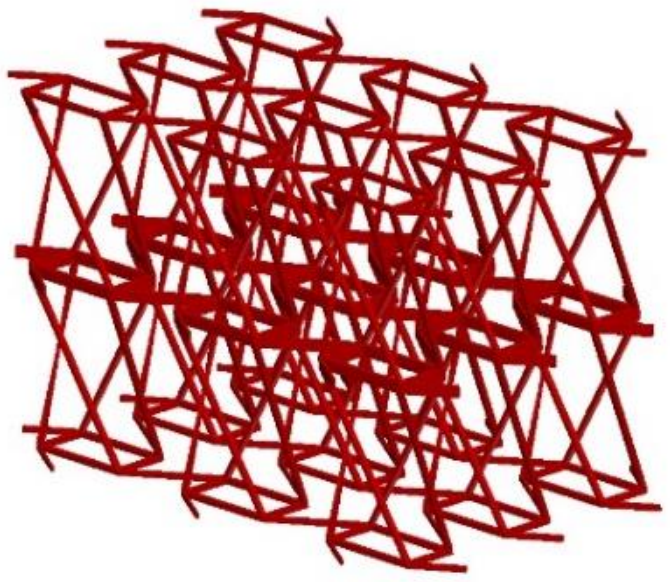 | [257] |
| 3D shell anti-tetra chiral auxetic structure | 3D | 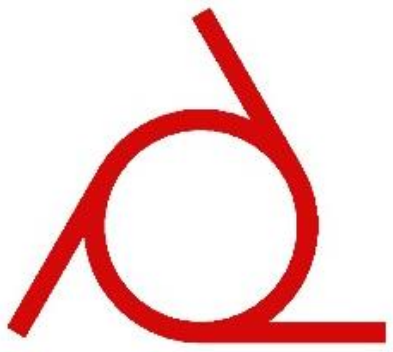 | 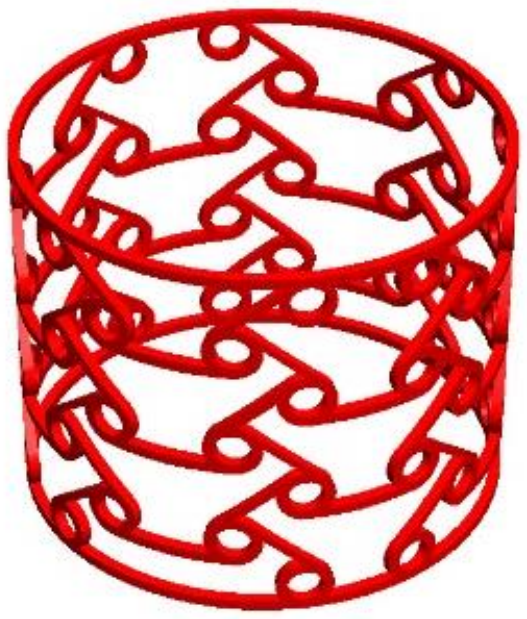 | [249] |
| 3D anti-tetra chiral auxetic structure with cubic core | 3D | 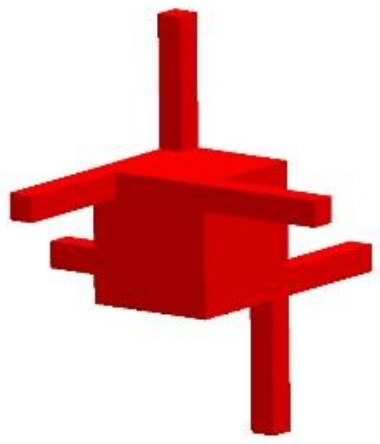 | 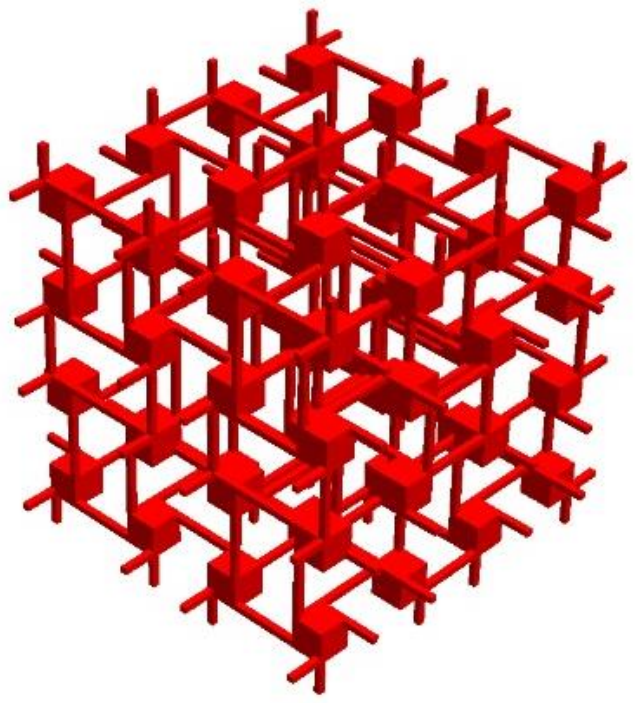 | [250] |
| 3D anti-tetra chiral auxetic structure | 3D | 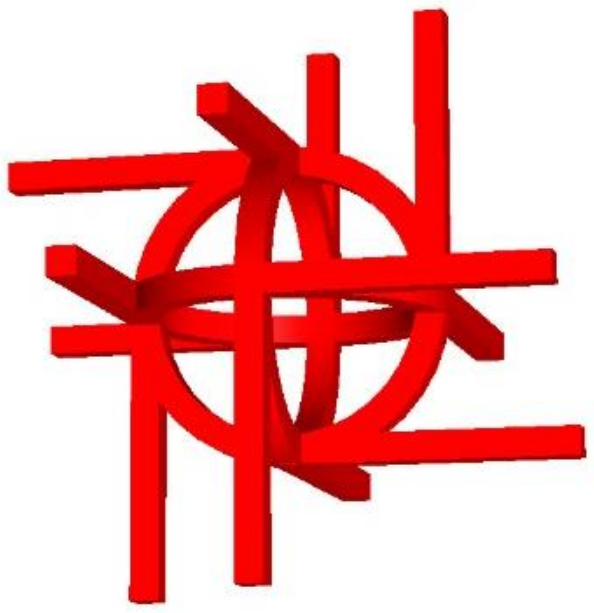 | 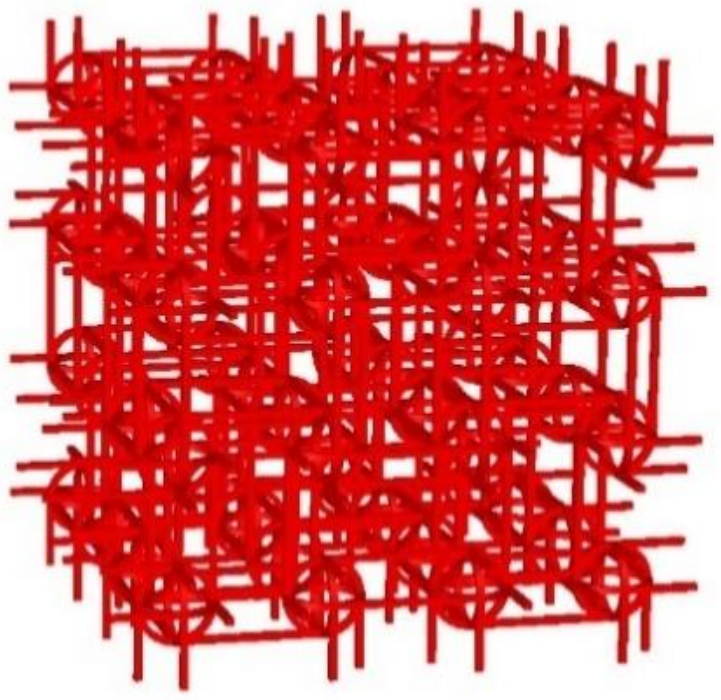 | [250] |
| 3D bio-inspired chiral auxetic structure | 3D | 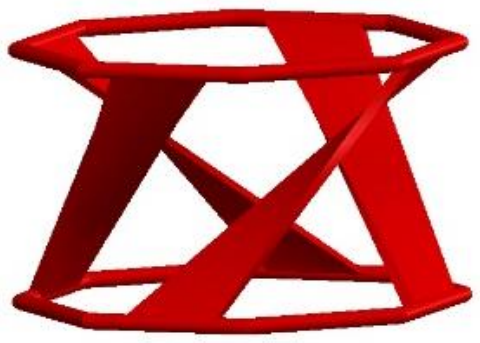 | 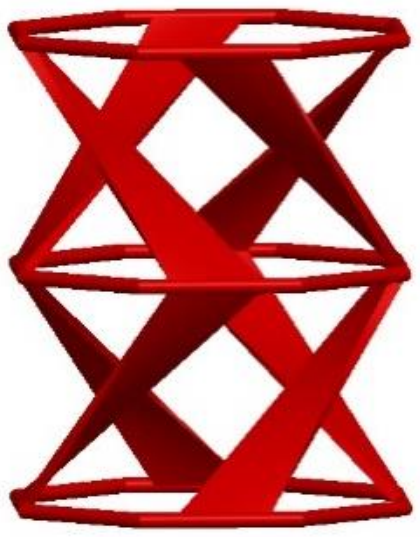 | [258] |

## *2.4. Arc-shaped AMS*

Arc-shaped AMS, in simple terms, are structures whose cells are like a square with wavy sides. The idea of this category of superstructures, unlike the previous categories, has been created and developed in recent years. Similar metastructures were first observed by Neville et al. in 2014 [259]. They were looking for a structure capable of controlling Poisson's ratio, and in their analysis, they examined structures similar to the arc-shaped structure. Then, in 2019, Choi et al. [260] proposed an auxetic structure with Poisson's ratio and optimal controllability. Their research work was similar to the first proposed unit cell of the arc-shaped structure. Then, in 2020, various researchers drew their attention to this category of metastructures. In 2020, Novak et al. [261] proposed a metastructure that can experience different strain rates. They actually proposed the 3D arc-shaped unit cell and introduced it as a chiral unit cell. Necemer et al. [262] did the same thing in 2020.

There are different names for arc-shaped metastructures. Chiral can be mentioned among them because some researchers believe that these metastructures are formed if the chiral unit cell is without a central cell [261–264]. Others have attributed the name bulking-inspired metastructure to these structures because they consider this category of superstructures to be the result of buckling the sides of a square [265,266]. In recent years, the name arc-shaped has been used more because researchers have used more than one wave in a single cell in different directions, and they may have used private relationships for the waveform used [267]. Finally, this name was selected for this category.

Arc-shaped AMS, characterized by their wavy-sided unit cells, demonstrate exceptional energy absorption and mechanical resilience. Studies indicate that these structures achieve SEA values in the range of 28–40 kJ/kg [267], significantly surpassing conventional honeycomb structures with SEA values of 15–25 kJ/kg [261]. Under impact loading, arc-shaped AMS exhibit energy absorption efficiencies exceeding 90%, making them highly effective for crashworthiness applications [264]. Their compressive strength varies depending on geometric parameters and material selection, with reported values ranging from 7 to 14 MPa, whereas standard cellular structures typically demonstrate 4–9 MPa in similar loading conditions [268]. Additionally, dynamic testing has shown that arc-shaped structures can withstand peak impact forces between 3 and 7 kN, indicating their superior energy dissipation capacity. Furthermore, experimental cyclic compression tests reveal that arc-shaped metastructures retain more than 85% of their initial energy absorption capacity after multiple deformation cycles, highlighting their durability and long-term efficiency under repeated loading [260]. These unique mechanical properties make arc-shaped AMS promising candidates for advanced protective systems, biomedical applications, and aerospace components.

A schematic of the behavior of this metastructure under axial crushing is presented in Fig. 8. The collapse mechanism mainly depends on the wavy-sided unit cells.

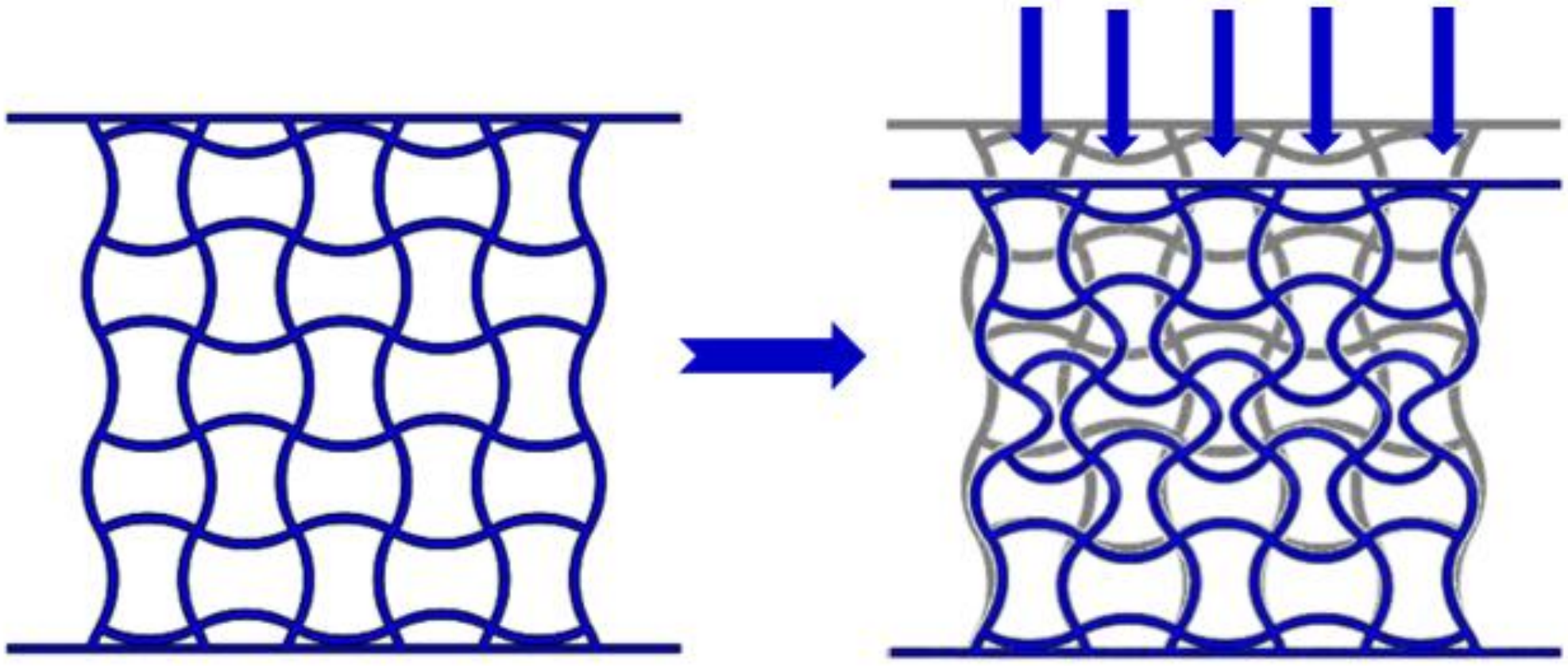

Fig. 8 Schematic auxetic behavior of an arc-shaped structure under axial compression.

Arc-shaped structures have the potential to be applied to other frameworks, resulting in the creation of distinctive properties. The arc-shaped model has an acceptable behavior under pressure. Researchers have obtained Poisson's ratio of -0.3 to -1 for this class of structures. Table 4 summarizes the developed unit cells of this structure so far.

Table 4 Classification of arc-shaped AMS

| Structure name | Model dimension | Unit cell | Aggregate of unit cells | Reference |
|---|---|---|---|---|
| Arc-shaped auxetic structure | 2D | 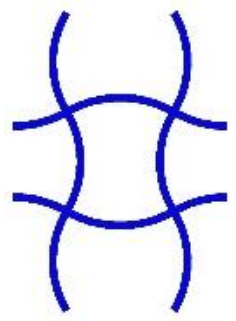 | 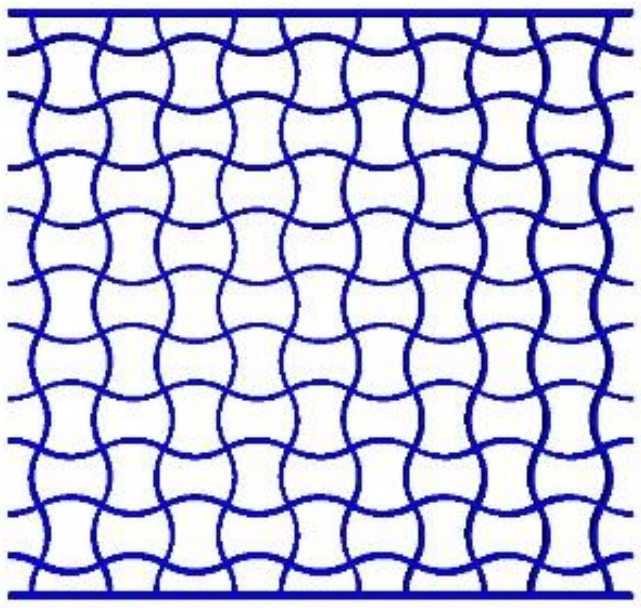 | |
| N1N3 arc-shaped auxetic structure | 2D | 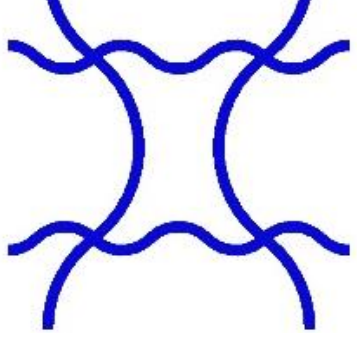 | 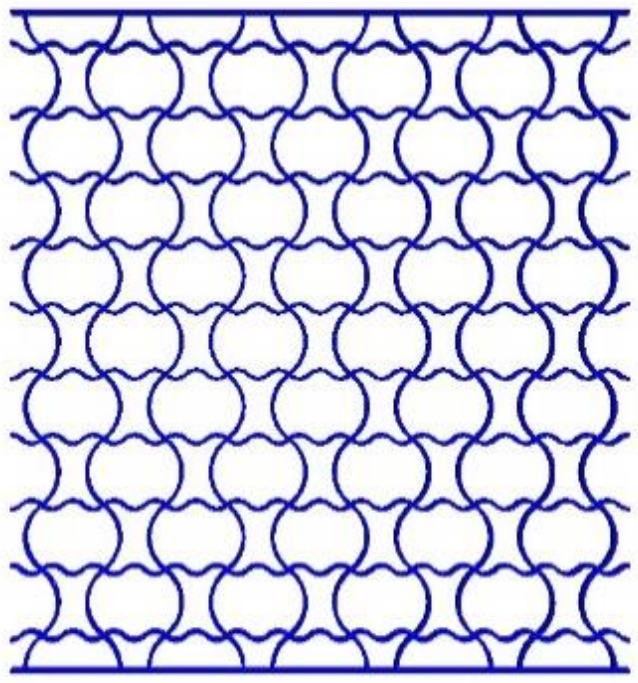 | [267] |
| N3N1 arc-shaped auxetic structure | 2D | 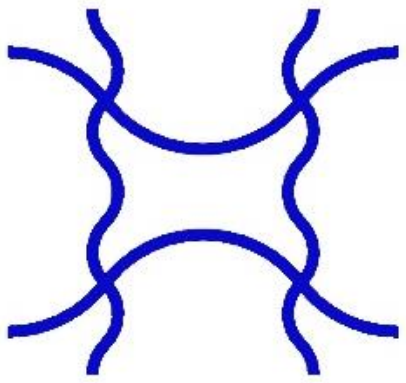 | 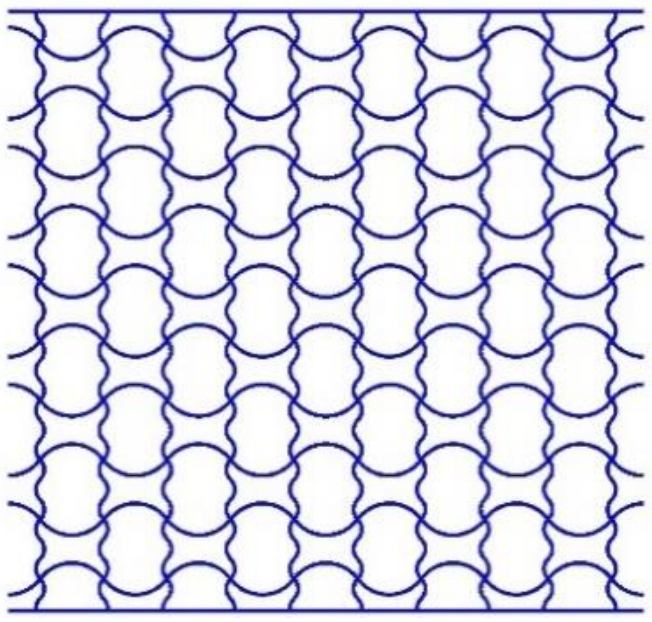 | [267] |
| N3N3 arc-shaped auxetic structure | 2D | 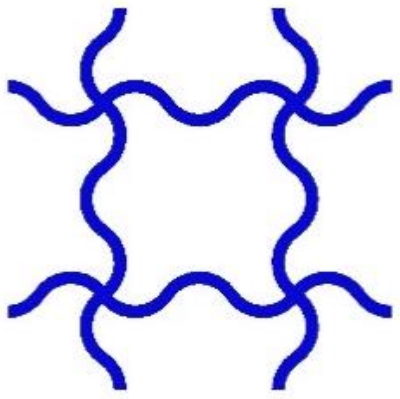 | 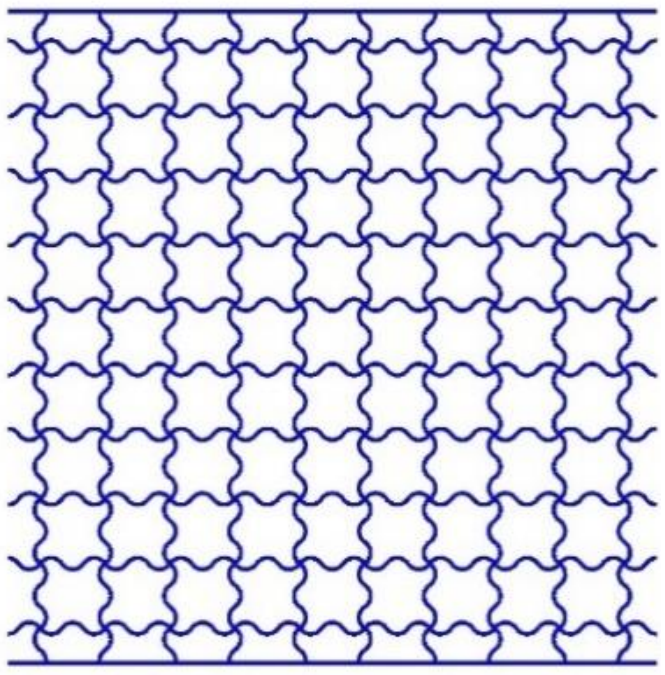 | [267] |
| Arc-shaped auxetic structure with different thicknesses | 2D | 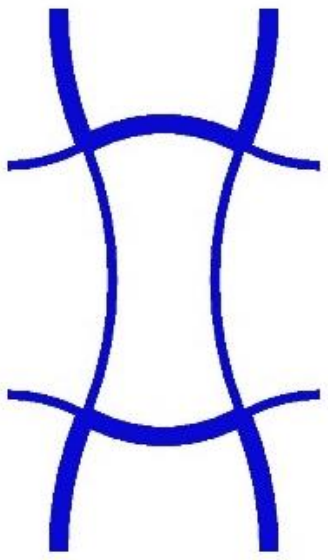 | 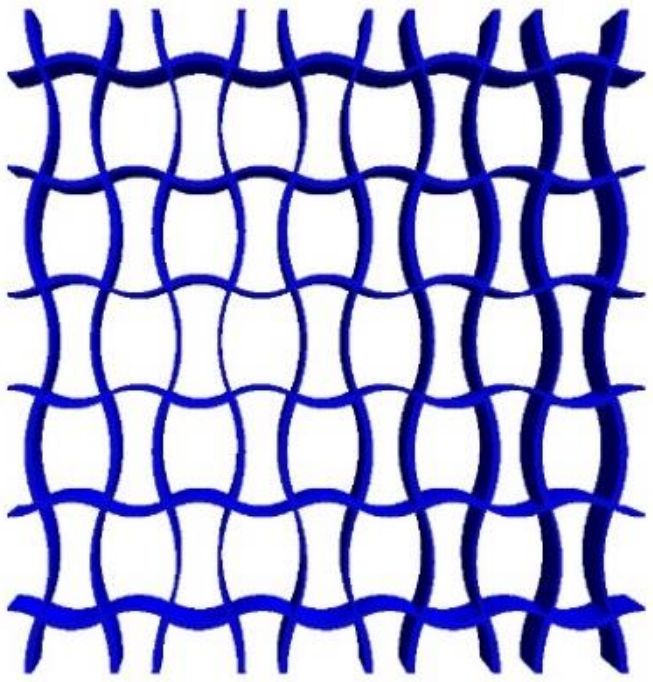 | [268] |
| 3D arc-shaped auxetic structure | 3D | 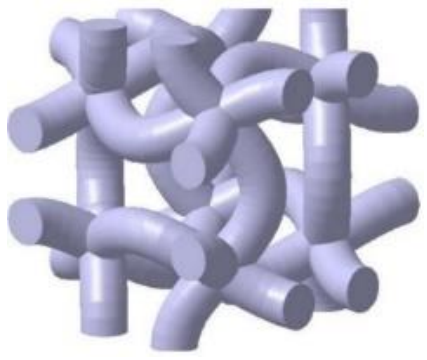 | 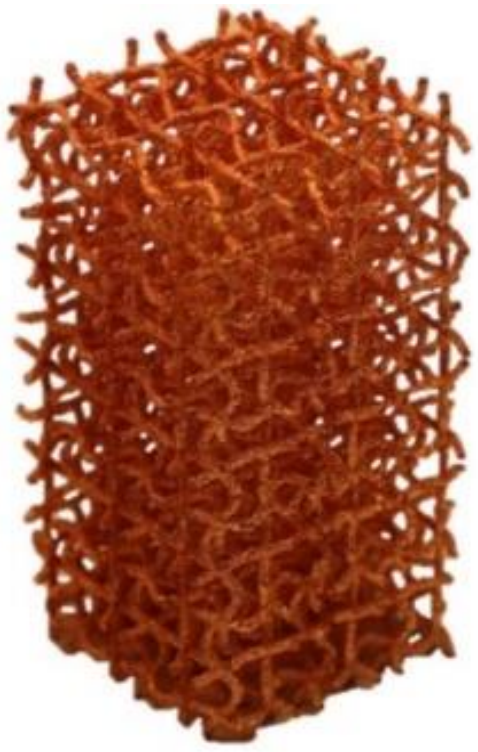 | [261] |

| | | | |
|---|---|---|---|
| 3D cylindrical arc-shaped auxetic structure | 3D | 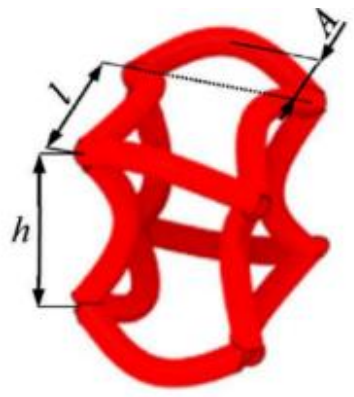 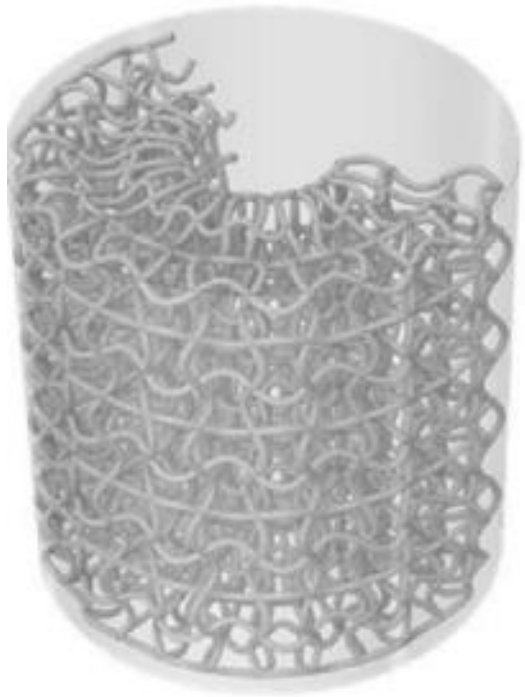 | [264] |

### *2.5. Arrow AMS*

The main idea of the arrow metastructures design is to use topology optimization techniques to design micro-scale structures with an NPR. In fact, unlike other metastructures that are generally inspired by looking at nature, this category of metastructures is developed based on mathematical relationships. The beginning of research in this field goes back to Larsen et al. [269] in 1997. They designed and developed sensors with special mechanical capabilities for MEMS systems. In 2006, Grima [270] mentioned these structures as a field of research in AMS. In 2018, Zhao et al. [271] investigated the effect of impact in this category of structures.

This category of metastructures was identified as a separate family due to the fact that they required a new and different geometry from the previous classes. In 2D, the arrow structure can be defined as a triangle with the lower side of the triangle bent inward. They did not have variety, but this geometry in 3D has been able to present different situations and it has been more interesting for researchers. These structures have demonstrated SEA ranging from 35 to 45 kJ/kg, significantly outperforming conventional auxetic and non-auxetic metamaterials, which typically exhibit SEA values of 20–30 kJ/kg [274]. Impact resistance studies reveal that arrow metastructures exhibit energy absorption efficiencies exceeding 92%, making them particularly useful in high-performance protective applications such as aerospace and automotive crashworthiness [275]. The compressive strength of these structures varies between 10 to 18 MPa, depending on material properties and geometric configurations, while standard AMS typically ranges from 6 to 12 MPa under similar loading conditions [276]. Additionally, dynamic impact testing indicates that arrow metastructures can sustain peak impact forces between 4 and 8 kN, highlighting their exceptional ability to dissipate mechanical energy efficiently [277]. Long-term cyclic compression tests show that these structures retain over 87% of their original energy absorption capacity even after 1,500 deformation cycles, making them highly durable under repeated mechanical loads [278]. Given their outstanding mechanical properties and adaptability in both 2D and 3D configurations, arrow metastructures offer promising potential for applications in impact-resistant systems, MEMS devices, and advanced lightweight structural components.

In recent years, researchers have focused on stiffness, Poisson's ratio, compressive strength, vibration behavior, ideal proportions for this structure and the development of structures with geometries that have improved mechanical behavior. The arrow-shaped model has an acceptable behavior under pressure. Researchers have obtained Poisson's ratio of -0.3 to -0.5 for this class of structures. Table 5 presents the different types of introduced unit cells of this structure so far.

Table 5 Classification of arrow AMS

| Structure name | Model dimension | Unit cell | Aggregate of unit cells | Reference |
|---|---|---|---|---|
| Arrow auxetic structure | 2D | 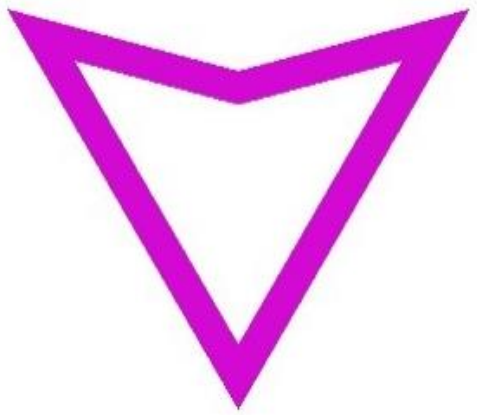 | 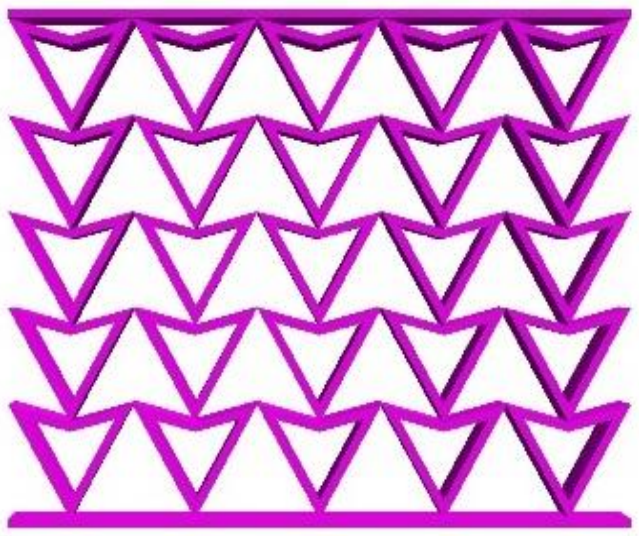 | |
| Double arrowhead auxetic structure | 2D | 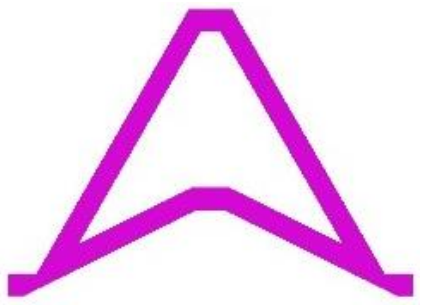 | 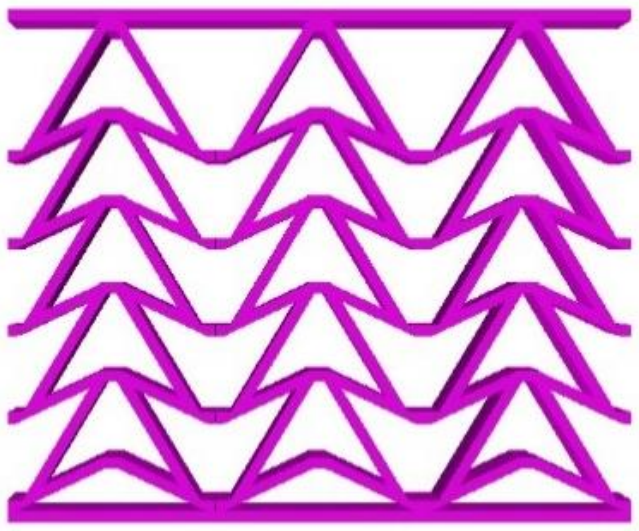 | [272] |
| 3D arrow auxetic structure | 3D | 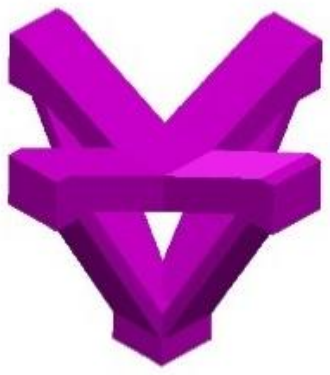 | 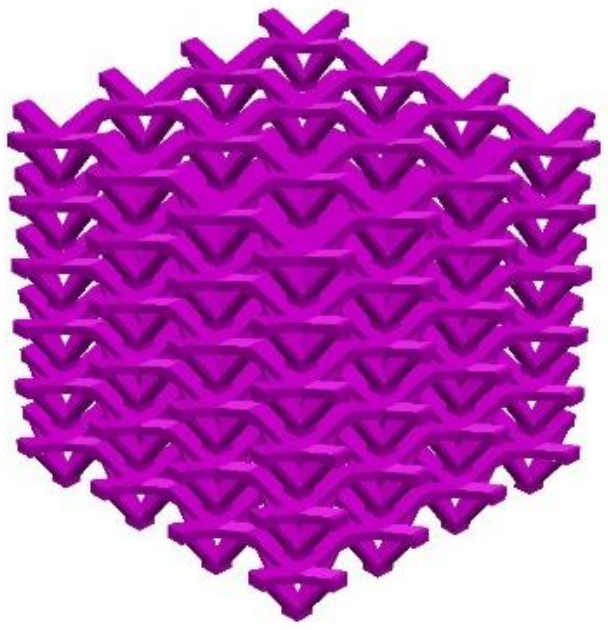 | [273] |
| Triangle tessellated double arrowhead auxetic structure | 3D | 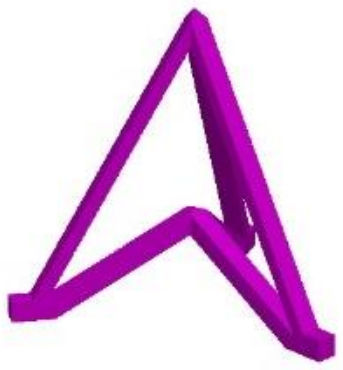 | 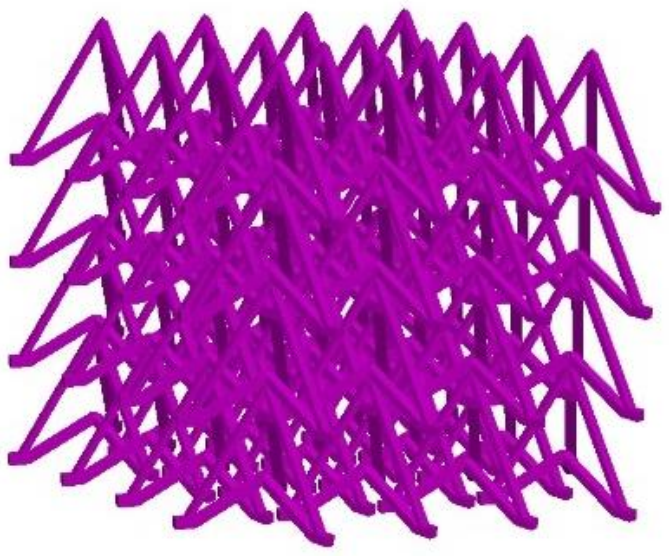 | [274] |
| Hexagon tessellated double arrowhead auxetic structure | 3D | 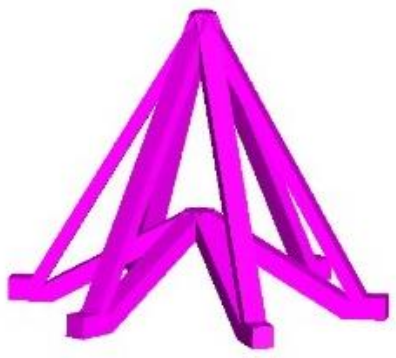 | 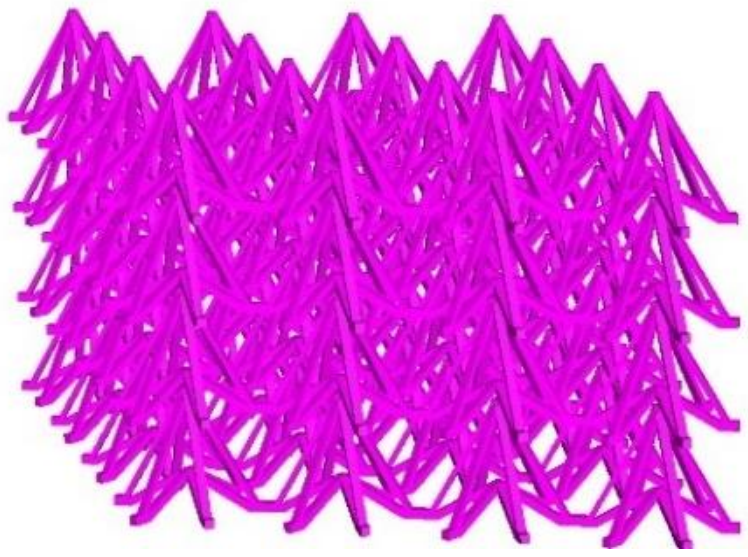 | [274] |
| TCU arrow auxetic structure | 3D | 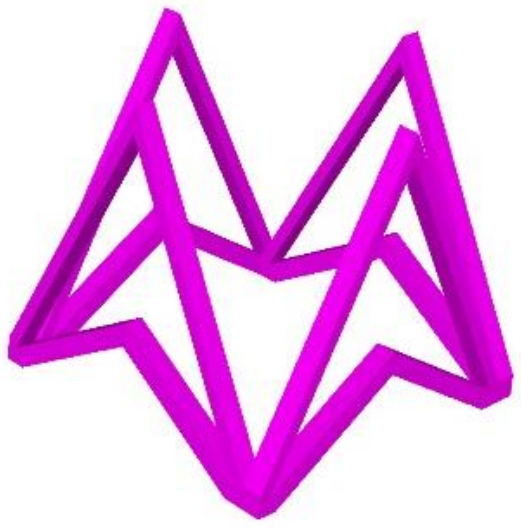 | 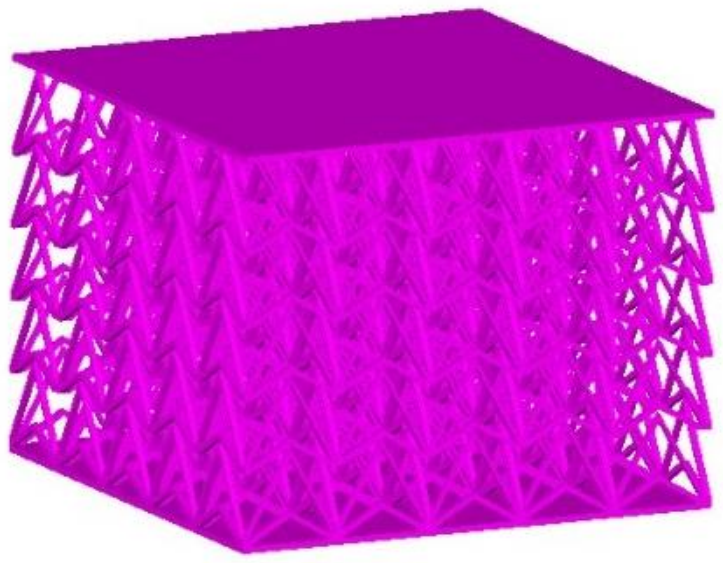 | [275] |
| TCU arrow auxetic structure | 3D | 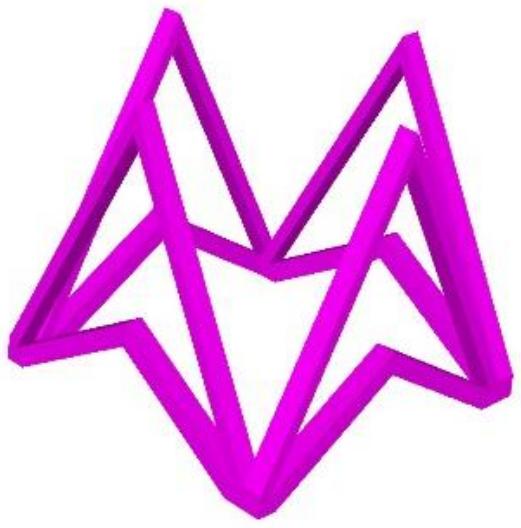 | 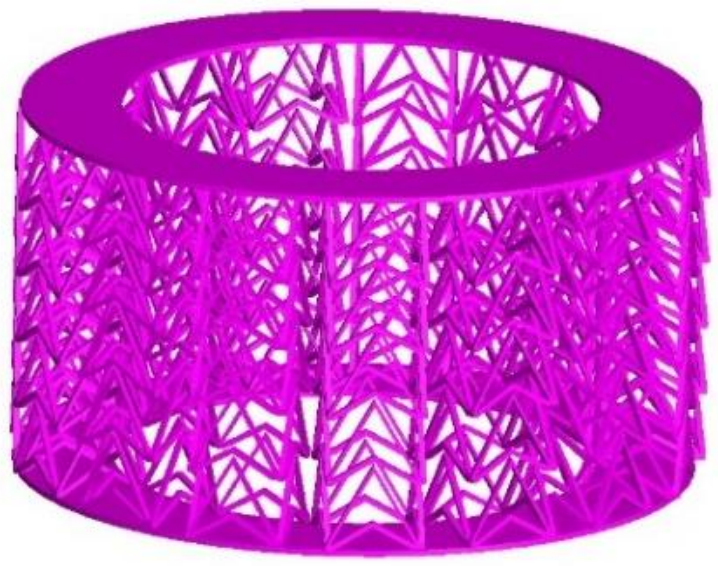 | [275] |

### *2.6. Peanut AMS*

This group of metastructures had an interesting attitude in their creation. The basis for creating a single cell of this category was based on the creation of holes with a specific geometry in a specific geometric cell. Maybe this basic idea is compatible with different categories of AMS, but the obvious difference between this category of metastructures and other metastructures is that in this category, the design point of view is not based on a specific geometry unit cell, but any geometry cell with arbitrary holes can be designed in this category. The variable parameters in the unit cells of this category can also be set based on the holes. This variable parameter allows designers to apply their desired changes more freely.

The first research according to the unit cells of this category goes back to 2014, when Shen et al. [276] were looking for a metamaterial with intelligent behavior. They proposed the first example of this class of metastructures and presented it as an auxetic cube. Then, Carta et al. [277] presented a different unit cell from this model of metastructures in 2016. In 2016, Hung Ho et al. [278] also presented a model that creates auxetic behavior based on cavity geometry change. In 2020, Yao et al. [279] investigated the effects of different cavities on a square unit cell. In 2021, Zhang et al. [280] investigated the circular holes and found the peanut shape in their analysis. Wang et al. [281] also presented a new model in 2021 using the peanut model and rotation in holes. In 2021, Hanet al. [282] developed the peanut model and investigated the effects of the number of holes. In 2022, Han et al. [283] created the peanut model in a tubular state and studied its properties.

In recent years, this category of metastructures has been the focus of researchers. They have tried this category of metastructures on different materials [284]. Also, the investigation of holes with different geometries [285–290], cylindrical and rectangular cube models of the peanut auxetic structure [291–294] and the development of new and improved geometries [295–297] have been on the agenda of the researchers. Researchers have obtained Poisson's ratio of -0.1 to -0.5 for this class of structures. Various types of introduced unit cells of this structure have been collected and summarized in Table 6.

Table 6 Classification of peanut AMS

| Structure name | Model dimension | Unit cell | Aggregate of unit cells | Reference |
|---|---|---|---|---|
| Peanut auxetic structure | 2D | 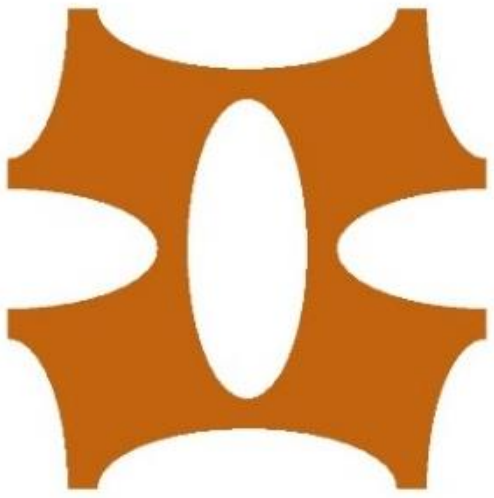 | 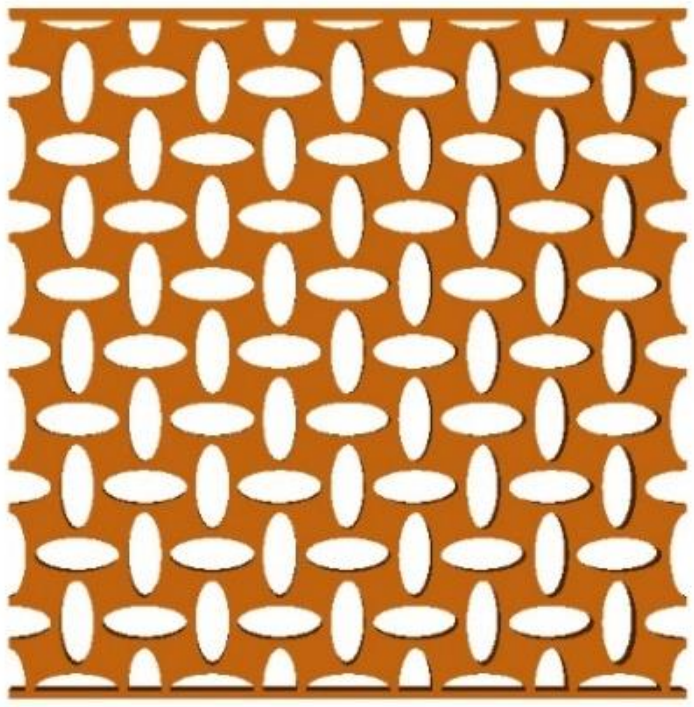 | [282] |
| Developed peanut auxetic structure | 2D | 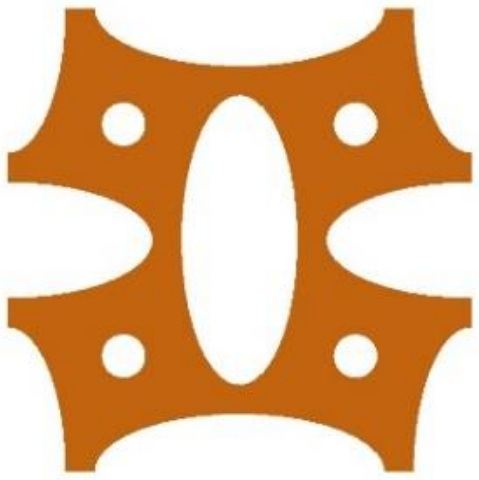 | 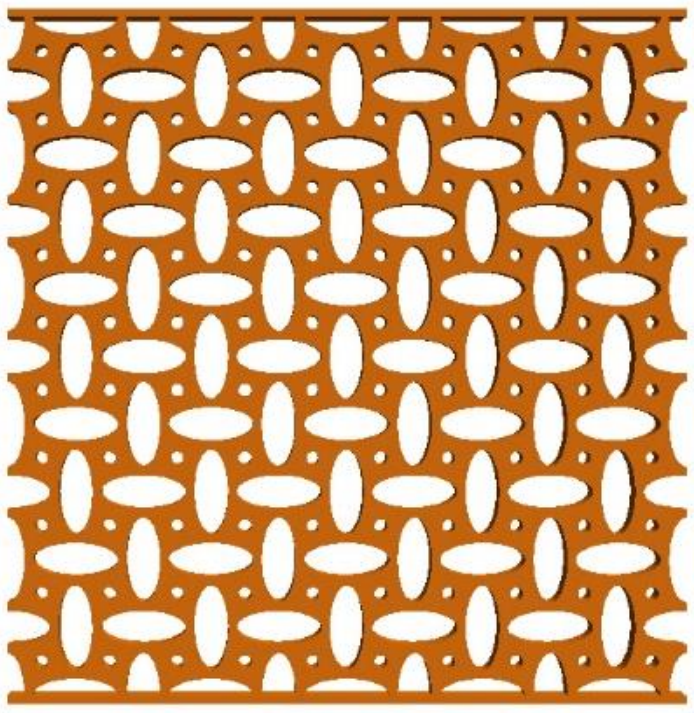 | [282] |
| Developed peanut auxetic structure | 2D | 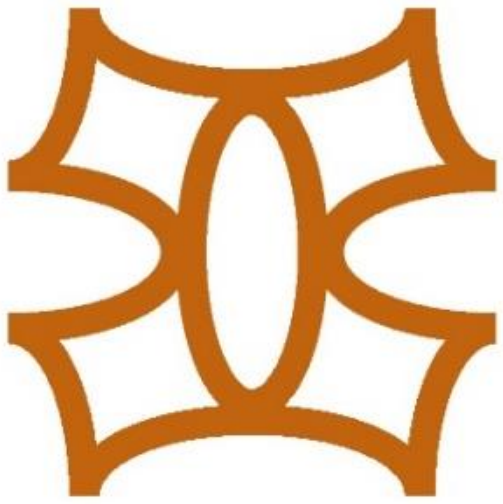 | 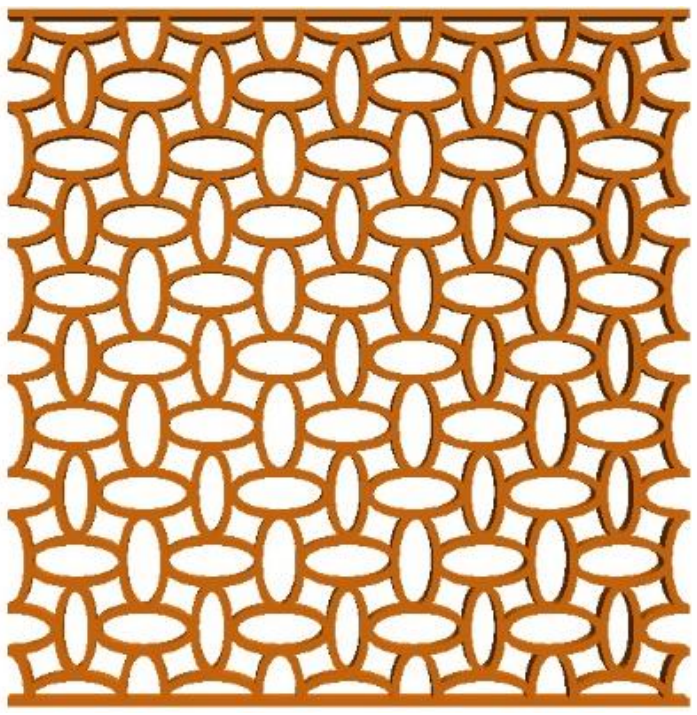 | [282] |
| Developed peanut auxetic structure | 2D | 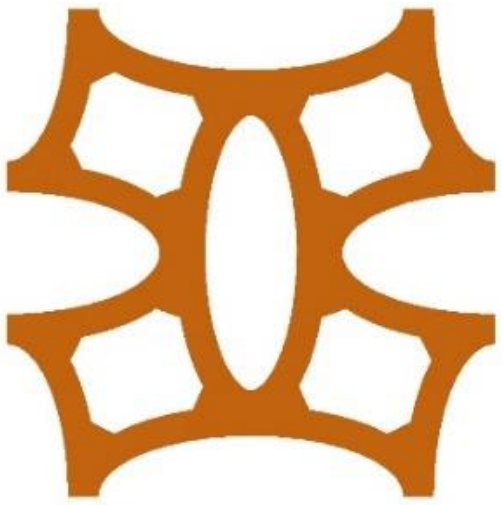 | 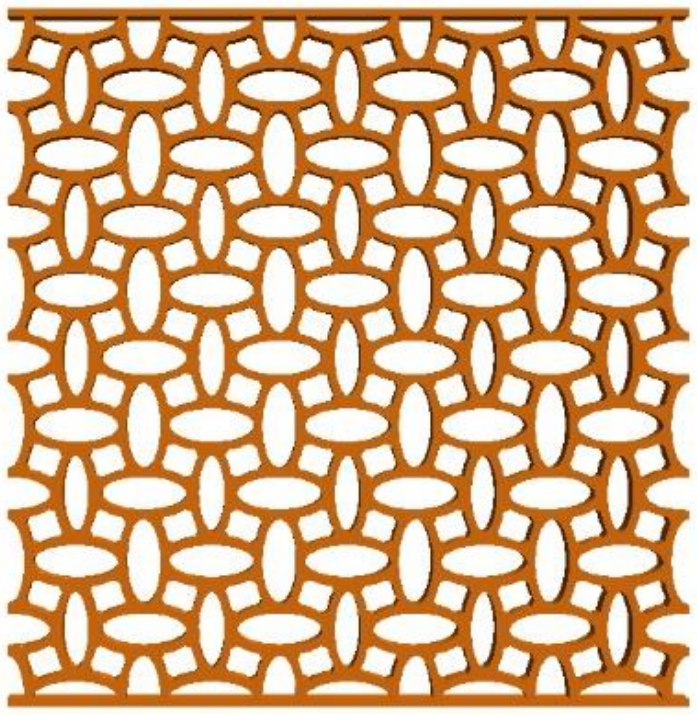 | [282] |
| Liner peanut auxetic structure | 2D | 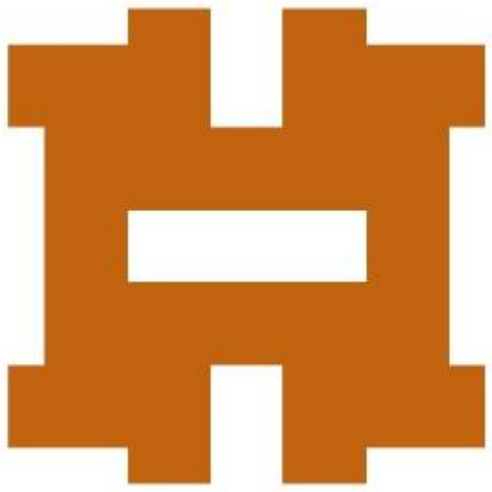 | 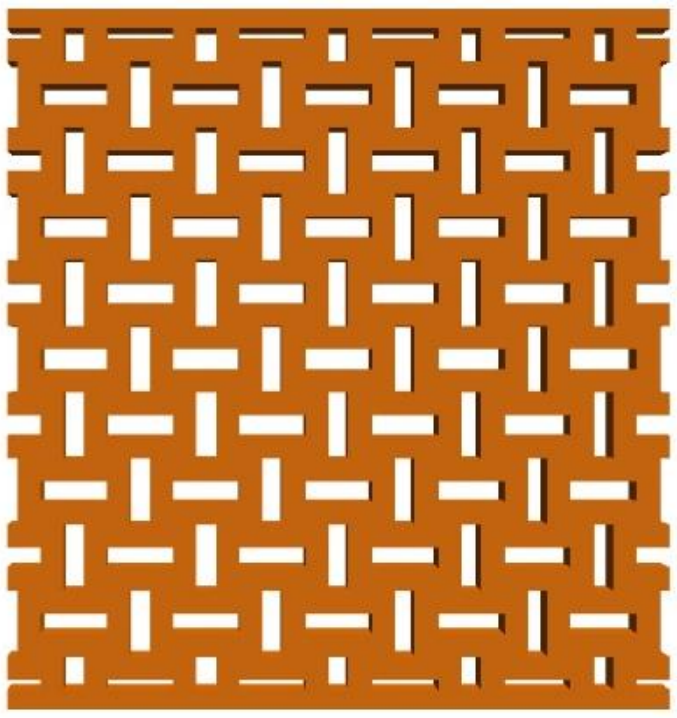 | [279] |
| Circle peanut auxetic structure | 2D | 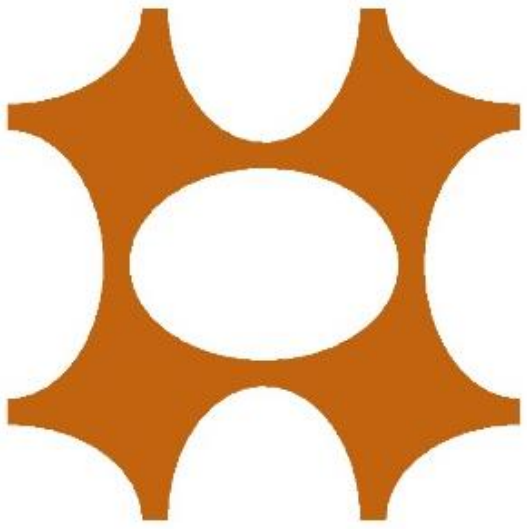 | 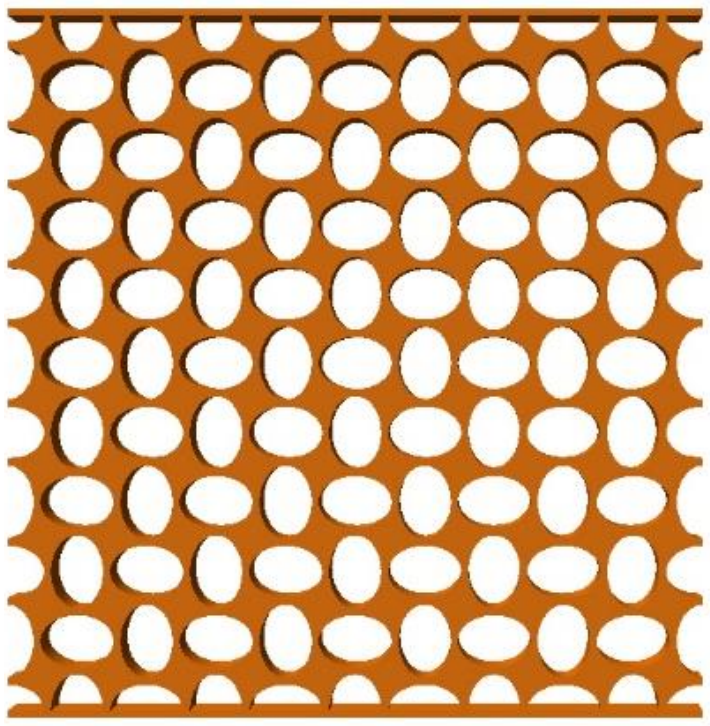 | [280] |
| Liner peanut auxetic structure | 2D | 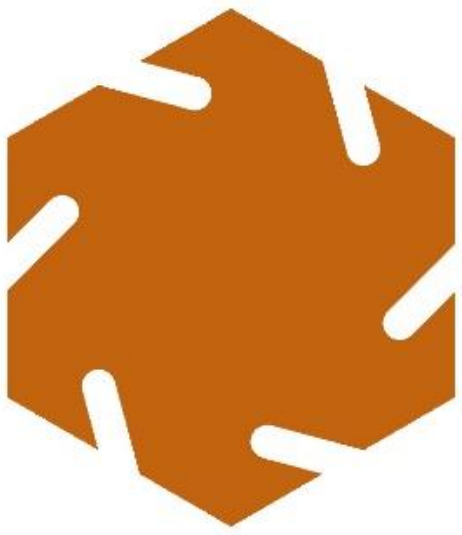 | 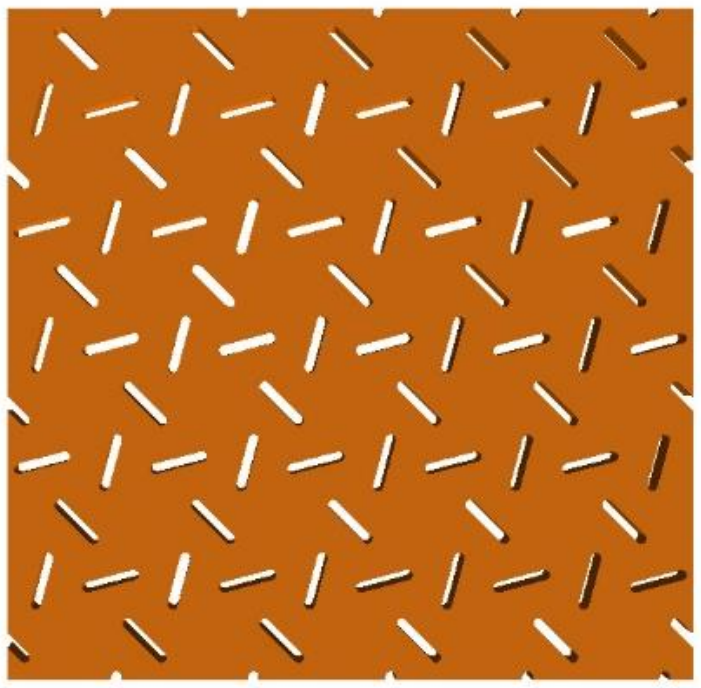 | [277] |
| Developed peanut auxetic structure | 2D | 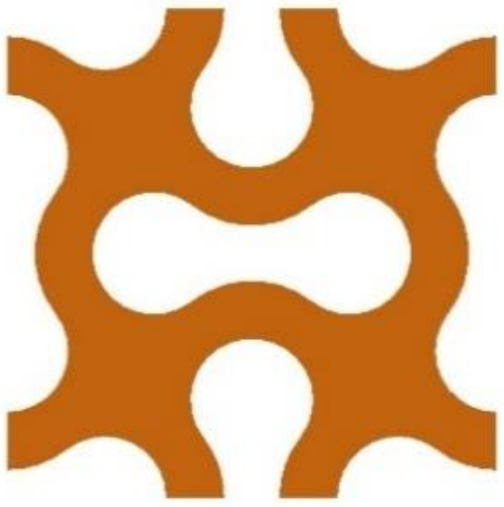 | 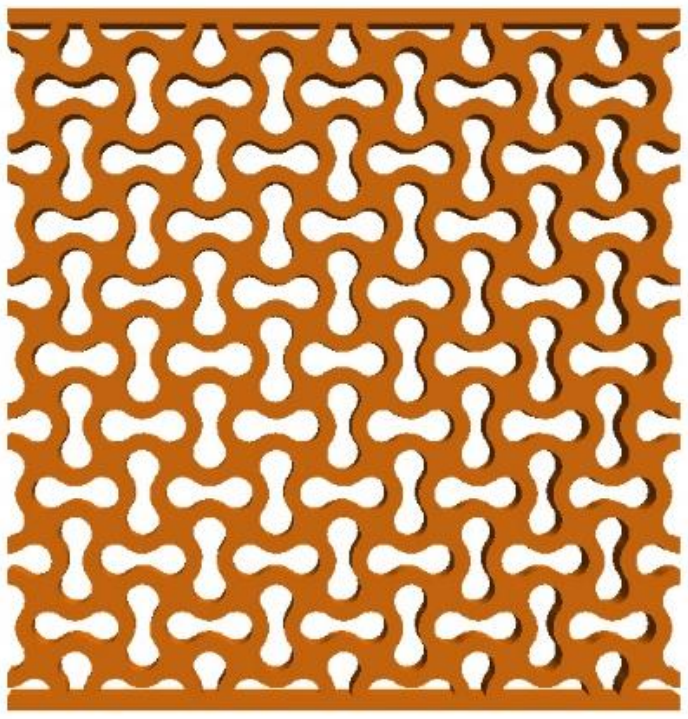 | [281] |
| Triangle peanut auxetic structure | 2D | 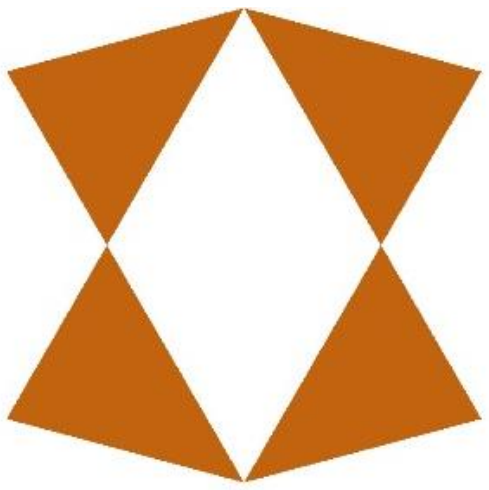 | 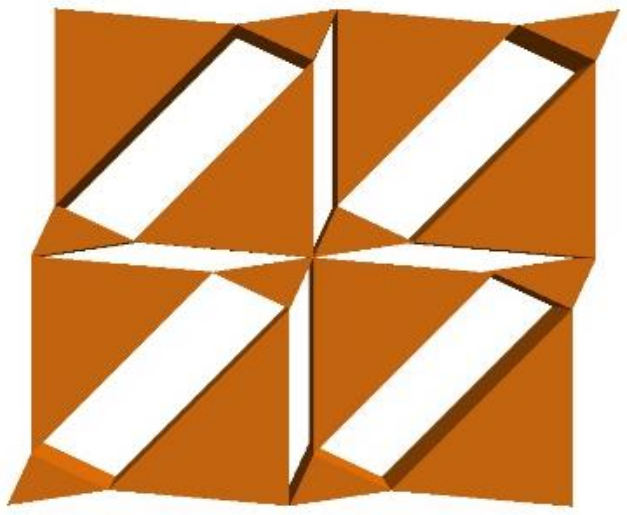 | [298] |

| | | | |
|---|---|---|---|
| Rectangular cube peanut auxetic structure | 3D | 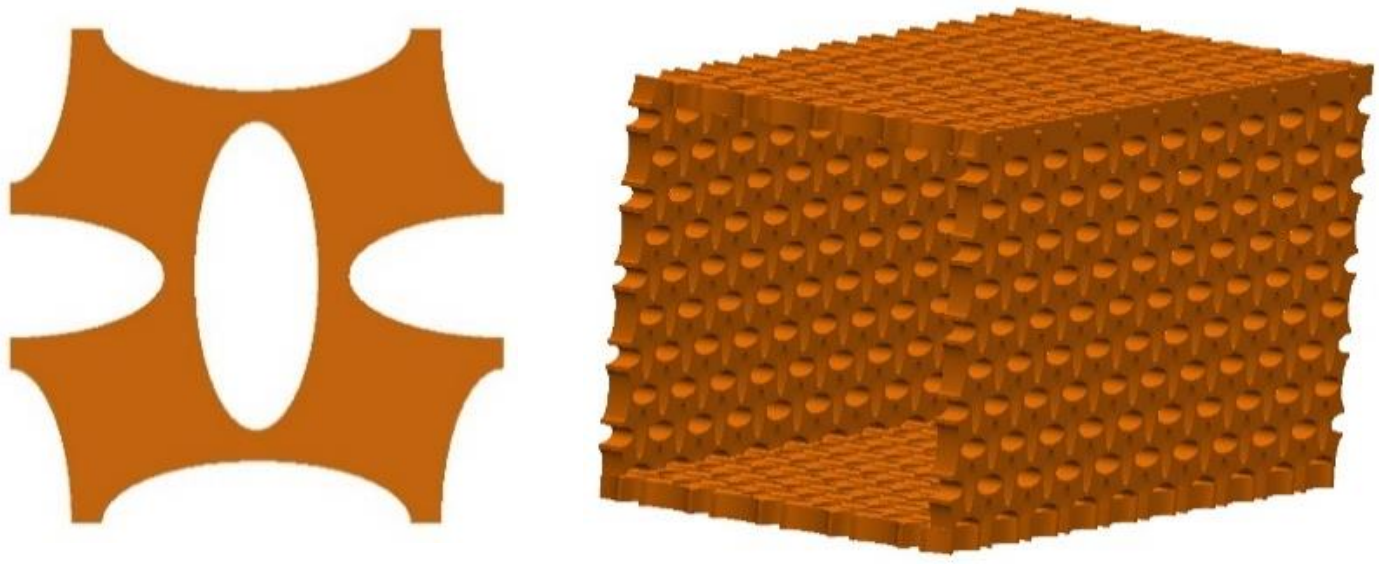 | [283] |
| 3D shell peanut auxetic structure | 3D | 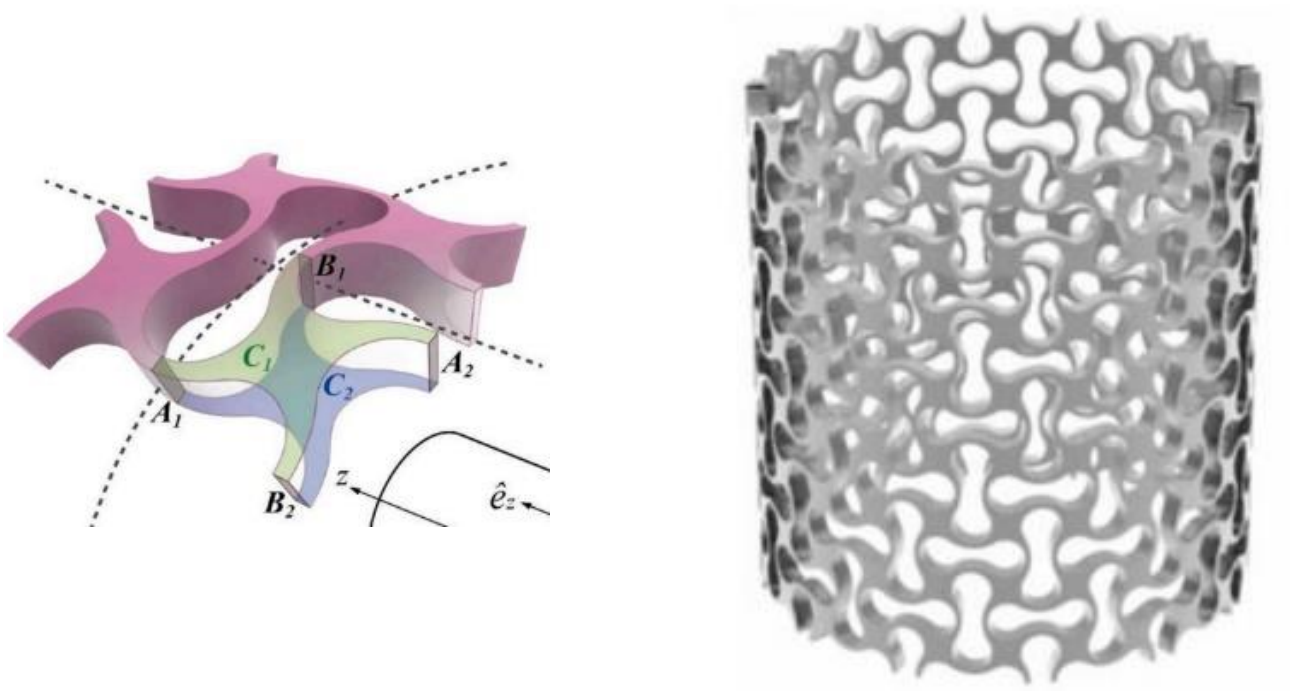 | [292] |
| Cubic peanut auxetic structure | 3D | 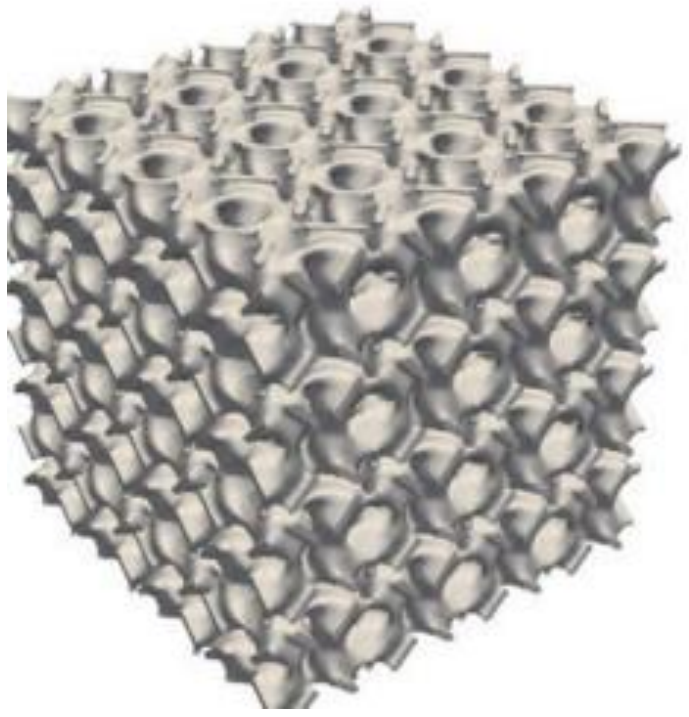 | [295] |

***2.7. Rotating rigid body AMS***

Rotating rigid body AMS has an interesting and relatively simple basic idea. This type of metastructures is created when an initial geometry can be repeated, each geometry rotates with a certain angle to its adjacent geometry to create a new structure. The initial geometry can be square, star, rectangle, and any other geometry that can be rotated. The initial idea of these structures to drive the auxetic behavior was born when rotation occurs in the initial geometry, the geometries return to their place during mechanical pressure, and in fact, by using this method, the auxetic behavior can be controlled.

The beginning of this category of metastructures dates back to 2004 when Grima et al. [233] were looking for new and promising ideas in the field of auxetic structures. For the first time, they presented these metastructures by using an initial square lattice structure, rotating each square unit to a certain amount, and analyzing its behavior. Then, in 2006, Grima and Evans [270] presented the rotating rigid body model on triangular geometries. In 2008, Attard and Grima [299] did their research on the way of rotation and the model of rotation in these structures and presented theoretical relationships. In recent years, this research has continued. In 2022, Grima-Cornish et al. [300] presented their research on models with auxetic behavior under thermomechanical effect. In 2023, Chen et al. [301] presented this model in an updated form as a programmable auxetic structure. Research in this field has been pursued more on the ability to control Poisson's ratio against a specific applied load [302,303]. The name of this category of metastructures has generally been “Rotating rigid body” in the articles, so the same name has been used for this category. The mechanical properties of this group of structures have been reported differently. These structures depend on the angle of rotation of these structures. Generally, Poisson's ratio is about -1 for these structures. Table 7 shows the types of introduced unit cells of this structure.

Table 7 Classification of rotating rigid body AMS

| Structure name | Model dimension | Unit cell | Aggregate of unit cells | Reference |
|---|---|---|---|---|
| Triangle rotating rigid body auxetic structure | 2D | 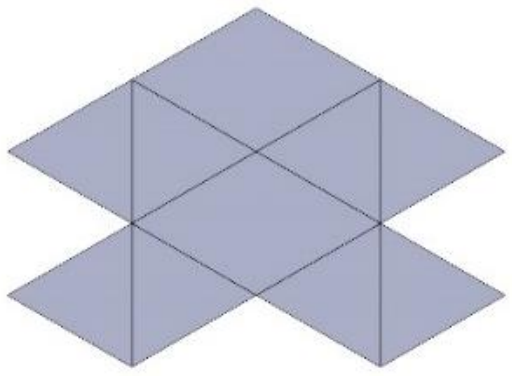 | 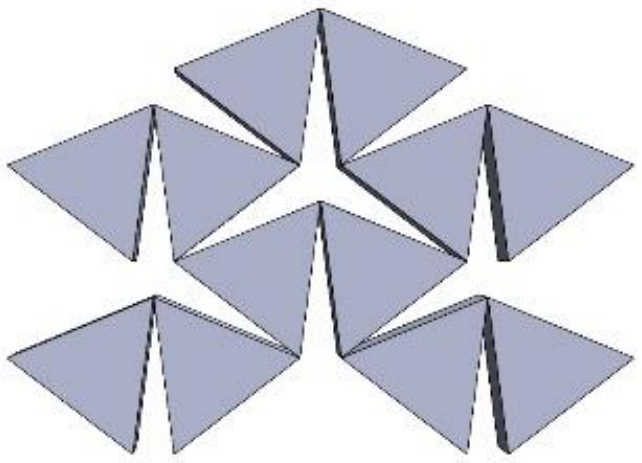 | [233] |
| Square rotating rigid body auxetic structure | 2D | 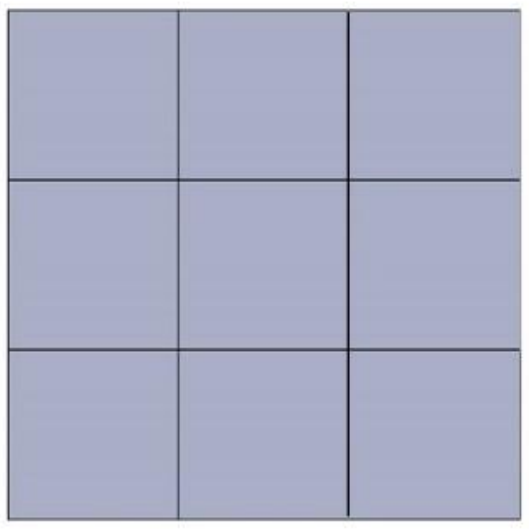 | 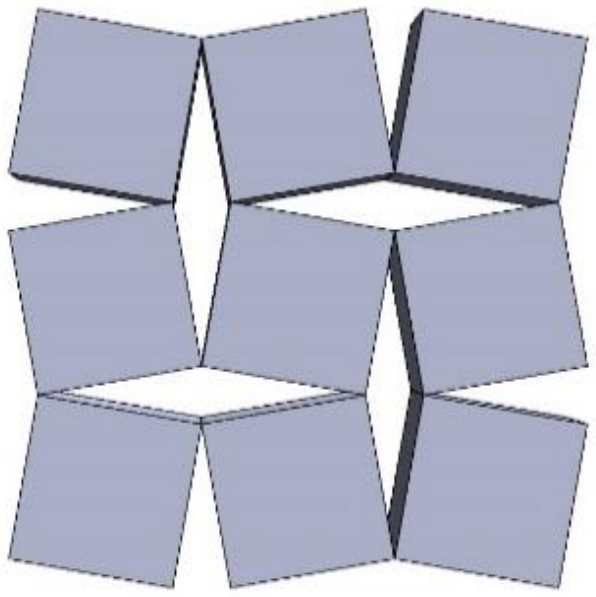 | [233] |
| Parallelogram rotating rigid body auxetic structure | 2D | 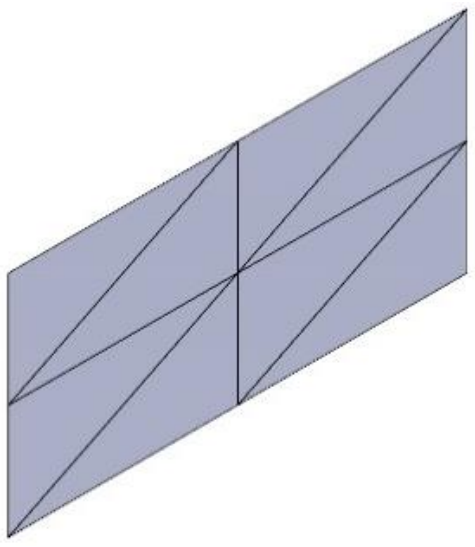 | 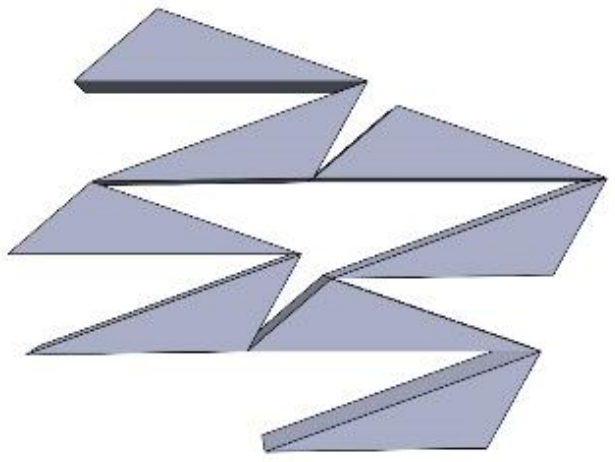 | [233] |
| Rectangle rotating rigid body auxetic structure | 2D | 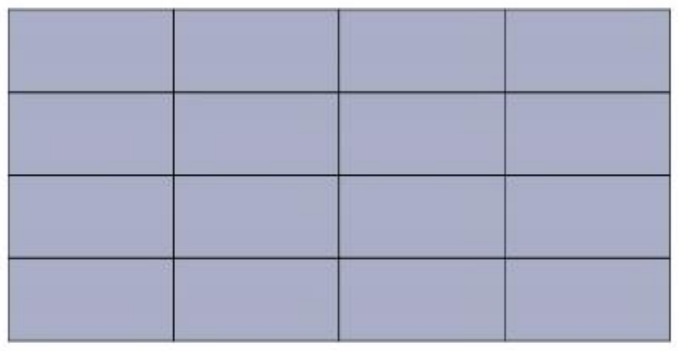 | 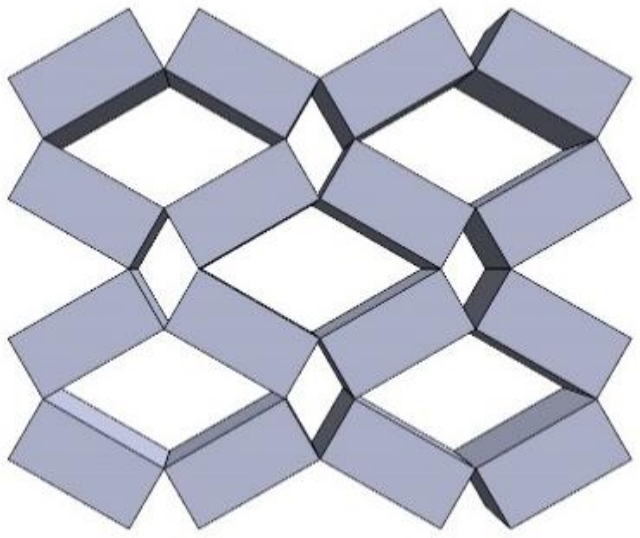 | [233] |
| Rectangle rotating rigid body auxetic structure | 2D | 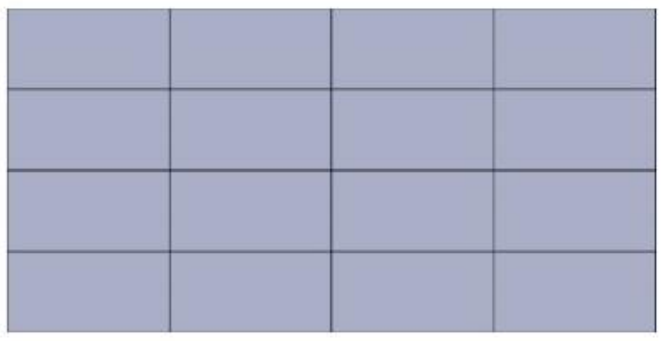 | 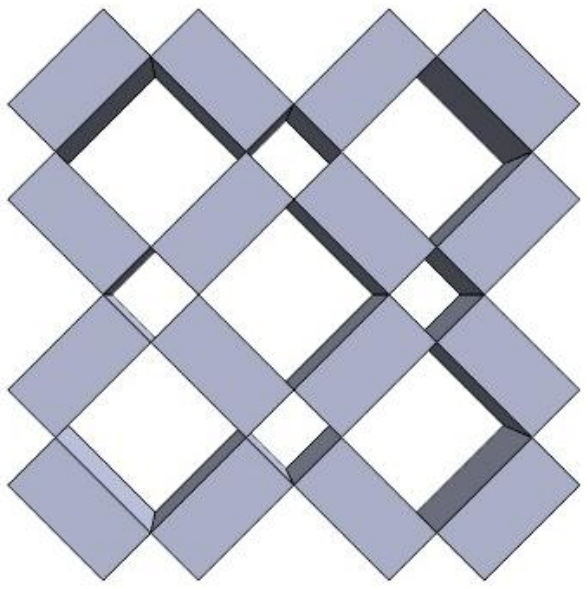 | [233] |
| Wavy Square rotating rigid body auxetic structure | 2D | 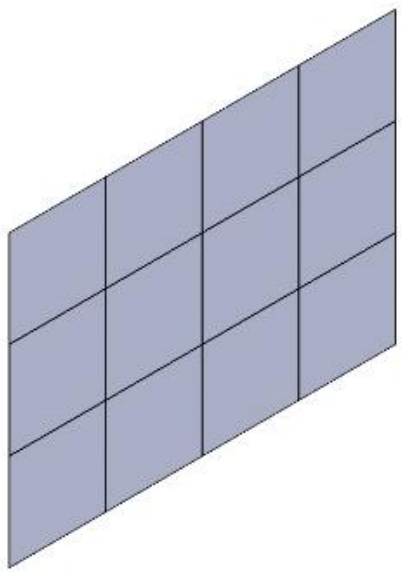 | 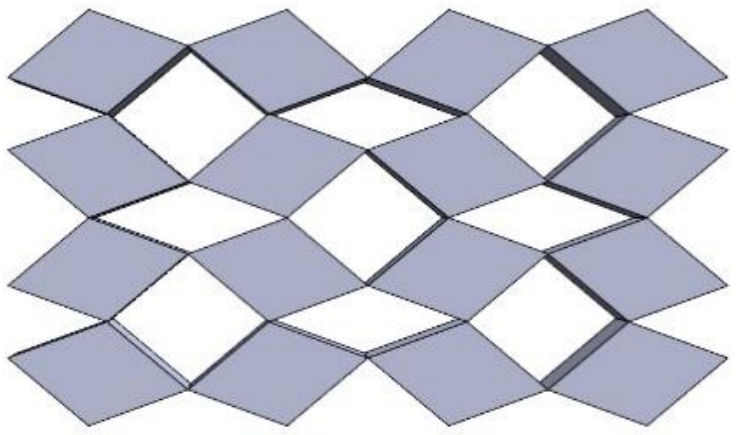 | [299] |
| Rectangle rotating rigid body auxetic structure | 2D | 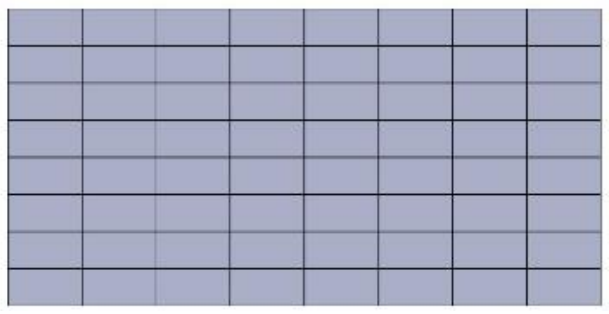 | 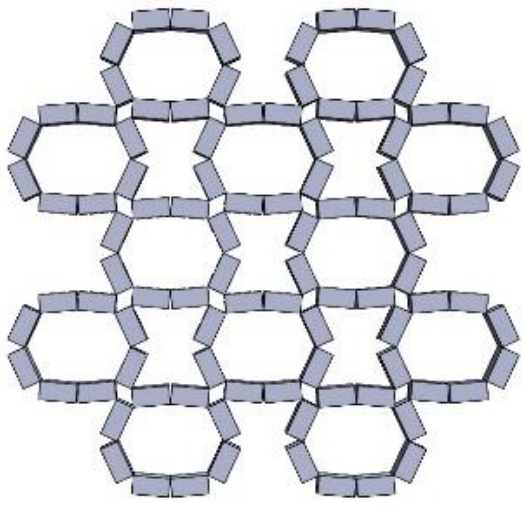 | [304] |
| Star rotating rigid body auxetic structure | 2D | 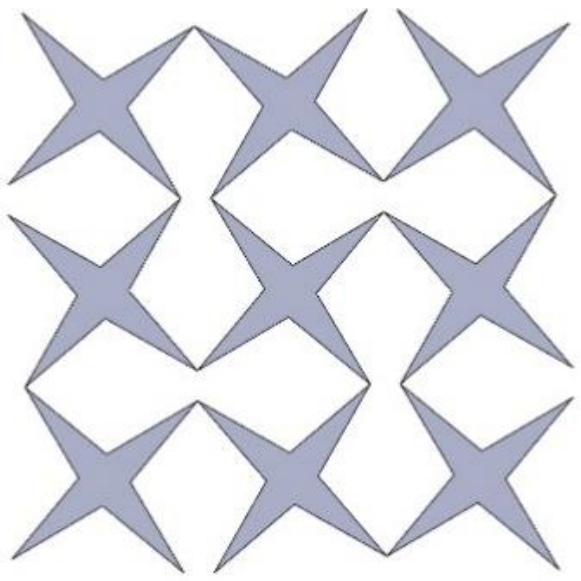 | 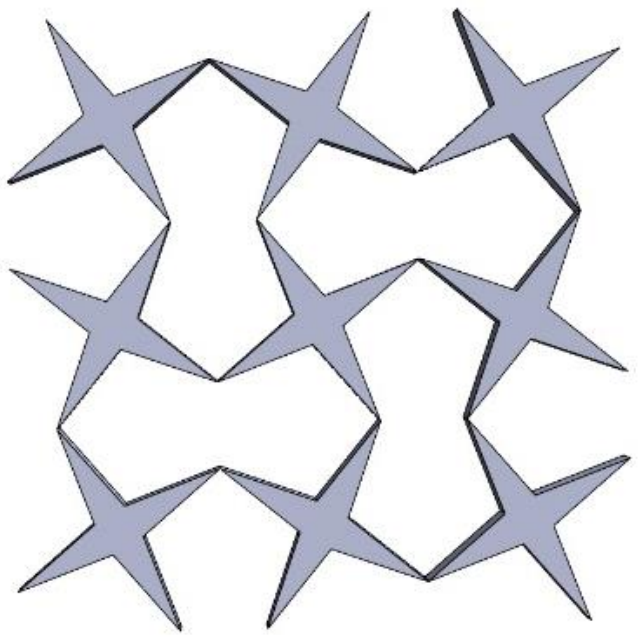 | [300] |
| Star rotating rigid body auxetic structure | 2D | 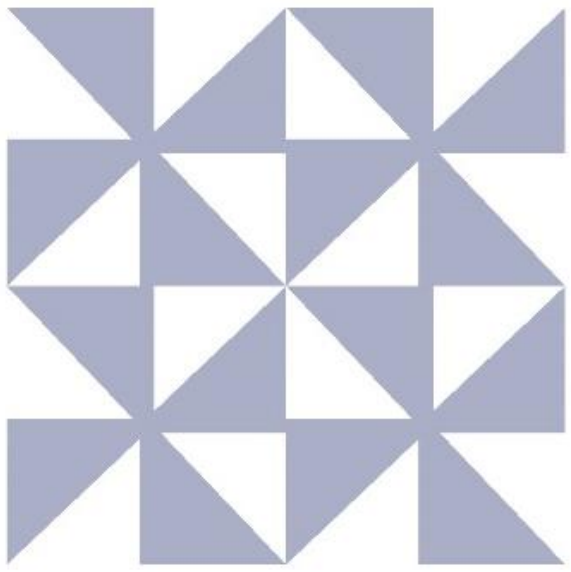 | 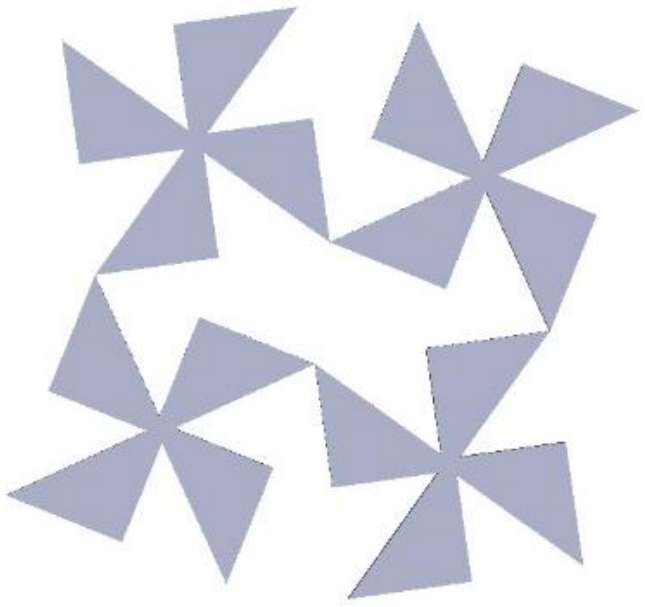 | [300] |
| Hierarchical rquare rotating rigid body auxetic structure | 2D | 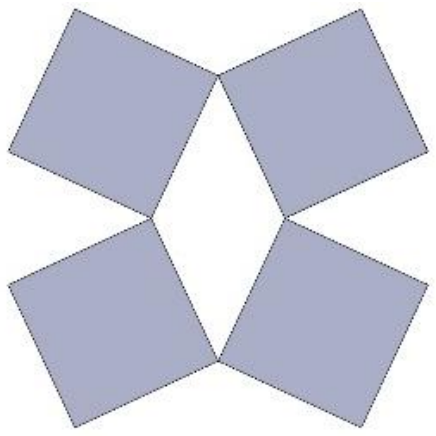 | 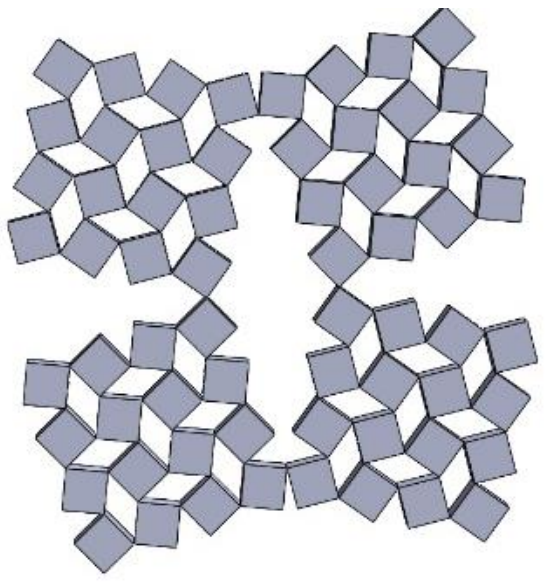 | [300] |
| Triangle rotating rigid body auxetic structure [301] | 2D | 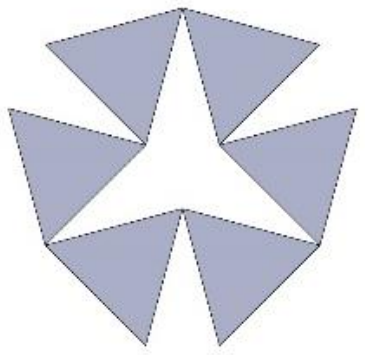 | 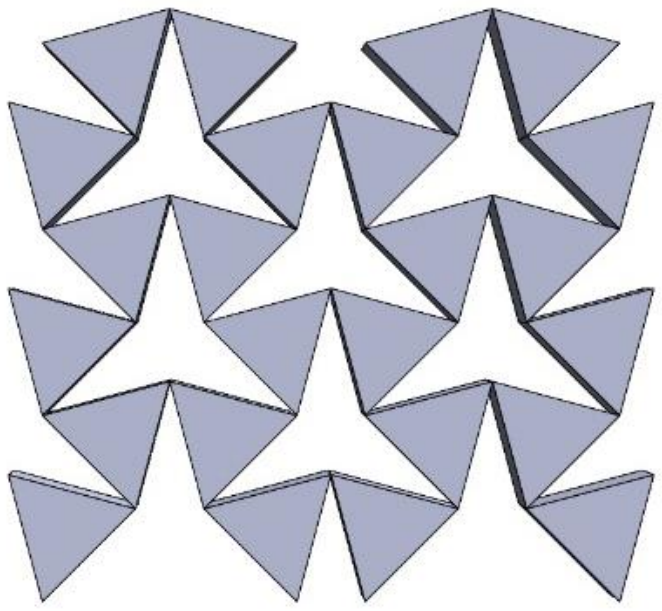 | [292] |
| Rectangle rotating rigid body auxetic structure | 2D | 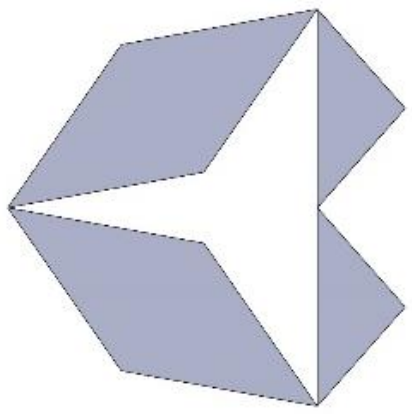 | 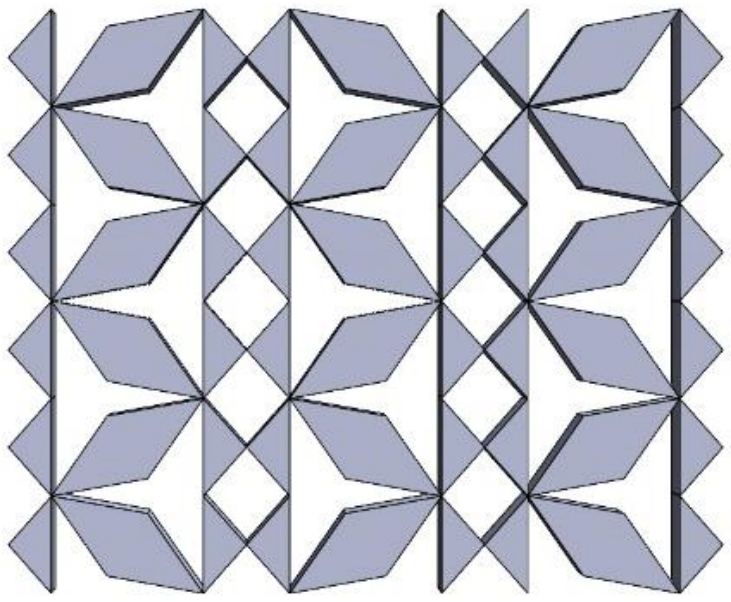 | [302] |

### *2.8. Origami AMSs*

Origami, rooted in the Japanese tradition of paper folding, has become a transformative approach in engineering and materials science. The integration of origami techniques with auxetic principles has led to innovative structures and materials with NPR properties. This concept can be classified into two primary categories: origami AMS and origami auxetic metamaterials. [305].

*Origami AMS*: Origami AMS consists of patterned sheets that, when folded using origami techniques, exhibit auxetic behavior. The NPR effect in these structures arises from the specific arrangement of two types of folds: mountain folds (convex) and valley folds (concave). Unlike traditional AMS that rely on mechanisms such as re-entrant geometries or rotating units, the auxeticity of origami structures is derived from their unique folding kinematics and geometric constraints [306–309]. A key example of origami AMS is the Miura-ori pattern, which exhibits auxetic behavior through its in-plane deformation when stretched or compressed. Another example, the Yoshimura pattern, achieves auxeticity via buckling mechanisms under axial loading. These structures are characterized by their ability to uniformly expand or contract in multiple directions due to their inherent fold geometries.

*Origami Auxetic Metamaterials:* Origami auxetic metamaterials take the concept further by employing origami-inspired folding principles to arrange and manipulate material layers. This approach often focuses on nanoscale or microscale materials, such as graphene, to create auxetic properties through precise structural arrangement. Sun et al. proposed a classification for kirigami-based metamaterials, identifying five primary groups that reflect the diversity of designs achievable with this methodology [310]. Recent studies have highlighted graphene as a key material for origami auxetic metamaterials, with research demonstrating its potential for achieving auxeticity at the atomic scale. These metamaterials are particularly significant for applications requiring lightweight, flexible, and tunable mechanical properties. Additionally, the exploration of thermomechanical behaviors has broadened the scope of origami-inspired metamaterials for multifunctional applications [311–318].

The primary distinction between origami AMS and other AMS lies in their principles of deformation. While metastructures such as re-entrant honeycombs rely on material-level deformation or rotational mechanisms, origami AMS derive their NPR from the geometric interplay of folds. This distinction underscores the structural focus of origami-inspired designs, where auxetic behavior is achieved through geometry rather than material composition [319–322]. For this reason, a centralized table for auxetic origami metastructures is not provided. By providing a comprehensive description of the configurations and principles behind typical origami and kirigami metastructures, this section addresses the growing interest in origami-inspired auxetics. The field's recent advancements highlight its potential for developing lightweight, flexible, and efficient materials and structures for diverse applications [323–326].

## 3. Comparison of NPR of various AMSs

The above literature review shows that properties such as energy absorption and crushing behaviour depend on the auxeticity of metastructure. This auxeticity originates from the internal architecture of a metastructure. Although establishing a relationship between internal architecture and Poisson's ratio is complex, understanding the NPR range of each category of AMS may help engineers in making informed design choices. This knowledge is crucial for optimizing structures in applications such as biomedical implants, and energy-absorbing materials.

To address this, a comprehensive review of NPR values for seven widely used metastructures available in the literature has been conducted, and a quantitative comparison is presented in Fig. 9. Among these, arc-shaped metastructures exhibit the broadest range of Poisson's ratios, making them highly versatile for applications requiring tunable mechanical properties. Peanut-shaped metastructures follow, demonstrating the second-largest NPR variation range. The other categories including re-entrant honeycomb, star-shaped, chiral, and rigid-body rotating metastructures exhibit a more limited NPR range, making them suitable for applications requiring controlled deformation characteristics. Notably, rigid-body rotating metastructures achieve the highest NPR values, which can be advantageous in applications demanding extreme auxeticity.

The variations in NPR among different metastructures suggest that their performance in terms of impact resistance, flexibility, and energy absorption varies significantly. While arc- and peanut-shaped metastructures provide broad design flexibility, the other categories may be preferred for applications where predictable and stable auxetic responses are required.

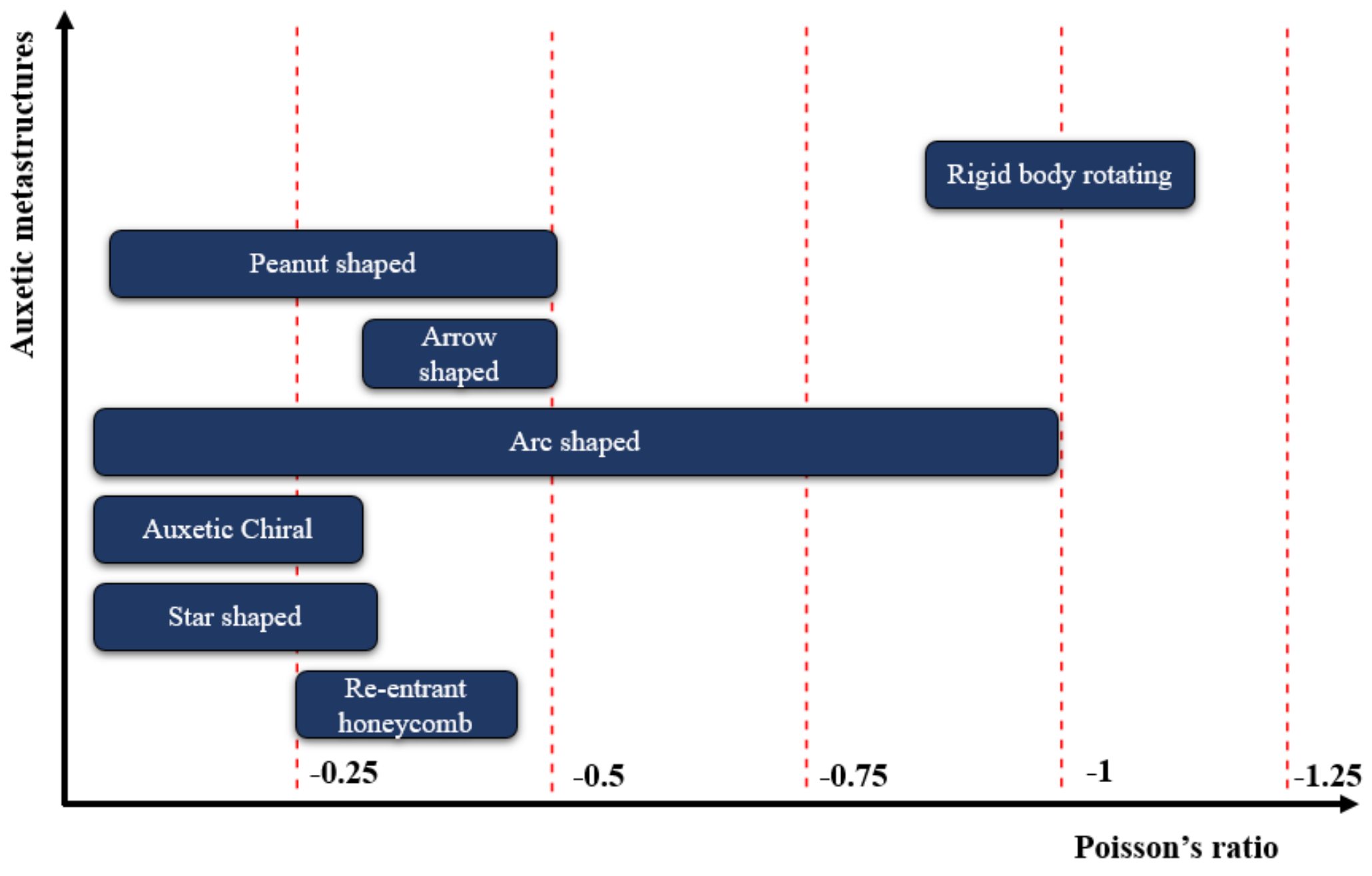

Fig. 9 Range of Poisson's ratio for various AMS.

## 4. Outlook and future work

The outlook for mechanical metastructures seems to be highly promising due to their unique properties. Since these advanced structures exhibit behaviors not typically found in natural materials, such as NPR, negative stiffness, zero stiffness, and high energy absorption, it is very likely to observe considerable research and development activities following the previous and current ongoing developments. Besides, advanced manufacturing techniques such as 3D printing will facilitate further development of innovative metastructures. In the following, we will elaborate on our opinion about future perspectives of this field in three main areas: applications, materials, and design for which a summary is shown in Fig. 10.

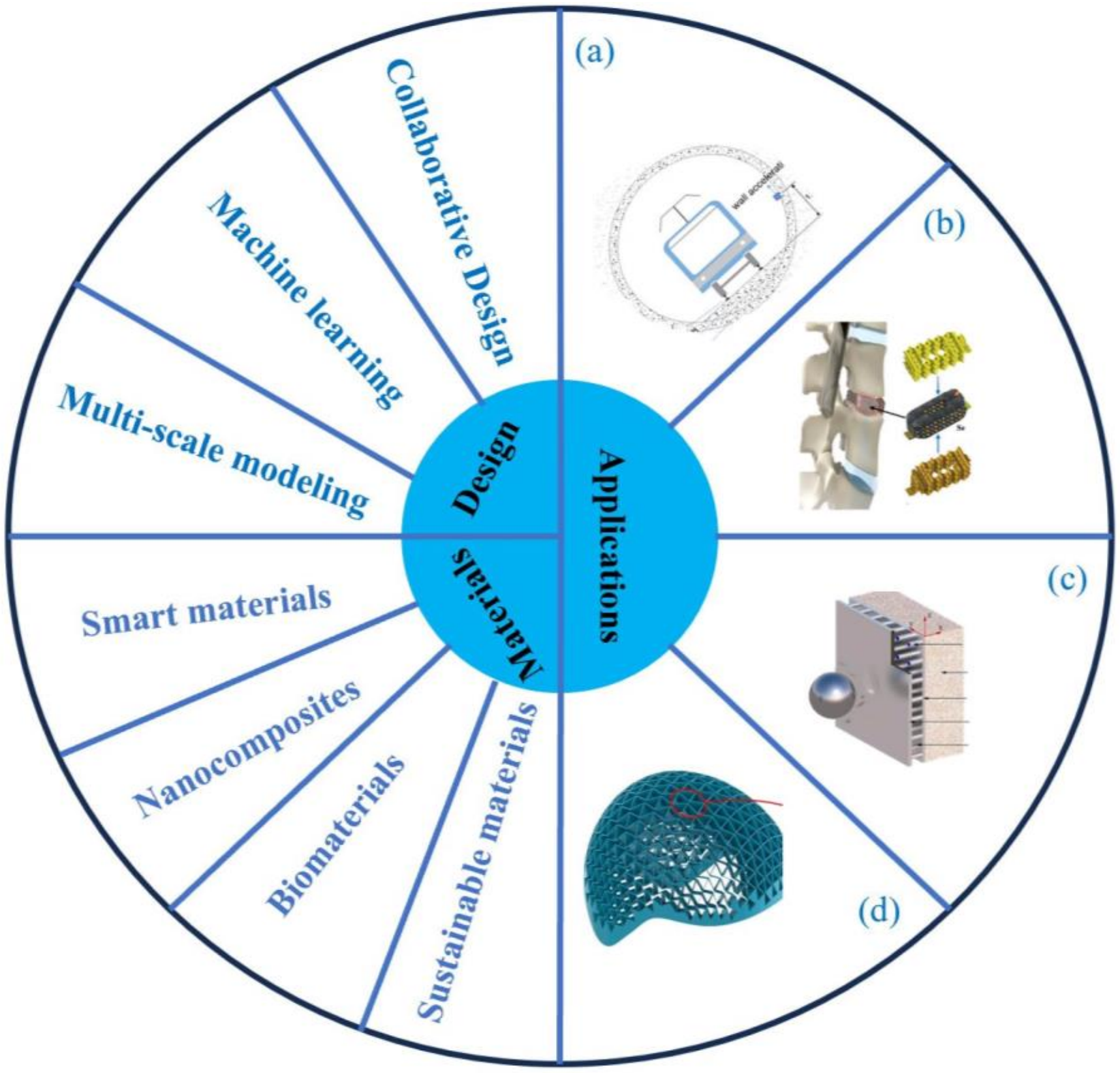

Fig. 10 Future of matastructures in design, materials, and applications: (a) Transportations [327], (b) Bio-mechanics applications [328], (c) Civil applications [329], and (d) Sport protective equipment [330].

### *4.1. Applications*

During the last decade, mechanical metastructures have found interesting applications in different industries (see e.g., [327,329]). We believe that further significant applications will be developed in different sectors, including:

- **Transportation:** Metastructures can improve the performance and safety of vehicles by offering lightweight yet strong materials that can absorb impact energy effectively. In automotive safety, for example, integrating auxetic foams into vehicle crash structures has shown significant improvements in energy absorption per unit mass compared to traditional impact-absorbing materials shown in fig 11. Studies have shown that auxetic foams can absorb up to 50% more energy per unit mass compared to traditional impact-absorbing materials [331]. Designing a new crash box with auxetic metastructures can enhance vehicle crash performance by significantly improving energy absorption. This advancement could lead to lighter, more efficient crash box designs, improving overall vehicle safety.

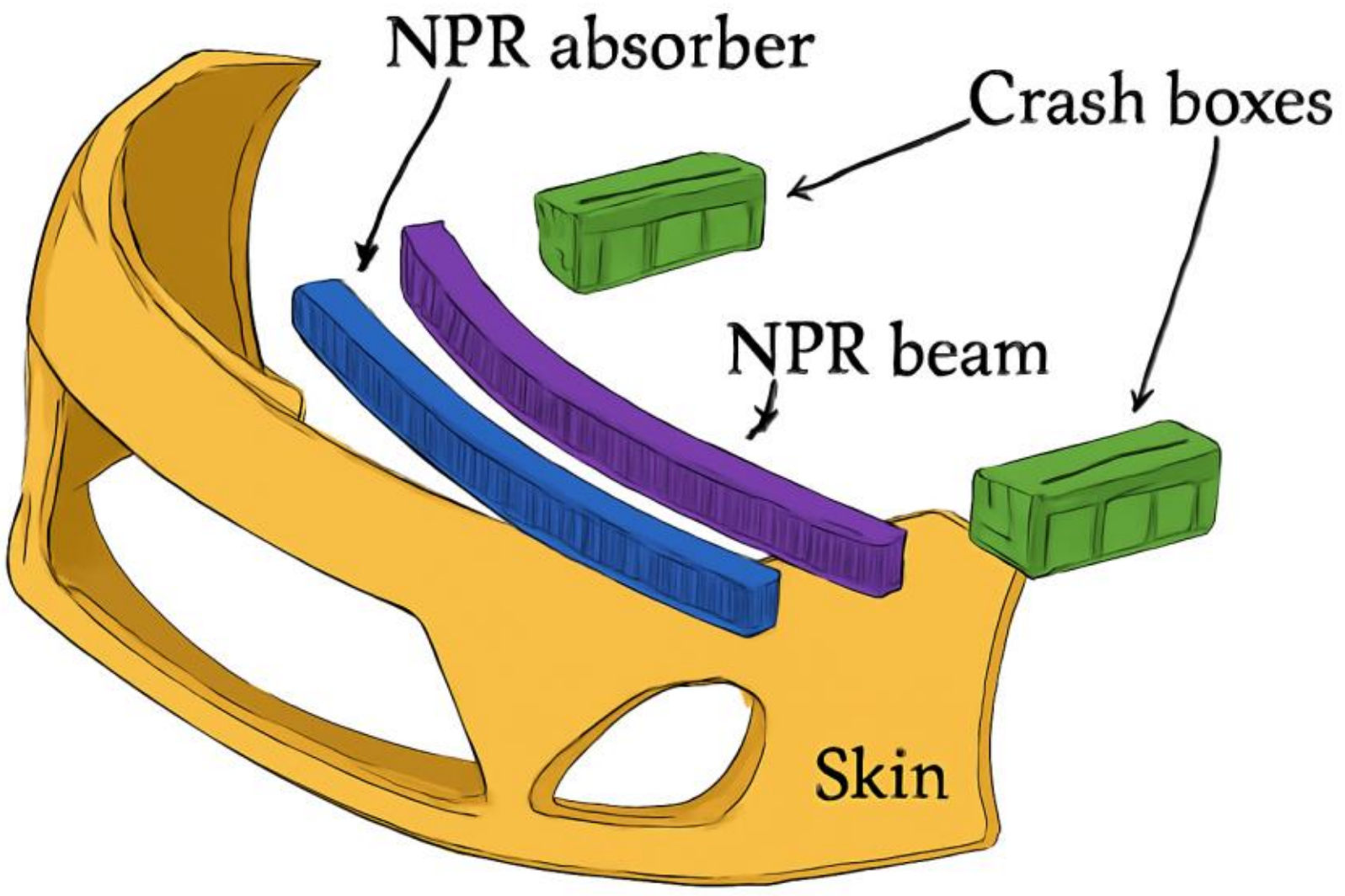

Fig. 11 Using NPR absorber and beam in bumper for incrasing safety [332]

- **Biomedical engineering**: Customizable mechanical properties make metastructures ideal for implants and prosthetics, providing better integration with biological tissues and enhanced durability (see [333] for a review of recent developments). Personalized auxetic implants with optimized mechanical gradients have demonstrated increased biomechanical compatibility with human bone structures shown in fig 12, resulting in reduced implant failure rates.

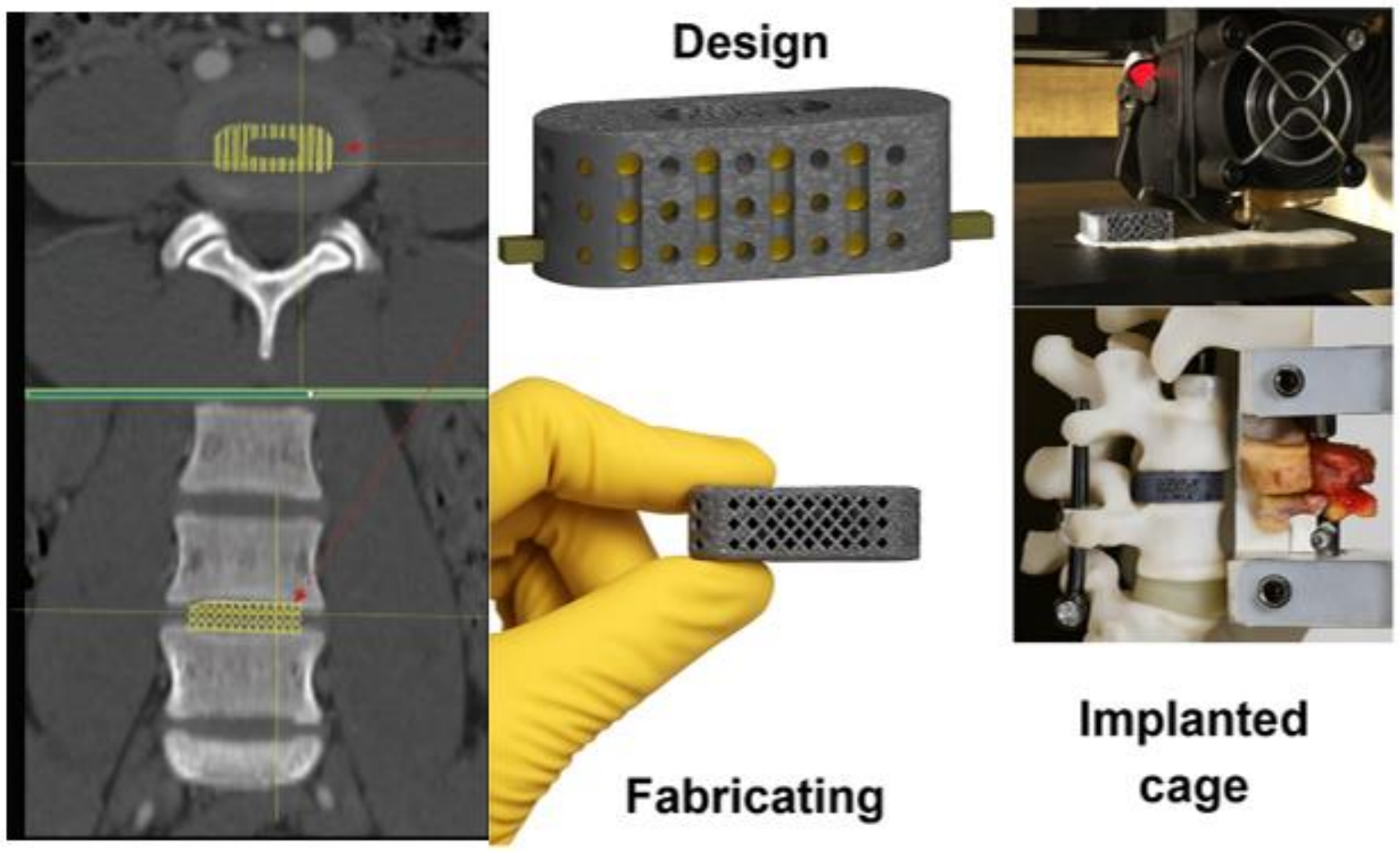

Fig. 12 AMS application in bone implanted structure [328]

- **Civil engineering**: Metastructures can be used in construction for earthquake-resistant buildings and structures, offering improved resilience and energy dissipation (see [329] for recently developed applications). As an example, adaptive auxetic panels, capable of altering their stiffness based on real-time external loads, offering an innovative solution for earthquake-resistant structures, are being investigated [334]. Auxetic road guardrails (e.g., [195]), and bridge pier protection (e.g., [335,336]) are other examples of AMS applications.
- **Consumer goods**: Enhanced material properties can lead to better performance in sports equipment and other consumer products. As an example, Zamani et al. [330] recently designed, fabricated and tested a bike helmet using an auxetic re-entrant metastructure for the helmet liner. An interested reader on this subject is referred to [337] for a review on sports helmets [338].

### *4.2. Materials*

Currently, different categories of materials including metals, ceramics, polymers, and composites are used to develop metastructures (see e.g., [339]). The following classes of materials will potentially be used for the next generation of metastructures:

- **Smart materials**: Incorporating materials that respond to external stimuli such as environmental changes (e.g., temperature and humidity) can help to create adaptive metastructures and multifunctional designs. By combining auxetic behavior with self-healing polymers, shape-memory alloys, and adaptive composites, researchers can create metastructures that respond dynamically to changing environmental conditions. Such integrations could lead to breakthroughs in self-regulating impact absorption systems, aerospace components, and bio-mimetic adaptive materials. See [340] for a review on smart materials and 4D printing.

- **Nanocomposites:** Using nanocomposites in metastructures could result in improved mechanical properties such as higher strength and flexibility. For an overview on polymer nanocomposites, an interested reader is referred to [341].
- **Biomaterials:** Developing bio-inspired and biocompatible materials for medical applications will offer superior integration and performance in biological environments. See [342] for some recent developments.
- **Sustainable materials**: Using eco-friendly and recyclable materials will result in reduced environmental impact, particularly important for large-scale industrial applications. There are some recent developments using biopolymers [343], and future applications using bio-composites will be potential further developments.

It should also be mentioned that recent breakthroughs in multi-material 3D printing (see e.g., [344]) have enabled the creation of auxetic structures with graded material properties, achieving significant improvements in mechanical performance compared to usual material.

### *4.3. Design*

The design process of metastructures is likely to evolve significantly due to advancements in several areas:

- **Multi-Scale modeling**: Advanced multi-scale modeling techniques will enable designers to understand and predict the behavior of metastructures at different length scales ranging from the nano- to the macro-structure. Using multi-scale models, it is possible to obtain not only accurate but also physics-based behavior of a material point considering its underlying micro-structure and nano-structure (see e.g., [345] for amorphous polymeric materials). This can be then used to design metastructures with desired properties across different scales. An interested reader is referred to [346] for an overview of modeling techniques that can be used to design metamaterials and optimize their performance.
- **Machine learning techniques**: Machine learning methods, such as Artificial Neural Networks (ANNs), have enabled the development of highly accurate and remarkably efficient surrogate models (see e.g., [347]). These models can generate and test thousands of design iterations quickly, finding optimal solutions that balance performance and weight. See [348] for a recent study which uses

Convolutional Neural Networks (CNN) for designing metastructures with desired anisotropic properties. Besides, generative models, such as Generative Adversarial Networks (GANs), can be sued for material design purposes [349]. We believe that the combination of machine learning-powered design methodologies and next-generation additive manufacturing techniques will play a crucial role in expanding the practical applications of AMS across multiple industries.

- **Collaborative design platforms**: Cloud-based collaborative platforms will allow designers, engineers, and researchers from different fields and locations to work together more efficiently, sharing data and insights in real time. There are already such platforms for designing mechanical objects, and architecture and construction. Extending currently active platforms, or developing new ones, for metastructures will be another possibility for innovative designs.

While laboratory-scale prototypes have demonstrated exceptional performance, large-scale manufacturing methods must still address cost-effectiveness, material sustainability, and production consistency. Future research should focus on optimizing material compositions, hybrid structures, and automated manufacturing processes to ensure that AMSs developed in laboratories find widespread industrial applications. Additionally, efforts in standardization and regulatory frameworks will be necessary to facilitate their adoption in safety-critical sectors such as aerospace, healthcare, and infrastructure. A clearer roadmap with specific performance targets and industry collaborations will provide meaningful direction for the future development of AMS.

## 5. Concluding remarks

This article provides a thorough overview of internal architecture of existing AMS with a unit cell repetition pattern. Reviewed AMSs are divided into eight categories: Re-entrant honeycomb, star-shaped, arc-shaped, arrow-shaped, chiral, peanut -shaped, rotating rigid body, and origami. In each category, the basic idea of each metastructure, which distinguishes them, was addressed. Then, the primary or basic unit cell of each category was examined and its development process was introduced. This study provides summarized and beneficial information on all unit cells proposed in the literature up to now. A qualitative comparison among the Poisson's ratios of seven widely used auxetic categories is performed. It helps designers and engineers to select appropriate metastructure for different applications. Also, an attempt is made to present an outlook of applications, materials, and design of AMS. The unit cells of AMS have been mostly developed by researchers in academia, but due to their very interesting properties, such as high energy absorption and improved toughness, an increasing trend in industrial applications is observed, as well. This review can be very helpful for industrial users to understand different metastructures and make an informed choice for their particular application.

So far, the development of AMSs has primarily relied on repeating unit cell models. However, it is also possible to design unit cells that do not follow a repeating pattern yet still exhibit auxetic behavior, which presents a promising direction for future research.

## Data availability

No data was used for the research described in the article.